\documentclass[fleqn,usenatbib]{mnras}

\usepackage{newtxtext,newtxmath}

\usepackage[T1]{fontenc}

\DeclareRobustCommand{\VAN}[3]{#2}
\let\VANthebibliography\thebibliography
\def\thebibliography{\DeclareRobustCommand{\VAN}[3]{##3}\VANthebibliography}

\usepackage{graphicx}	
\usepackage{amsmath}	

\usepackage{amssymb}
\usepackage{url}

\usepackage{booktabs}
\usepackage{tabularray}
\usepackage{threeparttable}
\usepackage{tablefootnote}
\usepackage{siunitx}

\usepackage{booktabs}
\usepackage{multirow}

\usepackage{placeins}

\usepackage{lscape}    
\usepackage{adjustbox} 
\usepackage{caption}   

\usepackage{hyperref}
\usepackage{natbib}

\usepackage{float}
\usepackage{subfigure}
\usepackage{placeins}

\usepackage{afterpage}

\title[Long-period low mass-ratio contact binaries]{Photometric and Spectroscopic Studies of Four Low Mass-ratio Contact Binaries with Period Longer than 0.7 days}

\author[Yi-Fan Wang et al.]{
Yi-Fan Wang,$^{1}$
Kai Li,$^{1}$\thanks{E-mail: kaili@sdu.edu.cn}
Fei Liu,$^{1}$
Xin Xu,$^{1}$
Mu-Zi-Mei Li,$^{1}$
Cheng-Yu Wu,$^{1}$
Yu-Tong Li,$^{2}$\thanks{E-mail: qiuwutongyu622@163.com}
Yan-Ke Tang,$^{2}$
\newauthor
Xing Gao$^{3}$
and Guo-You Sun$^{4}$\\
$^{1}$Shandong Key Laboratory of Space Environment and Exploration Technology, Institute of Space Sciences, School of Space Science and Technology, Shandong\\ University, Shandong, China\\
$^{2}$Shandong Key Laboratory of Space Environment and Exploration Technology, College of Physics and Electronic information, Dezhou University, 566 West\\ University Road, Decheng District, Dezhou 253023, China\\
$^{3}$Xinjiang Astronomical Observatory, 150 Science 1-Street, Urumqi 830011, People’s Republic of China\\
$^{4}$Xingming Observatory, Urumqi 830002, Xinjiang, People’s Republic of China}

\date{Accepted XXX. Received YYY; in original form ZZZ}

\pubyear{\the\year{}}

\begin{document}
\label{firstpage}
\pagerange{\pageref{firstpage}--\pageref{lastpage}}
\maketitle

\begin{abstract}
Photometric and spectroscopic studies of four long-period low mass-ratio contact binaries, V0508 And, V0844 Aur, V0699 Cep, and NSVS 6259046, are performed. V0508 And, V0844 Aur, and V0699 Cep are found to be A-type low-mass-ratio medium-contact binaries, while NSVS 6259046 is found to be an A-type deep-contact binary. $O - C$ analysis indicates no long-term variation in V0844 Aur. However, the orbital periods of the other three targets are increasing. We conclude that V0844 Aur, V0699 Cep and NSVS 6259046 are magnetically active, as evidenced by the presence and variable nature of the O’Connell effect in these systems. By analyzing the LAMOST spectroscopic data, we find out that NSVS 6259046 and V0508 And exhibit no chromospheric activity on the dates the LAMOST spectra were taken, while the low signal-to-noise ratio in LAMOST data for V0844 Aur prevents us from obtaining reliable results. We discover that V0699 Cep is an early-type contact binary with chromospheric activity.
Their initial masses and ages are calculated. All four systems are determined to be currently stable.
We collect 217 contact binaries with both spectroscopic and photometric observations, and compare the differences between short-period and long-period systems in terms of mass–luminosity relation and mass–radius relation, using 0.7 days as the period boundary.
\end{abstract}




\section{Introduction}
Contact binaries serve as fundamental astrophysical laboratories for exploring stellar structure, formation mechanisms, and evolutionary pathways \citep{2003MNRAS.342.1260Q,2005ApJ...629.1055Y,2007ApJ...662..596L,2008MNRAS.386.1756E,2012JASS...29..145E,2024RAA....24a5002P}. Numerous unresolved questions still exist after decades, including the orbital period cutoff \citep{2019MNRAS.485.4588L,2023MNRAS.523.1394Z}, mass ratio limit \citep{1976MNRAS.176...81W,1995ApJ...444L..41R,2007MNRAS.377.1635A,2009MNRAS.394..501A,2022AJ....164..202L,2024A&A...692L...4L}, and the O’Connell effect \citep{1951PRCO....2...85O}, all of which require more investigation.

In the galactic disk, one in every 500 F,G,K-type main-sequence stars is considered to be a contact binary (CB) \citep{2006MNRAS.368.1319R}. CBs are systems consisting of two components covered by a common convective envelope \citep{1968ApJ...151.1123L}, meaning that their surface temperatures are the same. The early known samples of CBs exhibited two key limitations: their numbers were substantially smaller, and due to the lower photometric precision at that time, the samples were biased toward systems with higher inclination angles. As clearly demonstrated in Figure 2 of \cite{2022ApJS..262...12K}, in such high-inclination configurations, the primary minimum is roughly equal to the second minimum \citep{1941ApJ....93..133K,Kraft_1967,1967AJ.....72Q.813L,1968ApJ...151.1123L}.
Due to the common convective envelope, there is energy and mass transfer between the two components, leading to changes in orbital period \citep[e.g.][]{2015AJ....149..120L,Liao_2017,2018PASP..130c4201L,2019RAA....19..147L}, and theories of thermal relaxation oscillations \citep[e.g.][]{1976ApJ...205..217F,1976ApJ...205..208L,1977MNRAS.179..359R}.
CBs are classified into two types \citep{1970VA.....12..217B}: type A and type W. For type A, the surface temperature of the more massive star is higher than that of the less massive star. Conversely, type W corresponds to the opposite situation compared to type A.

Magnetic activities in contact binaries include star spots, flares, and plages \citep{2003ChJAS...3..361P}. These phenomena are closely related to convective motions and rapid stellar rotation \citep{2009A&ARv..17..251S}.
Magnetic activity may cause noticeable asymmetry in the light curve maxima, a phenomenon known as the O'Connell effect \citep{1951PRCO....2...85O}. It is attributed to mechanisms such as cool spots due to magnetic activity \citep{2016Ap&SS.361...63L}, mass accretion \citep{shaw1994near,2013PASJ...65....1P}, circumstellar material \citep{2003ChJAA...3..142L}, and asymmetry of circumfluence from the Coriolis effect \citep{1990ApJ...355..271Z}.
Magnetic activities can influence the orbital period of CBs noticeable \citep{1992ApJ...385..621A}.
As for spectroscopic analysis, there are special emission lines indicating chromospheric activity, such as Ca~\textnormal{II} H\&K, Balmer series, and Ca~\textnormal{II} triplet above the continuum \citep{2018A&A...615A.120P,2019AJ....158..193C,2019MNRAS.487.5520L,2021MNRAS.506.4251Z}. Spectroscopic emission lines such as $H_\alpha$ provide direct evidence of intense magnetic activities on the stellar surface, often in conjunction with rapid rotation \citep{1978ApJ...226..379W}.

The study of mass ratio and orbital period is important for understanding the evolution of CB. \cite{2019AAS...23344805M,2022AAS...24030801M} developed angular momentum conserving models to explain the evolution of CBs, showing a relationship between orbital period and the lower limit of the mass ratio. Based on the models above, \cite{2022ApJS..262...12K} analyzed the relationship of mass ratios and orbital periods in CBs. They found that the critical instability mass ratio increases from 0.045 to 0.15 as the orbital period grows from 0.74 to 2.0 days. However, research on CBs with orbital periods longer than 0.7 days are currently rare \citep{2019AAS...23344805M,2022AAS...24030801M,2024MNRAS.527.6406L,2025MNRAS.537.2258L}. While the evolutionary processes of all CBs still present numerous unresolved questions, the evolutionary pathways of long-period systems are particularly poorly understood. Further research on long-period low mass-ratio CBs is necessary.

To study long-period low mass-ratio CBs, we select four totally eclipsing CBs based on specific criteria: orbital periods > 0.7 days and amplitudes < 0.4 mag. The amplitude criterion is motivated by \cite{2001AJ....122.1007R}, who established a correlation between amplitude and mass ratio, implying that small-amplitude systems are likely to have low mass ratios. Guided by this relationship, we identify four candidate systems from the All Sky Automated Survey for SuperNovae (ASAS-SN) \citep{2014ApJ...788...48S,2017PASP..129j4502K,2019MNRAS.486.1907J}. The fundamental parameters of the selected CBs are summarized in Table \ref{Tabel:The ASAS-SN name, period, mean V-band magnitude and color indices of the four investigated targets.}.

\begin{table*}
	\centering
	\caption{The ASAS-SN name, period, mean V-band magnitude and color indices of the four investigated targets.}
        \label{Tabel:The ASAS-SN name, period, mean V-band magnitude and color indices of the four investigated targets.}
	\begin{tabular}{lllllll}
		\hline
		Targets                       & ASAS-SN name      & Period(d) & Mean V(mag) & B-V(mag) & J-K(mag) & Reference  \\
		\hline
        V0508 And      & ASASSN-V J004744.15+360223.3 & 0.7752302 & 12.11       & 0.390    & 0.254    & \cite{2019MNRAS.486.1907J}  \\
        V0844 Aur      & ASASSN-V J054305.62+530235.7 & 0.8222160 & 12.04       & 0.910    & 0.310    & \cite{2019MNRAS.486.1907J}  \\
		V0699 Cep      & ASASSN-V J224600.69+574649.6   & 0.8050727 & 11.72       & 0.859    & 0.394    & \cite{2019MNRAS.486.1907J}      \\
        NSVS 6259046     & ASASSN-V J232745.98+245256.9 & 0.7344176 & 12.69       & 0.550    & 0.309    & \cite{2019MNRAS.486.1907J}  \\
		\hline
	\end{tabular}
\end{table*}

\section{Observations}

\subsection{Photometry}
\label{sec:Photometry} 
The photometric observations of four binary star systems are conducted using the Ningbo Bureau of Education Telescope and the Xinjiang Observatory Telescope (NEXT) at the Xingming Observatory.
The NEXT features a 60 cm aperture telescope and is equipped with an FLI PL23042 back-illuminated CCD camera that has a 2048 × 2048 pixel array, providing a field of view of approximately $22^\prime$ × $22^\prime$. All raw images are processed for bias, dark, and flat corrections using \href{https://c-munipack.sourceforge.net/}{C-MuniPack} software designed for photometric data processing. Aperture and differential photometry methods are employed for data analysis.
Details of our observation are presented in Table \ref{Table:The observating date, exposure time, comparison and check star of the four investigated targets.}. Subsequently, differential magnitudes are calculated between the target object and the comparision star, as well as between the comparision star and the check star. The photometric data are presented in Table \ref{Table:The photometric observation obtained by NEXT}, where $\Delta m$ represents the instrumental magnitude difference between the variable star and the comparison star. All four targets are total-eclipse CBs, with total eclipses lasting 0.123 of an orbit for V0508 And, 0.058 for V0844 Aur, 0.039 for V0699 Cep and 0.148 for NSVS 6259046.

To get a longer time baseline for $O - C$ analysis, we proceed to leverage the archival photometry. Our four systems have also been observed by various photometric surveys, including the Transiting Exoplanet Survey Satellite (TESS) \citep{2015JATIS...1a4003R}, the All-Sky Automated Survey for Supernovae (ASAS-SN) \citep{2014ApJ...788...48S,2019MNRAS.486.1907J}, the Super Wide Angle Search for Planets (SuperWASP) \citep{2010A&A...520L..10B}, and the Zwicky Transient Facility (ZTF) \citep{2019PASP..131a8002B,2019PASP..131a8003M}.

\begin{table*}                                          
    \centering                                                   
    \caption{The observating date, exposure time, comparison and check stars of the four investigated targets.}       
    \label{Table:The observating date, exposure time, comparison and check star of the four investigated targets.}      
    \begin{tabular}{lllll}                                                
        \hline                                                    
        Target & Observing Date & Exposure Time (s) & Comparison Star & Check Star \\
        \hline                      
        V0508 And         & 2022 Nov 02,03,12,13,14  & g22 i25 r20 & TYC 2288-1478-1 & Gaia DR3 364060257146400384 \\                       
        V0844 Aur         & 2021 Dec 26,30,31 2022 Jan 06,10,25  & B40 V25 g20 r17 i22 & TYC 3749-853-1 & 2MASS J05422180+5259148 \\          
        V0699 Cep         & 2022 Sep 09,10,15,17,21,24  & g15 i18 r12 & GSC 03992-00111 & TYC 3992-491-1 \\                       
        NSVS 6259046 & 2022 Sep 18-24, Oct 03  & g32 i37 r27 & 2MASS J232731.469 +245741.655 & Gaia DR3 2840468997644456192 \\                       
        \hline                                                           
    \end{tabular}                                                   
\end{table*}   

\begin{table*}
\centering
\begin{minipage}{\textheight}
    \caption{The photometric observation obtained by NEXT.}
    \label{Table:The photometric observation obtained by NEXT}
    \resizebox{0.75\textwidth}{!}{ 
    \begin{tblr}{
      column{even} = {c},
      column{3} = {c},
      column{5} = {c},
      column{7} = {c},
      column{9} = {c},
      column{11} = {c},
      hline{1,11} = {-}{0.08em},
      hline{3} = {-}{},
    }
    Target        & HJD         & $\Delta m$\_B   & HJD          & $\Delta m$\_V   & HJD          & $\Delta m$\_g   & HJD         & $\Delta m$\_r       & HJD         & $\Delta m$\_i       \\
     & (2400000+) & & (2400000+) & & (2400000+) & & (2400000+) & & (2400000+) & & \\
    V0508 And         & -             & -          & -             & -          & 59886.35012~  & -0.415~    & 59886.35057~  & -0.346~     & 59886.35099~  & -0.295~     \\
                      & -             & -          & -             & -          & …             & …          & …             & …           & …             & …           \\
    V0844 Aur         & 59575.34264   & -0.648~    & 59575.34325~  & -0.687~    & 59575.34373~  & -0.657~    & 59575.34418   & -0.715 & 59575.34459   & -0.716 \\
                      & …             & …          & …             & …          & …             & …          & …             & …           & …             & …           \\
    V0699 Cep         & -             & -          & -             & -          & 59832.30586~  & -0.429~    & 59832.30180~  & -0.443~     & 59832.30212~  & -0.494~     \\
                      & -             & -          & -             & -          & …             & …          & …             & …           & …             & …           \\
    NSVS 6259046 & -             & -          & -             & -          & 59841.34852~  & 0.137~     & 59841.34904~  & 0.512~      & 59841.34956~  & 0.659~      \\
                      & -             & -          & -             & -          & …             & …          & …             & …           & …             & …           
    \end{tblr}
    }
    \begin{tablenotes}   
        \footnotesize               	  
        \item[] Note. This table is available in its entirety in machine-readable form. 
    \end{tablenotes}
\end{minipage}
\end{table*}

\subsection{Spectroscopy}
The Large Sky Area Multi-Object Fiber Spectroscopic Telescope (LAMOST), which possesses a 4-meter aperture and is a unique reflecting Schmidt telescope with a 5° field of view, collects numerous stellar spectra \citep{2012RAA....12.1197C,2012RAA....12..723Z}. The resolution in low-resolution mode is approximately 1800 \citep{1996ApOpt..35.5155W}, with wavelength ranging from 3700 to 9000 Å \citep{2016ApJS..227...27D}. 
The telescope incorporates 4000 fibers on its surface, significantly enhancing the speed of spectrum collection \citep{2012RAA....12.1197C}. LAMOST records all of our targets in low resolution. We download the data in DR11. The spectral parameters of the four targets are demonstrated in Table \ref{Table:Spectral observations of our targets from LAMOST}.

\begin{table*}
	\centering
	\caption{Spectral observations of our targets from LAMOST.}
        \label{Table:Spectral observations of our targets from LAMOST}
	\begin{tabular}{lllllllll}
		\hline
		Target  & Observational date & Orbital Phase & Spectral type & $T_{\text{eff}}$(K) & SNRg  & [Fe/H]     \\
		\hline
        V0508 And & 2016-09-09    & 0.41      & F5            & $6588\pm13$     &  253.30 & $0.117\pm0.007$  \\
        V0844 Aur & 2014-02-10    & 0.98 & F5 & - &  2.50 & - \\
		V0699 Cep & 2014-12-02    & 0.35     & F0            & $6660\pm24$     &  73.12 & $0.243\pm0.021$   \\
        NSVS 6259046 & 2013-10-15   &  0.27      & F0            & $6642\pm17$     &  97.24 & $-0.083\pm0.014$  \\
		\hline
	\end{tabular}
\end{table*}

\section{Photometric Solution}
\begin{figure*}
\centering
\subfigure{
\includegraphics[width=9cm,height = 6.5cm]{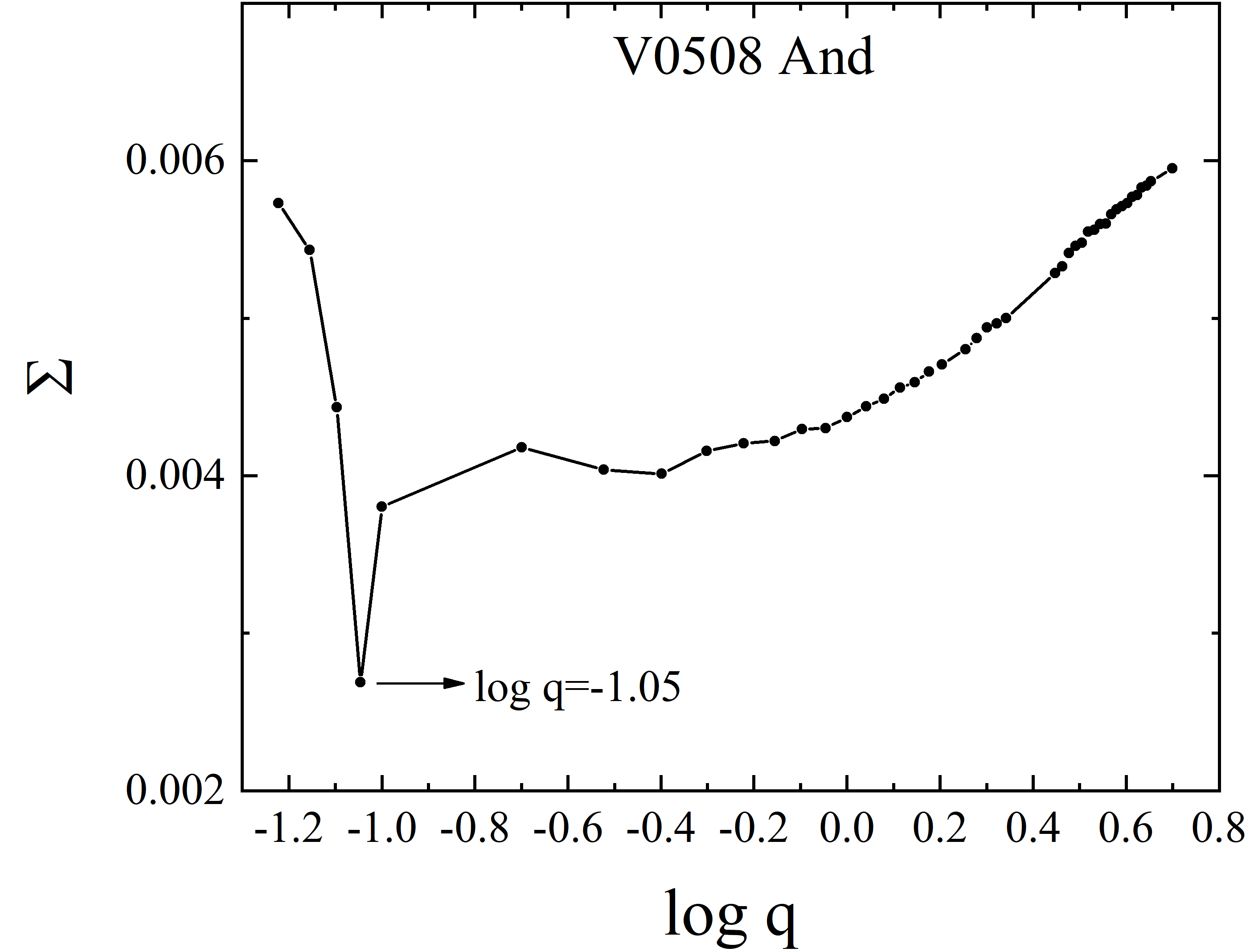}}\subfigure{
\includegraphics[width=9cm,height = 6.5cm]{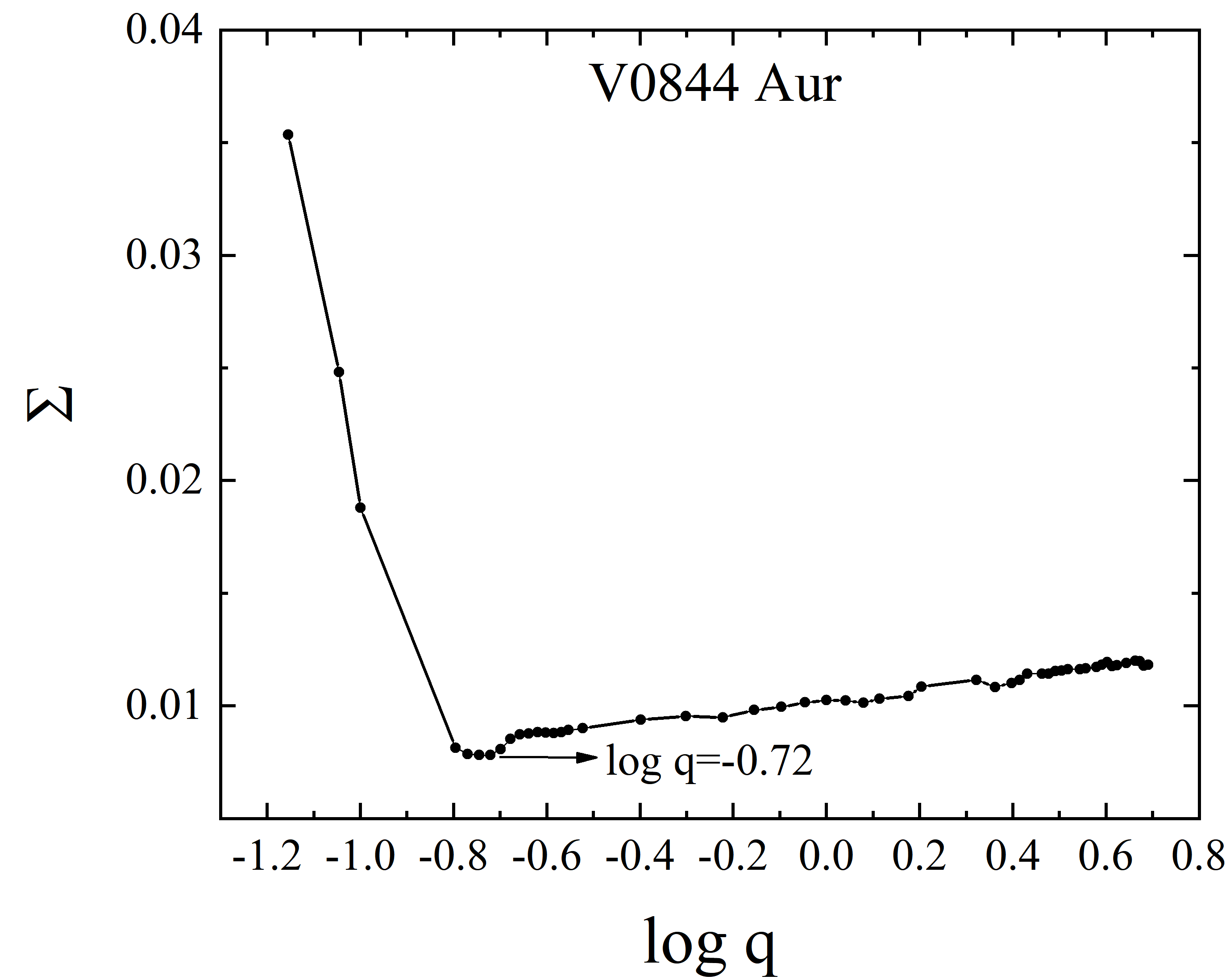}}
\subfigure{
\includegraphics[width=9cm,height = 6.5cm]{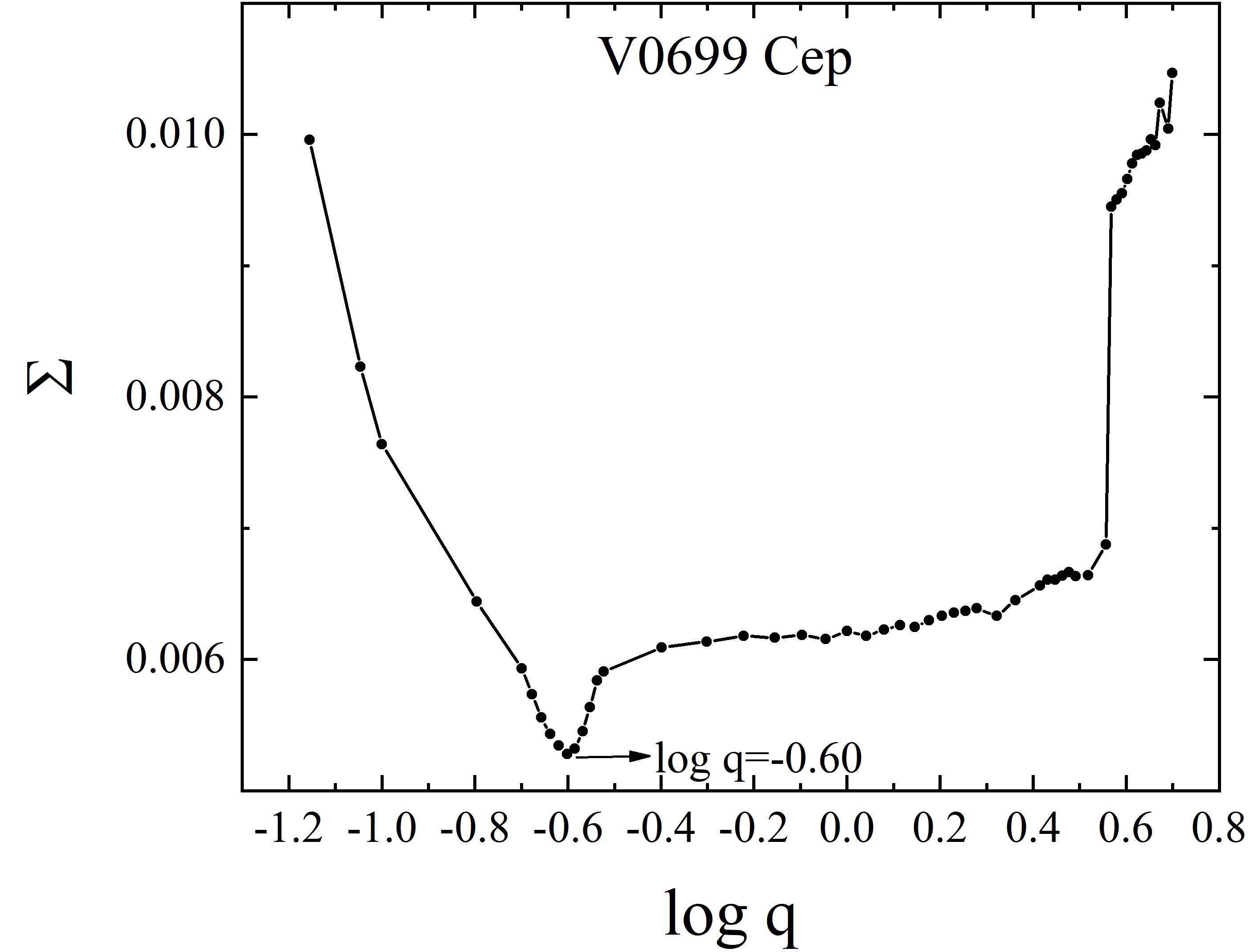}}\subfigure{
\includegraphics[width=9cm,height = 6.5cm]{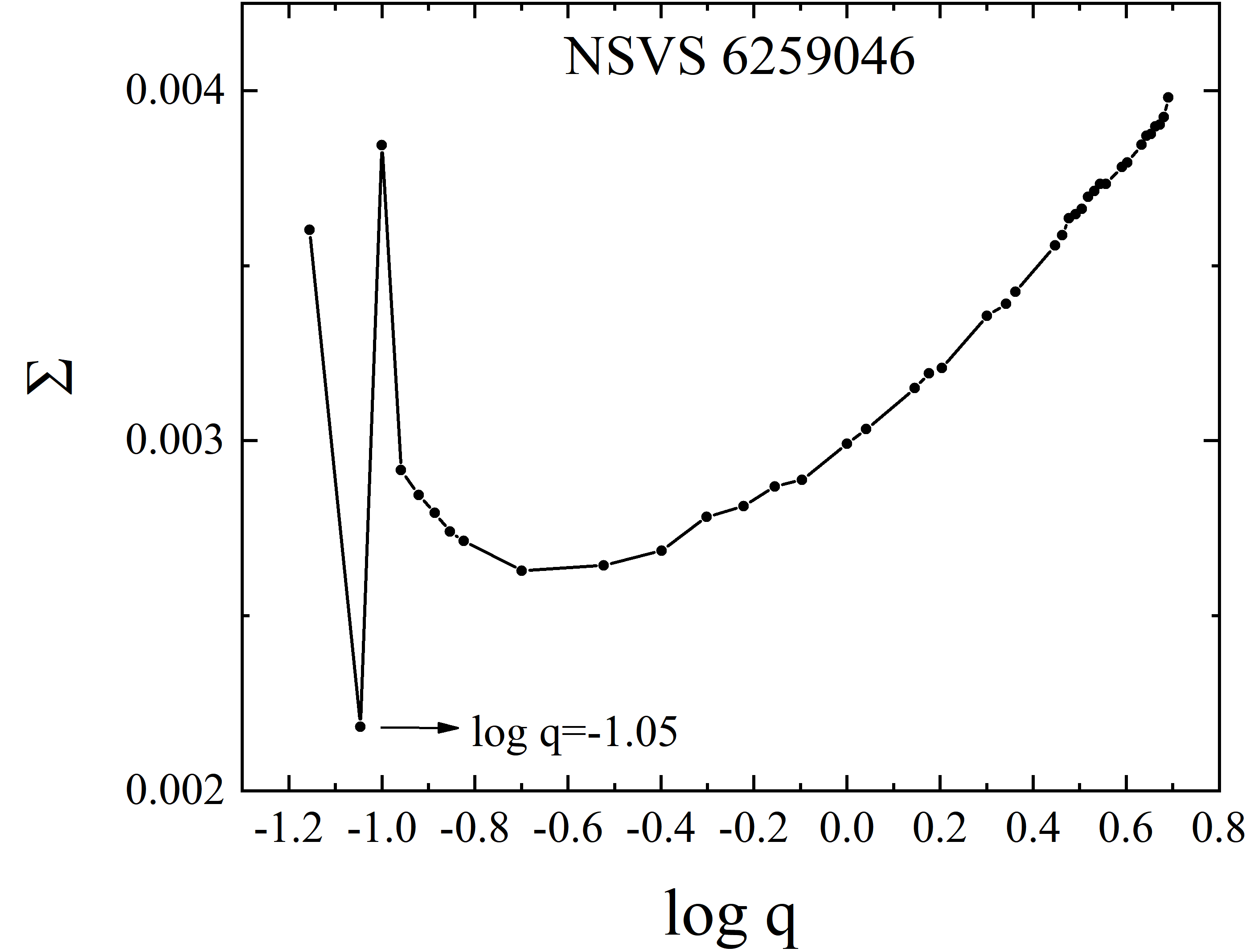}}
\caption{The relation between mass ratio \( q \) and mean residual $\varSigma$ generated by the W-D program for the NEXT light curves, the mass ratios are ranging from 0.06 to 5.00.}
\label{Figure:qsearch}
\end{figure*}

Photometric analysis involves the use of the Wilson-Devinney (W-D) software \citep{1971ApJ...166..605W,1979ApJ...234.1054W,1994PASP..106..921W}, version 2013, for solving the light curves. We use LAMOST $T_{\scriptsize \text{eff}}$ as the effective temperature of each target initially, which is fixed during the W-D procedure. However, for V0844 Aur, due to the low signal-to-noise ratio, the effective temperature is given by its F5 spectral type, which is 6550 K, according to the Table 5 of \cite{2013ApJS..208....9P}.
All $T_{\scriptsize \text{eff}}$ of our four targets are below 7200 K. When $T_{\scriptsize \text{eff}}$ < 7200 K, gravitational darkening is assumed to be \( g \) = 0.32 \citep{1967ZA.....65...89L}, and bolometric albedo coefficient is set at \( A \) = 0.5 for both components \citep{1969AcA....19..245R}. Using square root law, we interpolate \cite{1993AJ....106.2096V}'s table to obtain the limb-darkening coefficients.
The parameter \( q \) is computed using \( q \)-search method. Adjustable parameters include orbital inclination \( i \), effective temperature of the less massive star $T_{\scriptsize \text{2}}$, luminosity of the massive star $L_{\scriptsize \text{1}}$, and potential ($\mathit{\Omega}_1$ = $\mathit{\Omega}_2$ = $\mathit{\Omega}$). 
To check whether our four targets exhibit third light, we also consult the Gaia \footnote{\href{https://gea.esac.esa.int/archive/}{https://gea.esac.esa.int/archive/}} database and find that only V0699 Cep shows evidence of third light within our photometric aperture. Table \ref{Table:Sources within 5 magnitudes and 42 arcsecs of our targets in Gaia DR3. We use the mean g-band photometric flux to determine the third light contribution in Section 3.} presents data for objects within a 5-magnitude range of our targets based on Gaia DR3, mean g-band photometric flux of V0699 Cep is 348708.7 \(\mathrm{electron\cdot s^{-1}}\) , and the third light’s is 100942.6 \(\mathrm{electron\cdot s^{-1}}\), we calculate that the flux contribution of the third light is 22.4\%, therefore third light is also chosen as a free parameter for V0699 Cep. This method only weeds out certain kinds of tertiaries, such as more massive companions at large enough separations to produce astrometric deviations. It does not rule out the presence of tertiary stars in relatively small orbits.
\begin{table*}
	\centering
	\caption{Sources within 5 magnitudes and 42 arcsecs of our targets in Gaia DR3.}
        \label{Table:Sources within 5 magnitudes and 42 arcsecs of our targets in Gaia DR3. We use the mean g-band photometric flux to determine the third light contribution in Section 3.}
	\begin{tabular}{llllllll}
		\hline
        Target       & Gaia Source ID           & RA         & Dec       & RUWE  & g mean flux (e$^-$/s) & g mean flux error (e$^-$/s) & g mean mag \\
		\hline
        V0508 And    & 364073176407230592  & 11.93397~  & 36.03975~ & 1.26~ & 293920.45~         & 1073.60~                 & 12.02~     \\
             & 364061429672596992  & 11.93335~  & 36.02890~ & 0.98~ & 14057.75~          & 3.78~                    & 15.32~     \\
             & 364073180703117696  & 11.92798~  & 36.04014~ & 1.09~ & 6526.76~           & 2.81~                    & 16.15~     \\
V0844 Aur    & 263713672034656000  & 85.77345~  & 53.04316~ & 1.02~ & 319566.71~         & 2112.52~                 & 11.93~     \\
             & 263713672035152512  & 85.77299~  & 53.04748~ & 0.94~ & 9397.16~           & 3.07~                    & 15.75~     \\
             & 263713710690720640  & 85.78852~  & 53.04550~ & 2.09~ & 9243.48~           & 7.58~                    & 15.77~     \\
             & 263713706394886656  & 85.78283~  & 53.04506~ & 0.94~ & 9080.11~           & 17.68~                   & 15.79~     \\
             & 263713676330987264  & 85.76471~  & 53.03615~ & 1.24~ & 7285.53~           & 9.96~                    & 16.03~     \\
V0699 Cep    & 2007362057959414144 & 341.50272~ & 57.78044~ & 1.78~ & 348708.68~         & 2259.62~                 & 11.83~     \\
             & 2007362062263640960 & 341.50306~ & 57.78074~ & 1.98~ & 100942.60~         & 48.10~                   & 13.18~     \\
             & 2007361684306521344 & 341.51398~ & 57.78073~ & 0.98~ & 33056.56~          & 41.72~                   & 14.39~     \\
             & 2007361306349403008 & 341.50580~ & 57.76985~ & 1.00~ & 7903.99~           & 3.00~                    & 15.94~     \\
             & 2007362062263637888 & 341.49361~ & 57.78182~ & 0.99~ & 7077.66~           & 4.07~                    & 16.06~     \\
             & 2007361684306520448 & 341.51335~ & 57.78211~ & 1.04~ & 6272.52~           & 2.36~                    & 16.19~     \\
             & 2007362440220762112 & 341.51681~ & 57.78814~ & 0.98~ & 3891.32~           & 2.22~                    & 16.71~     \\
NSVS 6259046 & 2840468276089961728 & 351.94181~ & 24.88248~ & 0.96~ & 170505.10~         & 689.82~                  & 12.61~                  \\
		\hline
	\end{tabular}
\end{table*}
The relationship between \( q \) and the mean residual is shown in Fig. \ref{Figure:qsearch}.
\( q \) with minimum mean residuals is obtained, then we set it as the initial value and an adjustable parameter to obtain a convergent solution.
As $T_{\scriptsize \text{eff}}$ represents the integrated temperature of both primary and secondary components, we use the following equations to calculate the temperatures of both the primary and secondary stars \citep{2003A&A...404..333Z},
\begin{equation}
      T_1=\left( \frac{\left( 1+\left( \frac{r_2}{r_1} \right) ^2 \right) T_{\text{eff}}^{4}}{1+\left( \frac{r_2}{r_1} \right) ^2\left( \frac{T_2}{T_1} \right) ^4} \right) ^{0.25},
\end{equation}
\begin{equation}
      T_2 = T_1 \times \frac{T_2}{T_1},
\end{equation}
We set more accurate temperatures as initial values in W-D program, while the primary temperature is fixed and the secondary temperature is adjustable. By running the W-D program, convergent solutions are obtained after one iteration.

Three targets exhibit the O'Connell effect, which has four hypotheses to explain it. For circumstellar material and asymmetry of circumfluence from the Coriolis effect, the W-D code does not incorporate options to account for these mechanisms. Moreover, both hypotheses have been shown to suffer from significant shortcomings and are inadequate in explaining the O'Connell effect. For example, the authors who proposed the circumstellar material hypothesis, \cite{2003ChJAA...3..142L}, themselves acknowledged that the model requires a specific geometric configuration and fails to consistently reproduce the observed O'Connell effect across multiple photometric bands. As a result, subsequent studies have generally regarded this explanation as limited in its applicability.
Regarding the asymmetric circulation hypothesis, \cite{2009SASS...28..107W} pointed out that this mechanism is not universally applicable to all types of binary systems, especially in detached or semi-detached systems. Moreover, this model has failed to successfully reproduce the observed light curve features in numerical simulations, further casting doubt on its validity.
Now comes the hypothesis of cool spots and hot spot. For all targets, we test all possible combinations of cool spots and hot spots, and find that cool spots models provided the best explanation for their O'Connell effects. For V0699 Cep, we apply a cool spot to the secondary component. For V0844 Aur, we apply a cool spot to the primary component. In NSVS 6259046, initial attempts using a single cool or hot spot fail to provide a satisfactory fit to the light curve. Consequently, we introduce two spots and examine all four possible configurations involving cool and hot spots on either component. The best fit is achieved with a model incorporating a cool spot on the primary star and a hot spot on the secondary star.

\cite{2024NewA..11202270A} also performed photometric analysis on V0699 Cep. Apart from the third light contribution, the other parameters are generally consistent with our results. Since our value for the third light of V0699 Cep is consistent with the Gaia data shown in Table \ref{Table:Sources within 5 magnitudes and 42 arcsecs of our targets in Gaia DR3. We use the mean g-band photometric flux to determine the third light contribution in Section 3.}, this supports the accuracy of our study.
For the other three targets, even with GAIA or O-C analysis, considering that the former is based on astrometric measurements and thus insensitive to smaller third-body orbits, while the latter is insensitive to face-on orbits of third bodies, we also attempt to add third light to these three targets. However, the results show that the proportion of third light is either negative or fails to converge to a solution. Therefore, we believe it is highly likely that these three targets do not exhibit third light. Future radial velocity observations may reveal whether these systems harbor third conponent. The photometric solutions for our four targets are detailed in Tables \ref{Table:Photometric solutions of V0699 Cep}, \ref{Table:Photometric solutions of V0844 Aur}, \ref{Table:Photometric solutions of NSVS 6259046} and \ref{Table:Photometric solutions of V0508 And} respectively.
The light curves observed by ASAS-SN, SuperWASP, TESS and ZTF are also analyzed by the W-D program. V0699 Cep is analyzed using TESS and ASAS-SN light curves, while data from TESS, ASAS-SN, and SuperWASP are utilized for V0844 Aur. For NSVS 6259046 and V0508 And, we collect data from ASAS-SN, SuperWASP, TESS and ZTF.
For TESS data, V0699 Cep is observed in Sectors 16 and 17 and V0844 Aur is observed in Sectors 19, 59 and 73. Their light curves have significant differences between different sectors, so we analyze the different TESS sectors of these two targets separately. NSVS 6259046 is observed in Sector 56. V0508 And is observed in Sector 17 and 57. For V0508 And, there are no significant differences between the two light curves of Sectors 17 and 57, therefore we have merged them for analysis. We normalize the SAP-FLUX and convert the normalized flux to magnitude.

In the TESS, ASAS-SN, SuperWASP and ZTF studies, we initially set NEXT parameters as starting values. To improve program execution speed, TESS and SuperWASP data are binned into 200 normal points. Fig. \ref{Figure:Theoretical light curves} presents the theoretical light curves, while Tables \ref{Table:Photometric solutions of V0699 Cep} to \ref{Table:Photometric solutions of V0508 And} contain the photometric results. Additionally, all our four targets are totally eclipsing CBs, so the mass ratios obtained from the photometric observation is reliable \citep{2003CoSka..33...38P,2021AJ....162...13L}. Some parameters differ among different surveys, especially the contact degree \( f \), likely due to the difference in photometric accuracy among the surveys. We have selected the physical parameters derived from NEXT data as definitive solutions for our targets, due to the better multiband wavelength coverage. We also find that the mass ratios obtained from long-cadence TESS data are slightly lower than NEXT data, which is consistent with \cite{2017MNRAS.466.2488Z}. They demonstrated that long-cadence observations can distort the eclipse shape, particularly by altering the depth and width of minima, potentially leading to biased parameter estimates.

\begin{figure*}
\centering
\subfigure{
\includegraphics[width=5.75cm,height = 5.75cm]{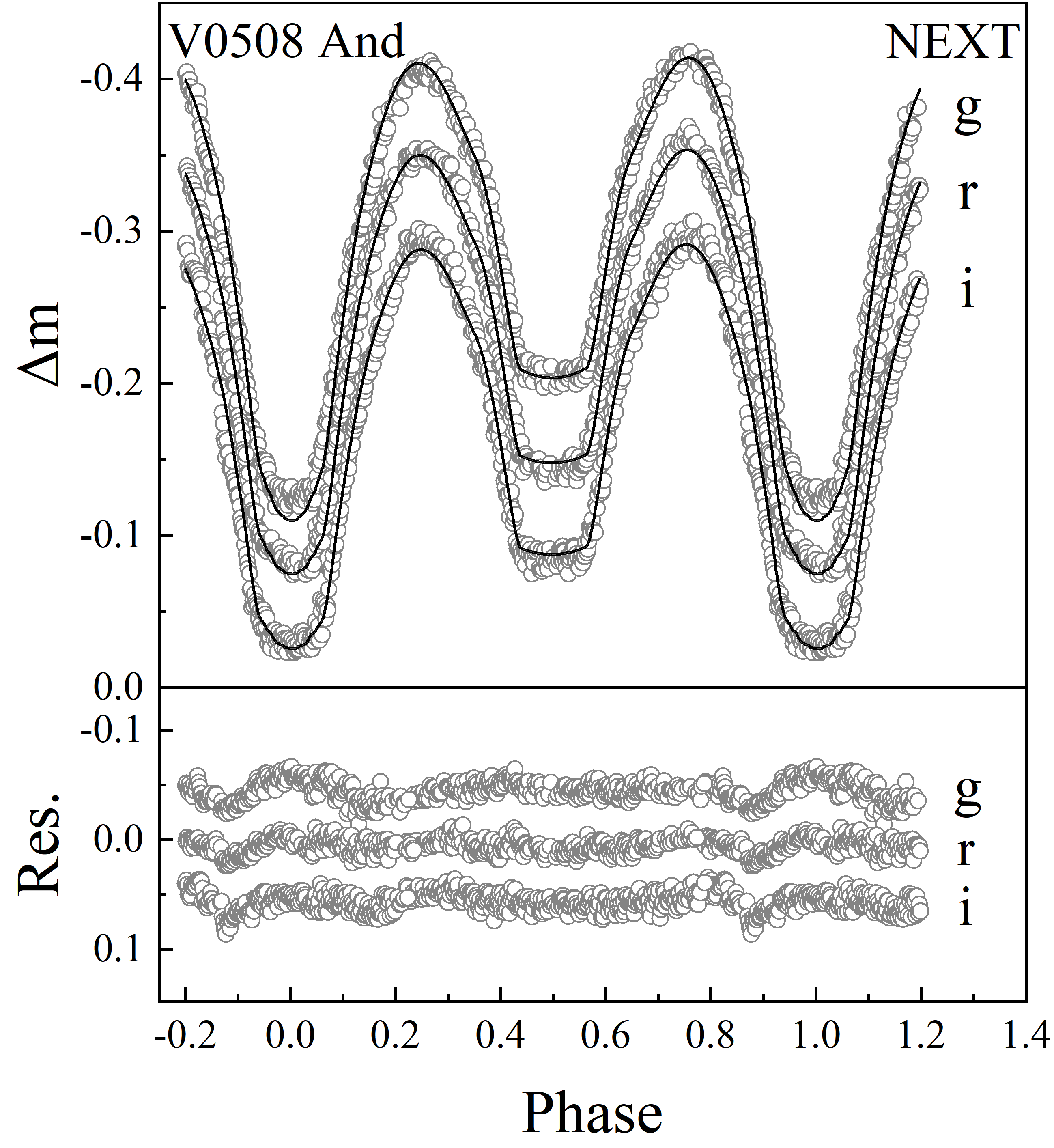}}\subfigure{
\includegraphics[width=5.75cm,height = 5.75cm]{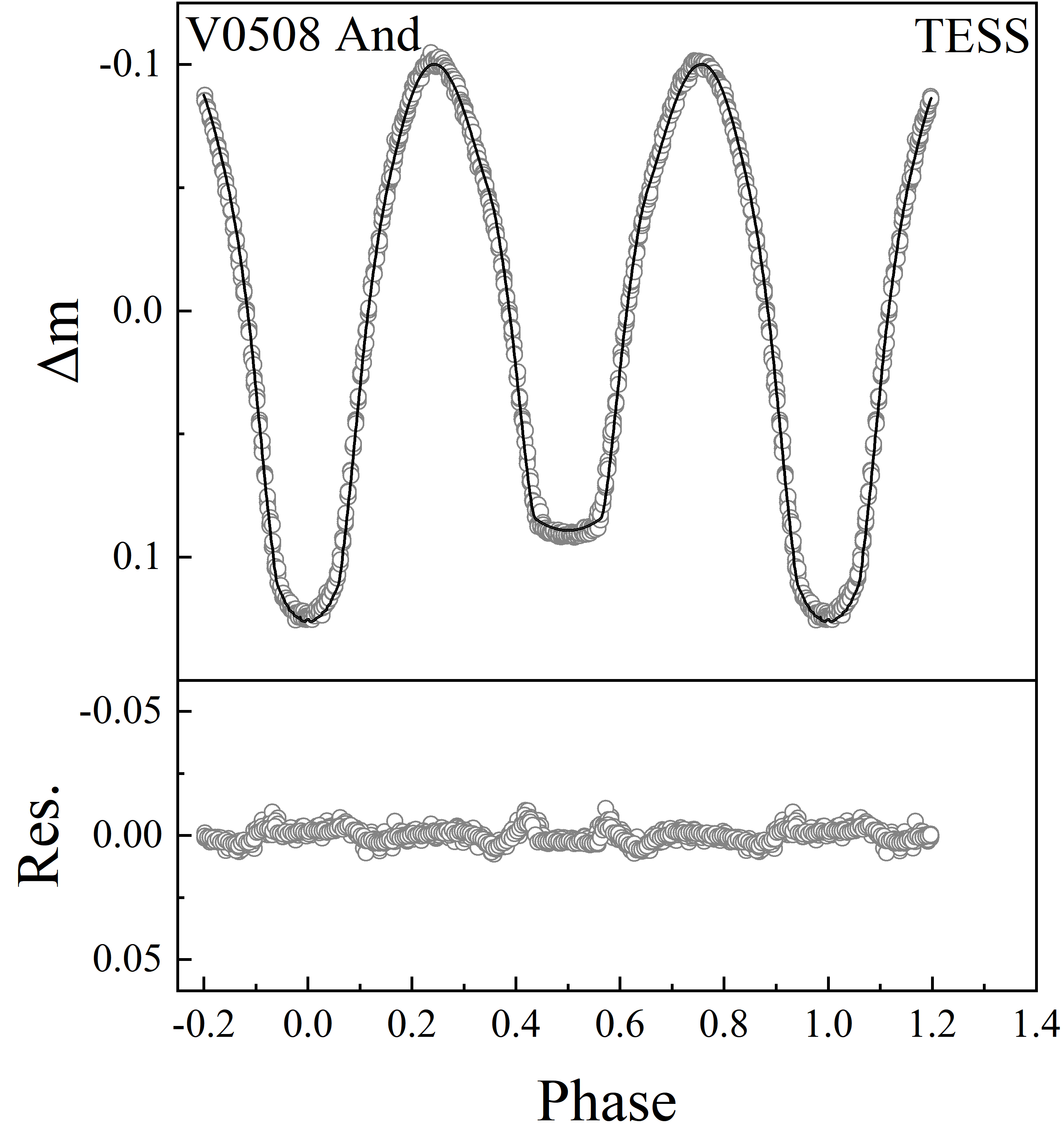}}\subfigure{
\includegraphics[width=5.75cm,height = 5.75cm]{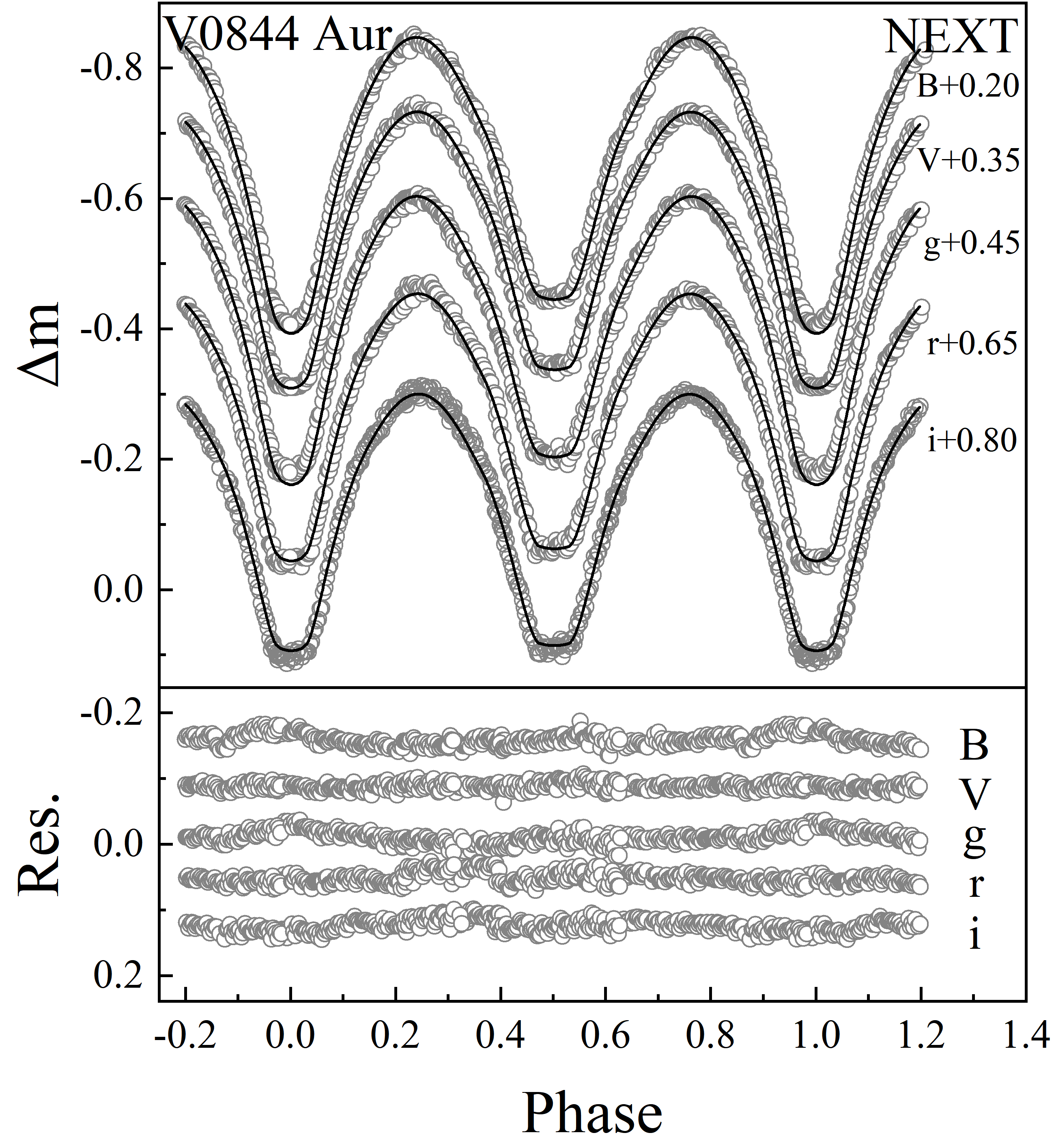}}
\subfigure{
\includegraphics[width=5.75cm,height =5.75cm]{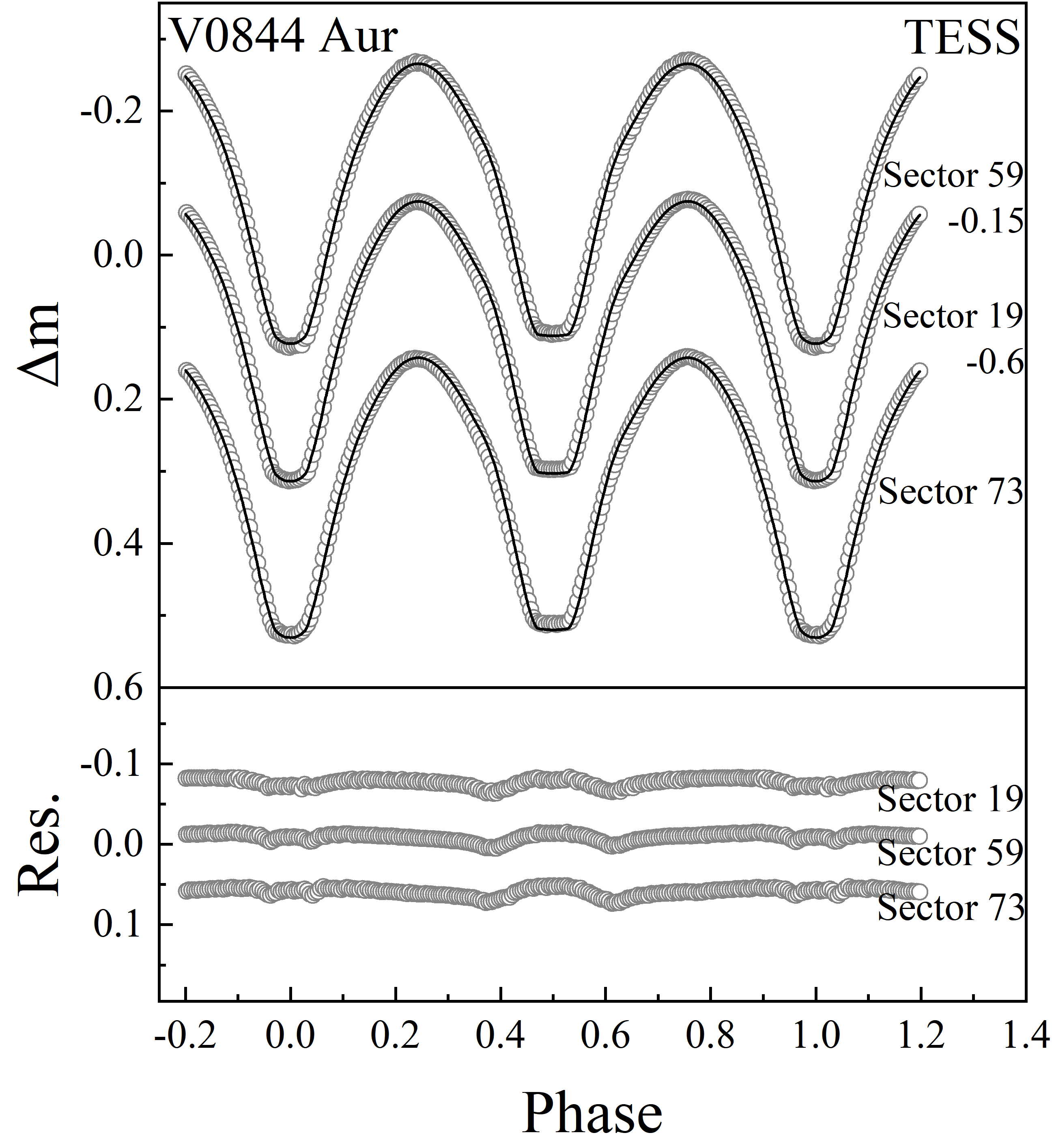}}\subfigure{
\includegraphics[width=5.75cm,height =5.75cm]{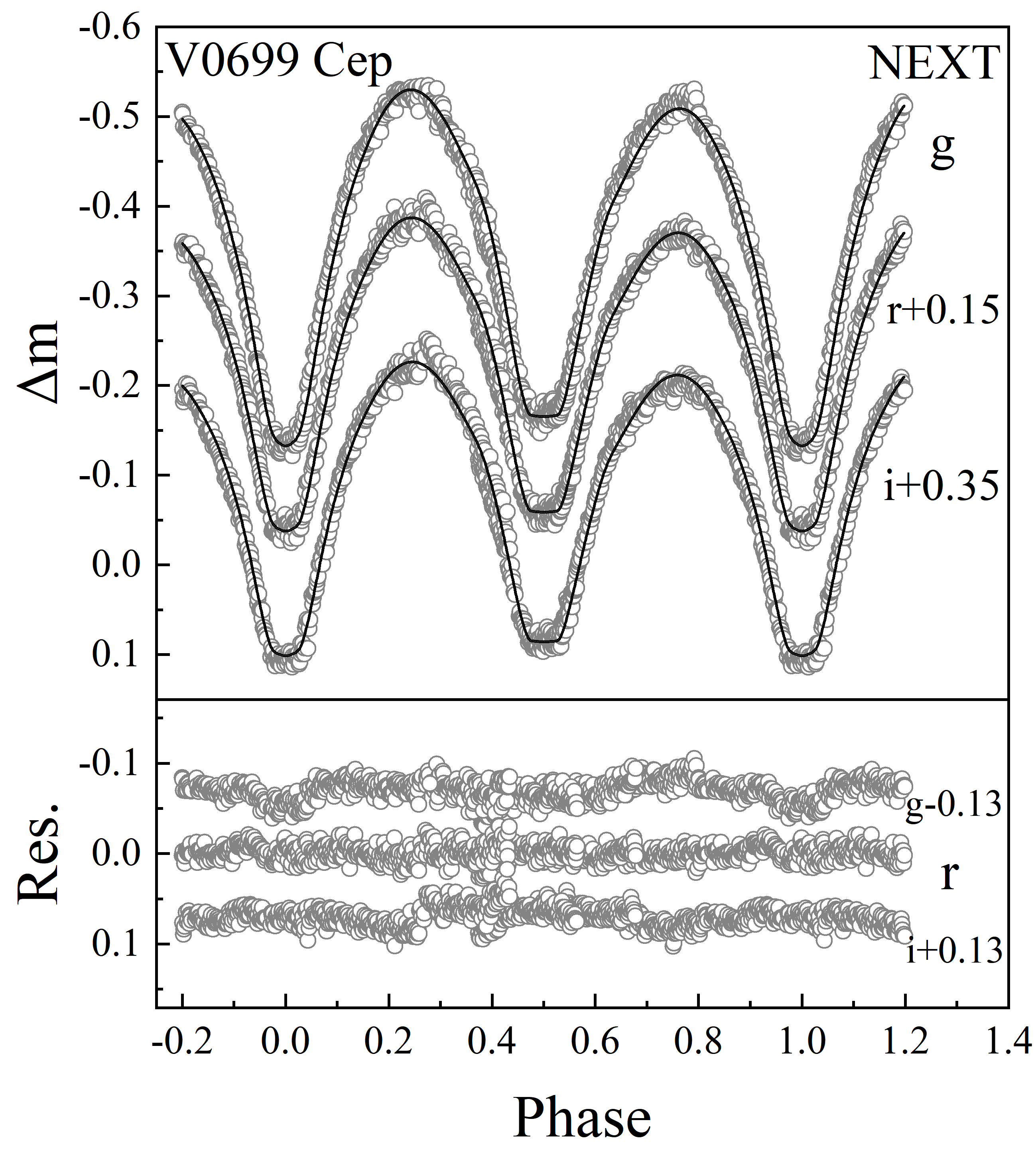}}\subfigure{
\includegraphics[width=5.75cm,height =5.75cm]{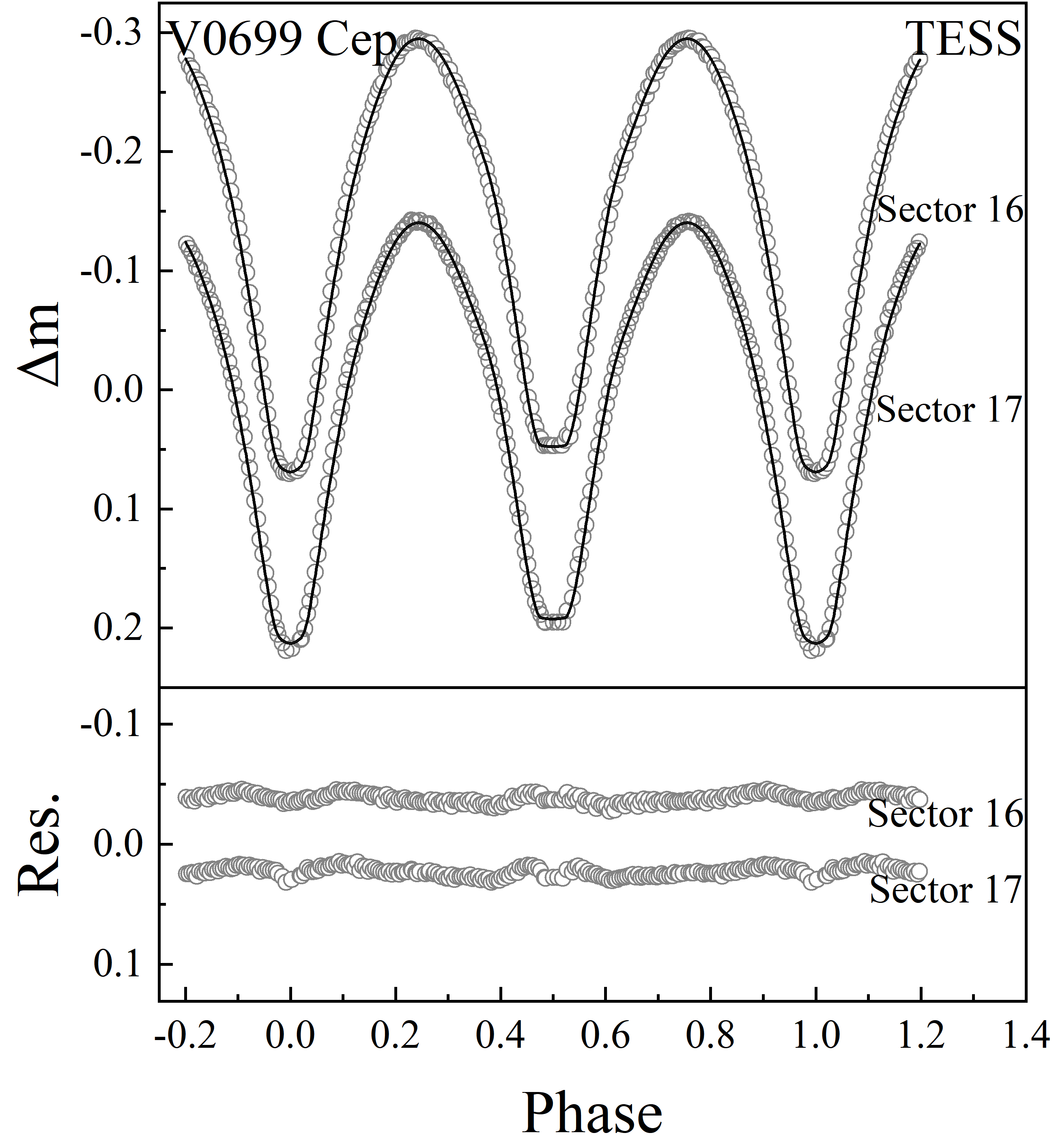}}
\subfigure{
\includegraphics[width=5.75cm,height =5.75cm]{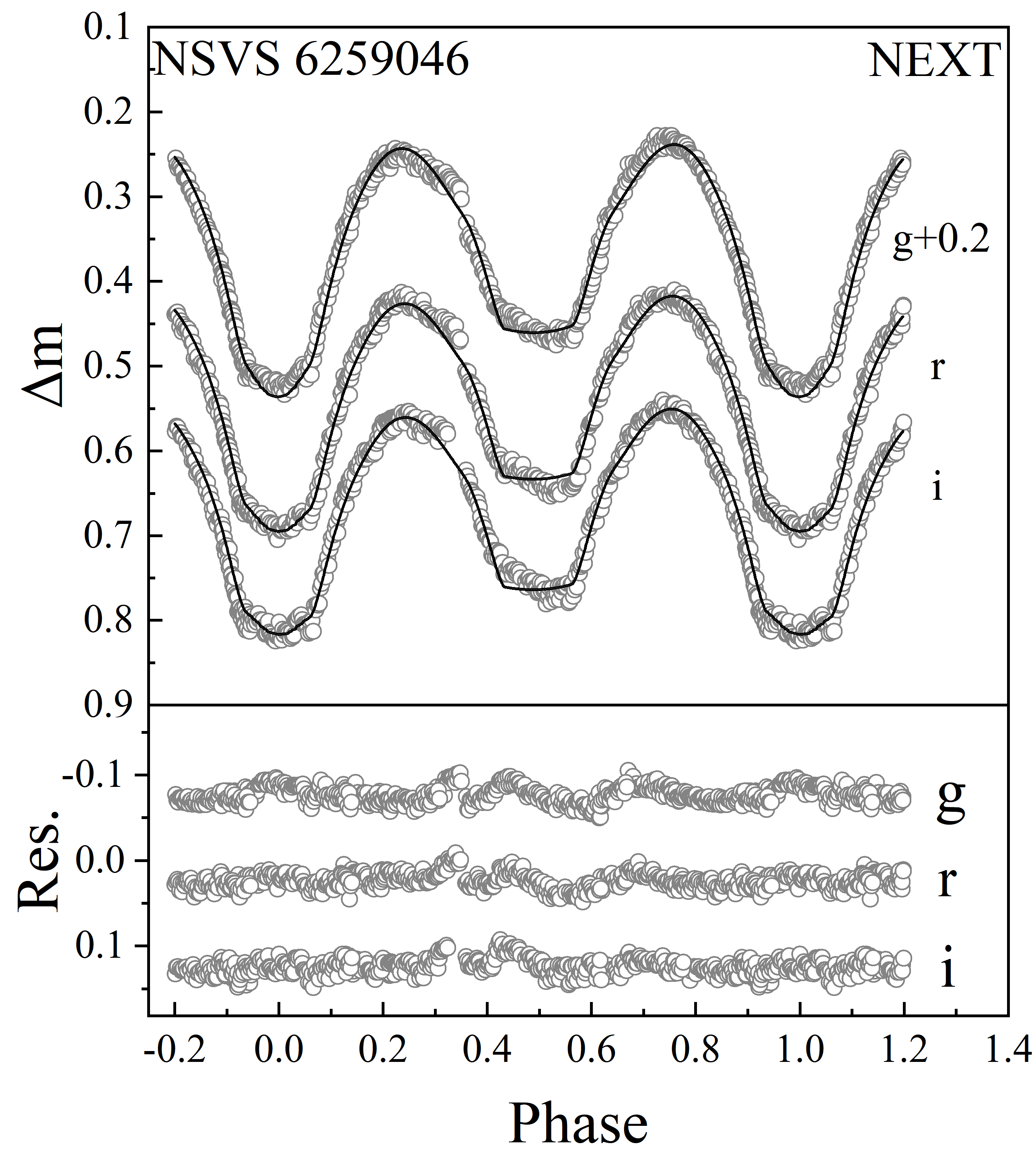}}\subfigure{
\includegraphics[width=5.75cm,height =5.75cm]{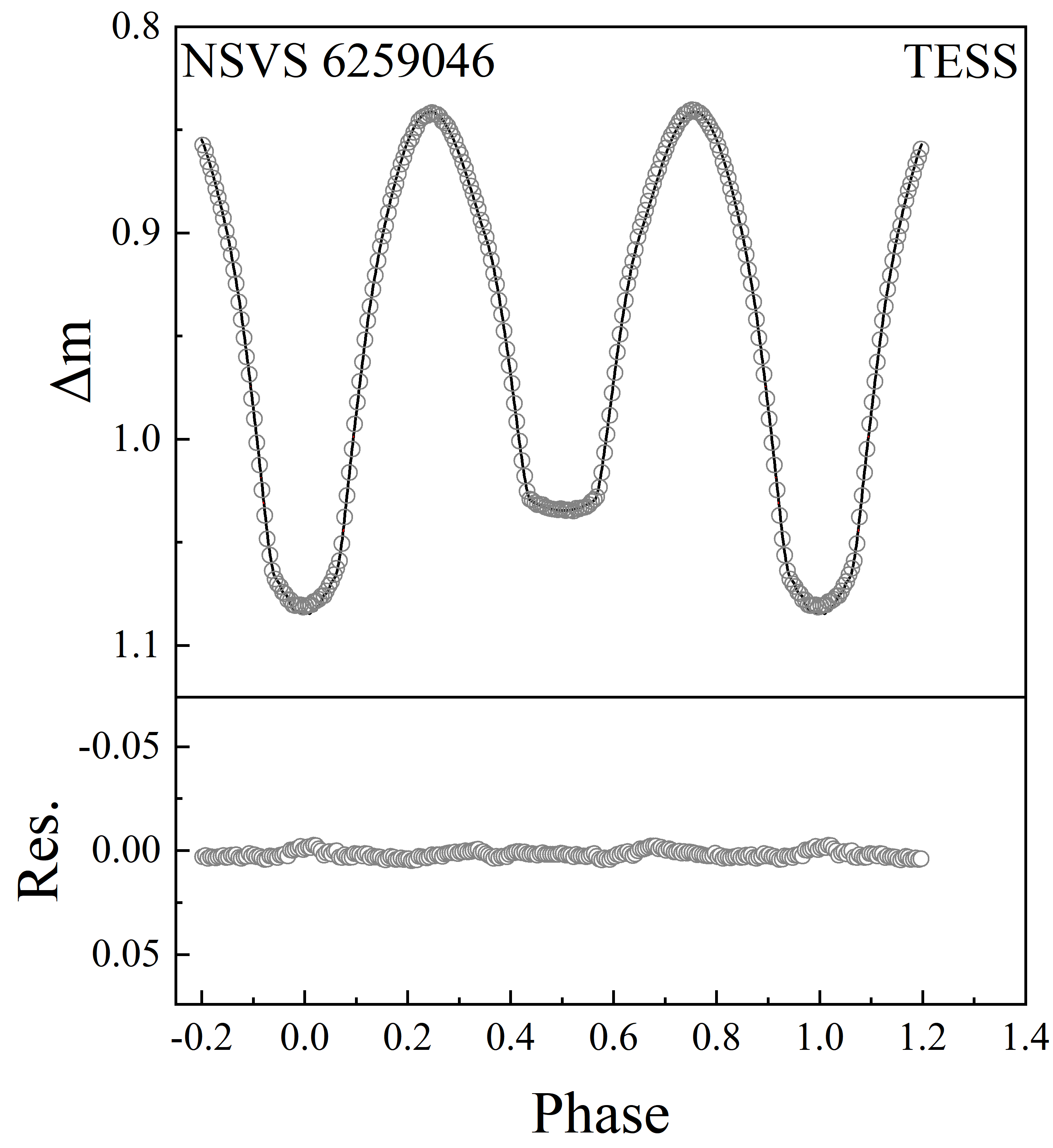}}
\caption{This figure presents a comparison between the observed (symbols) and theoretical (solid lines) light curves generated by the W-D program for the NEXT, TESS, ASAS-SN, SuperWASP and ZTF light curves. The residuals of the fit are also displayed.}
\label{Figure:Theoretical light curves}
\end{figure*}

\begin{figure*}
\ContinuedFloat
\centering
\subfigure{
\includegraphics[width=5.75cm,height = 5.75cm]{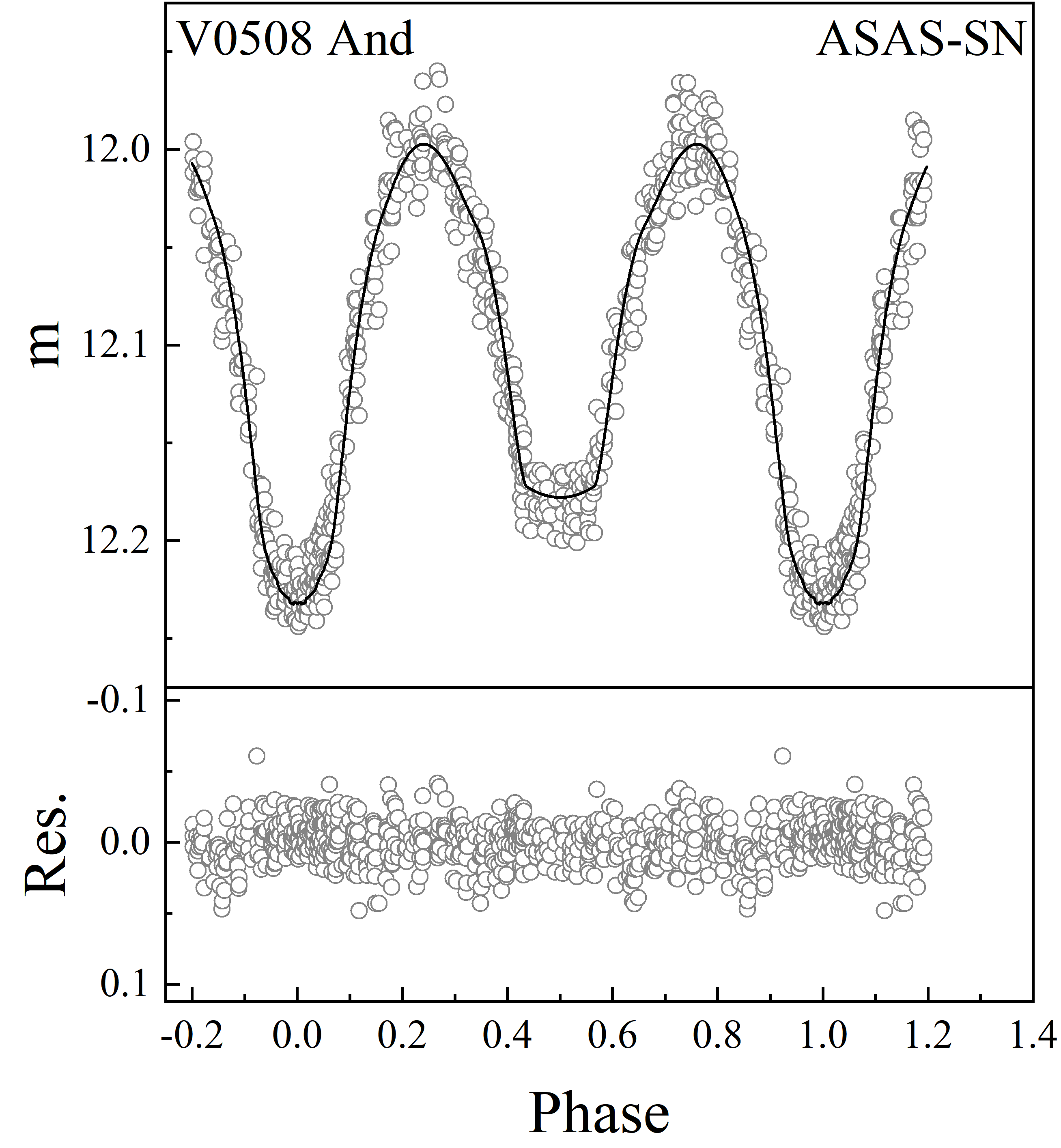}}\subfigure{
\includegraphics[width=5.75cm,height = 5.75cm]{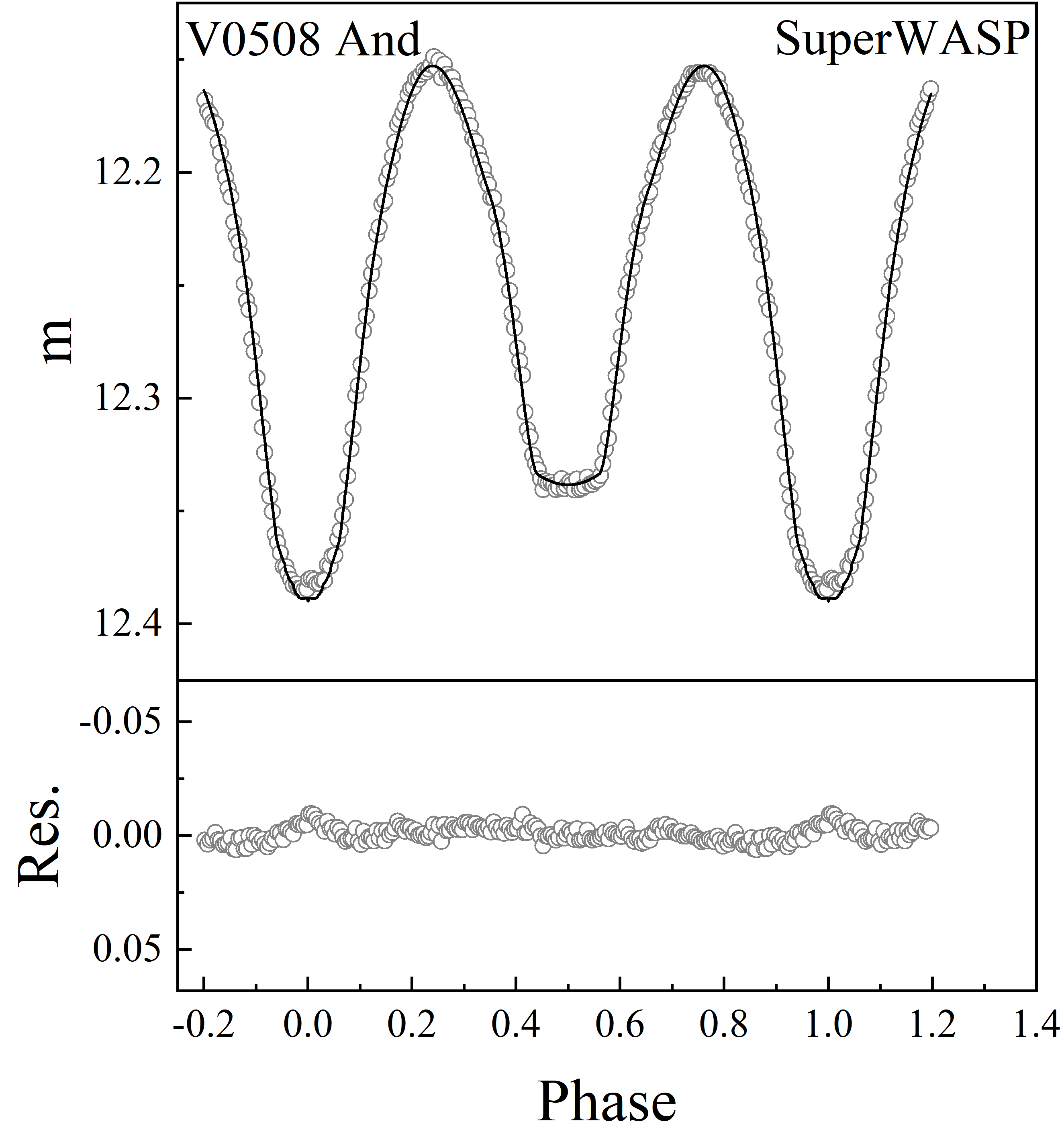}}\subfigure{
\includegraphics[width=5.75cm,height = 5.75cm]{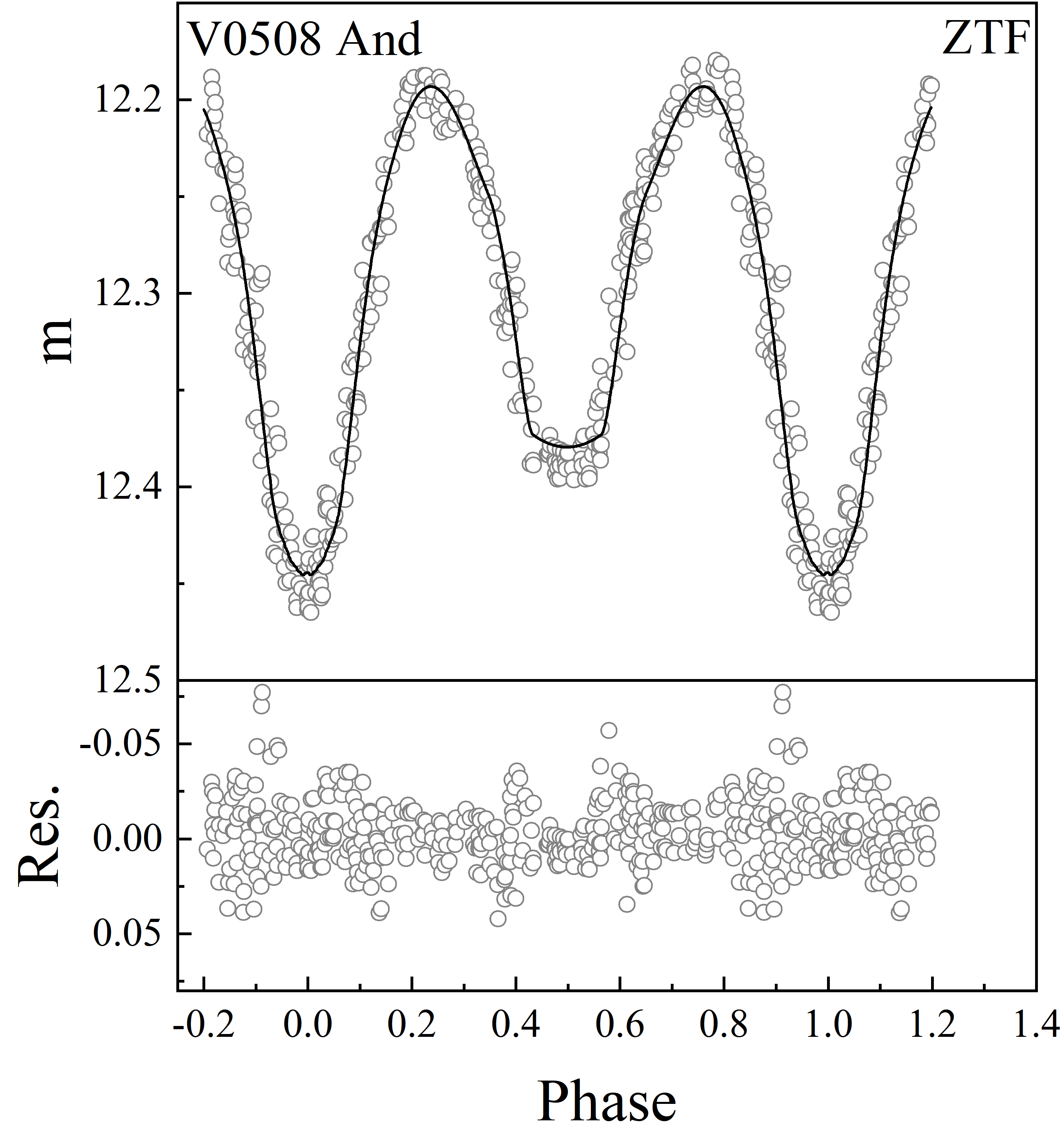}}
\subfigure{
\includegraphics[width=5.75cm,height = 5.75cm]{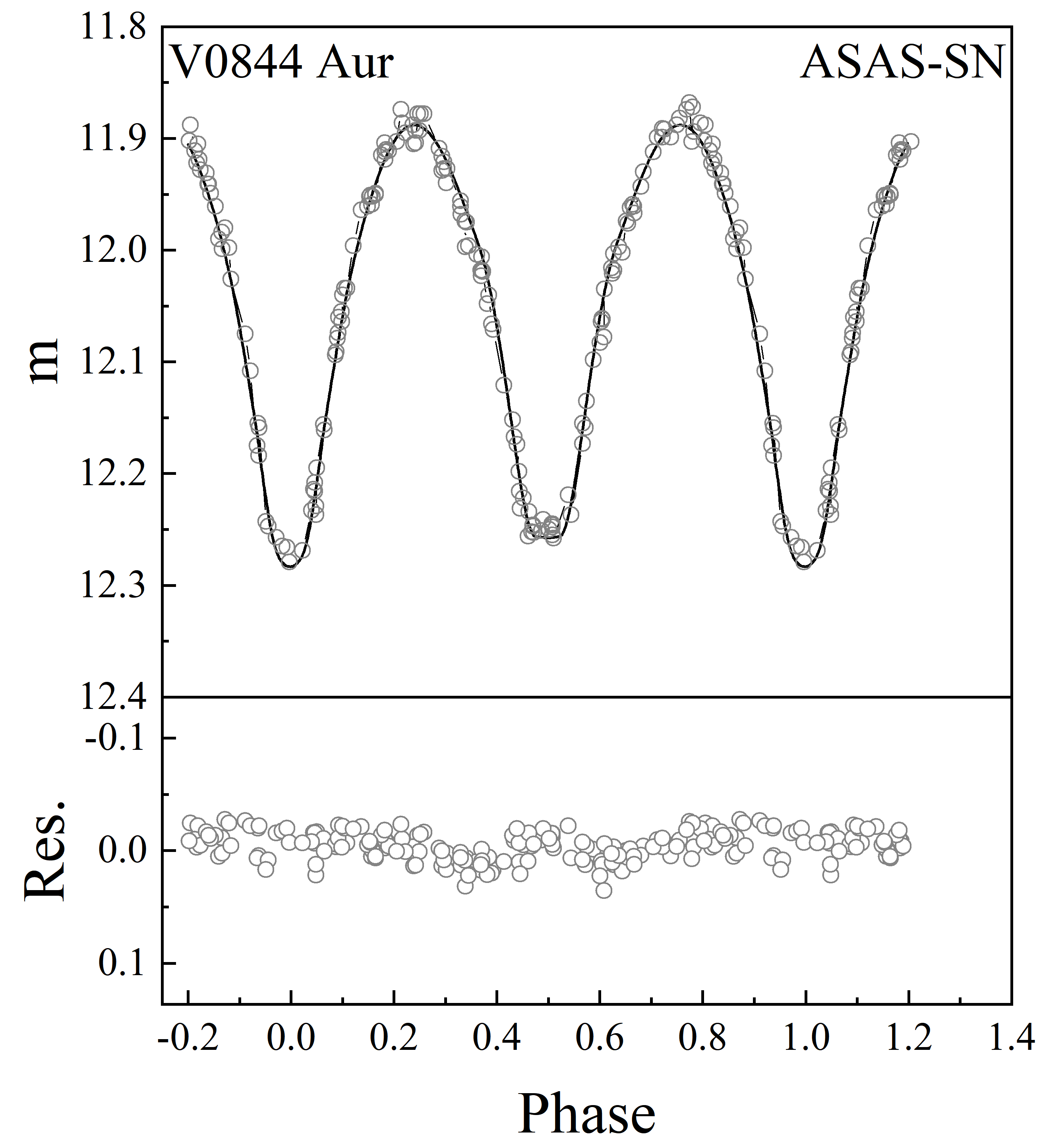}}\subfigure{
\includegraphics[width=5.75cm,height = 5.75cm]{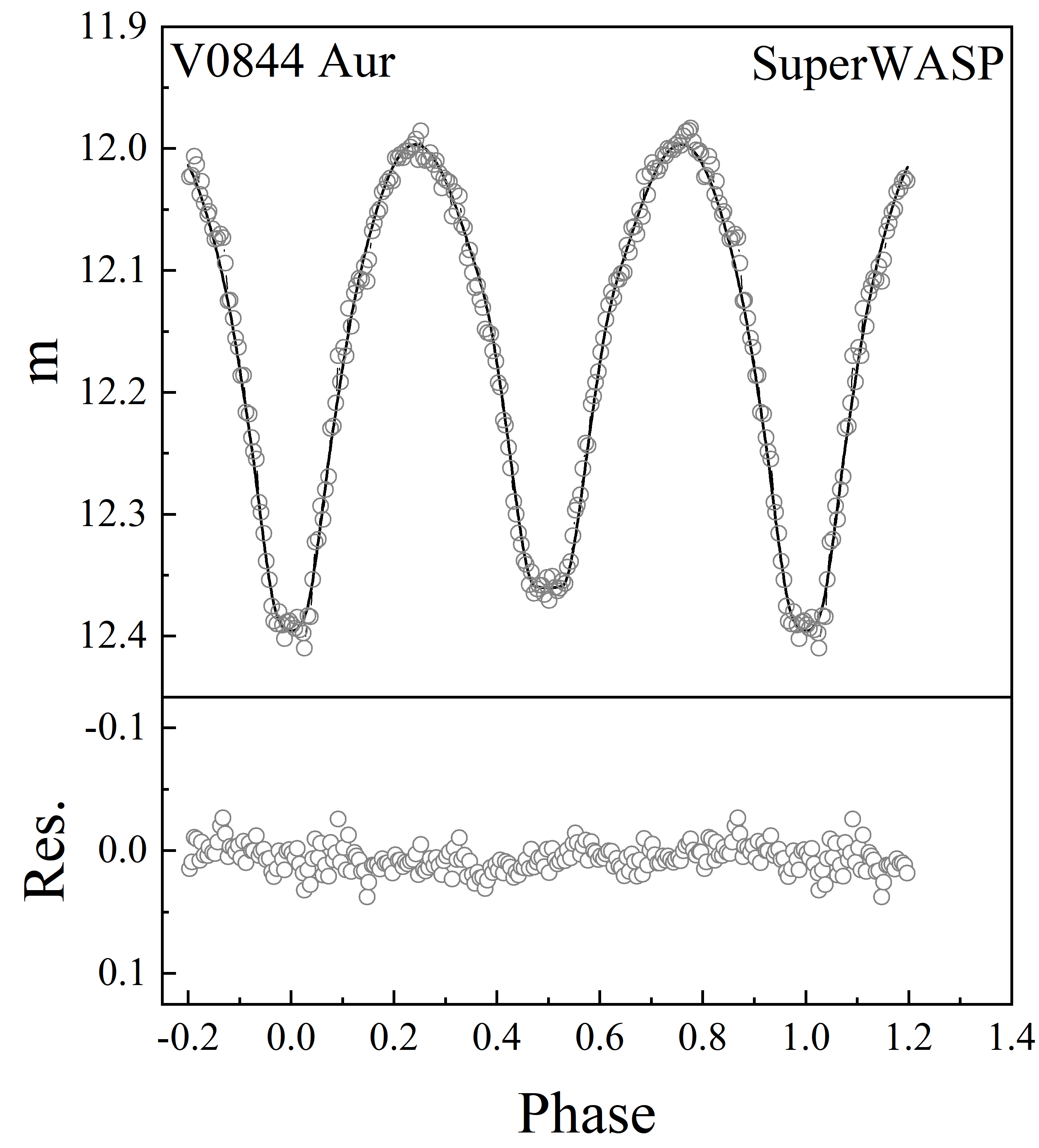}}\subfigure{
\includegraphics[width=5.75cm,height = 5.75cm]{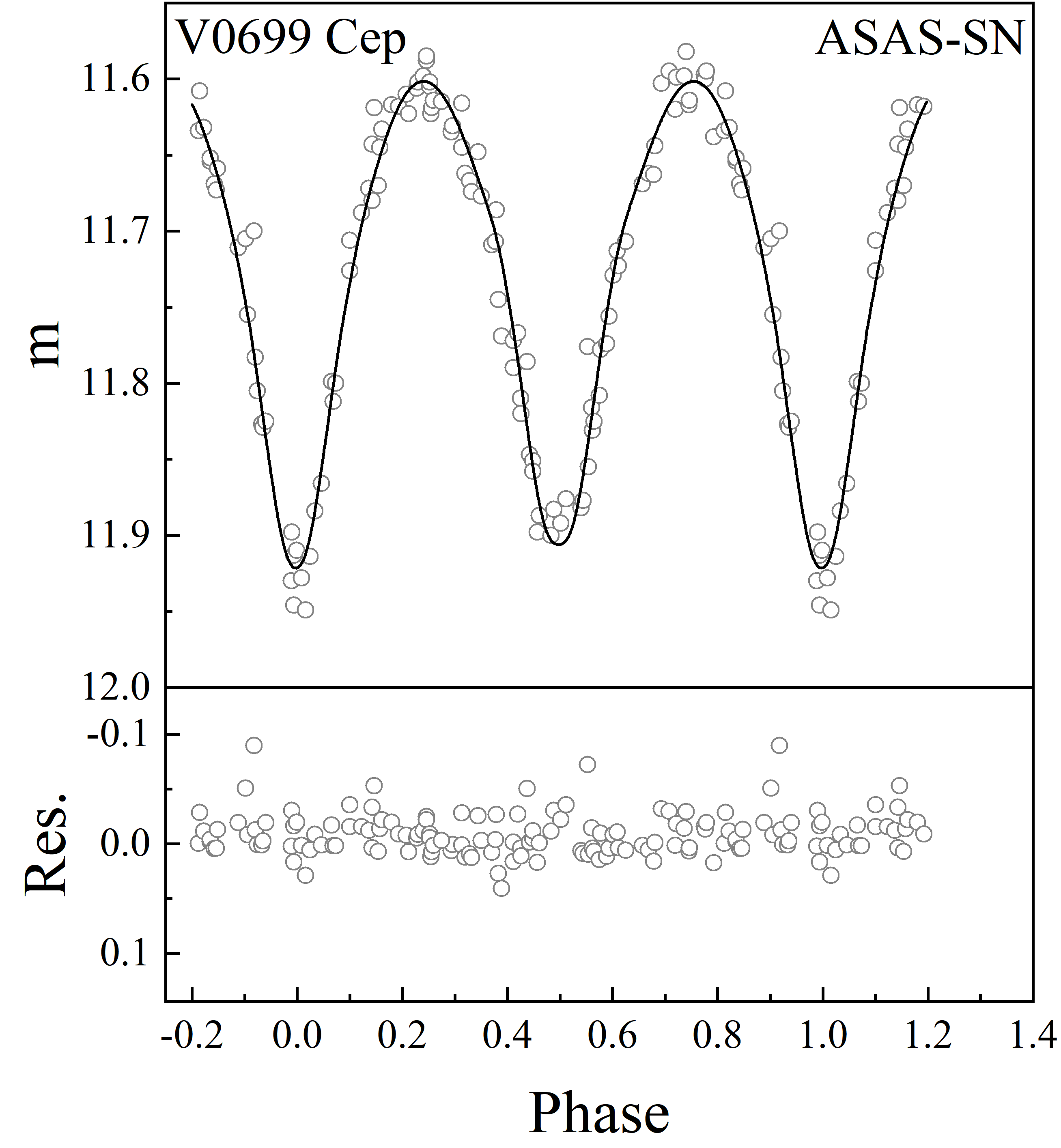}}
\subfigure{
\includegraphics[width=5.75cm,height = 5.75cm]{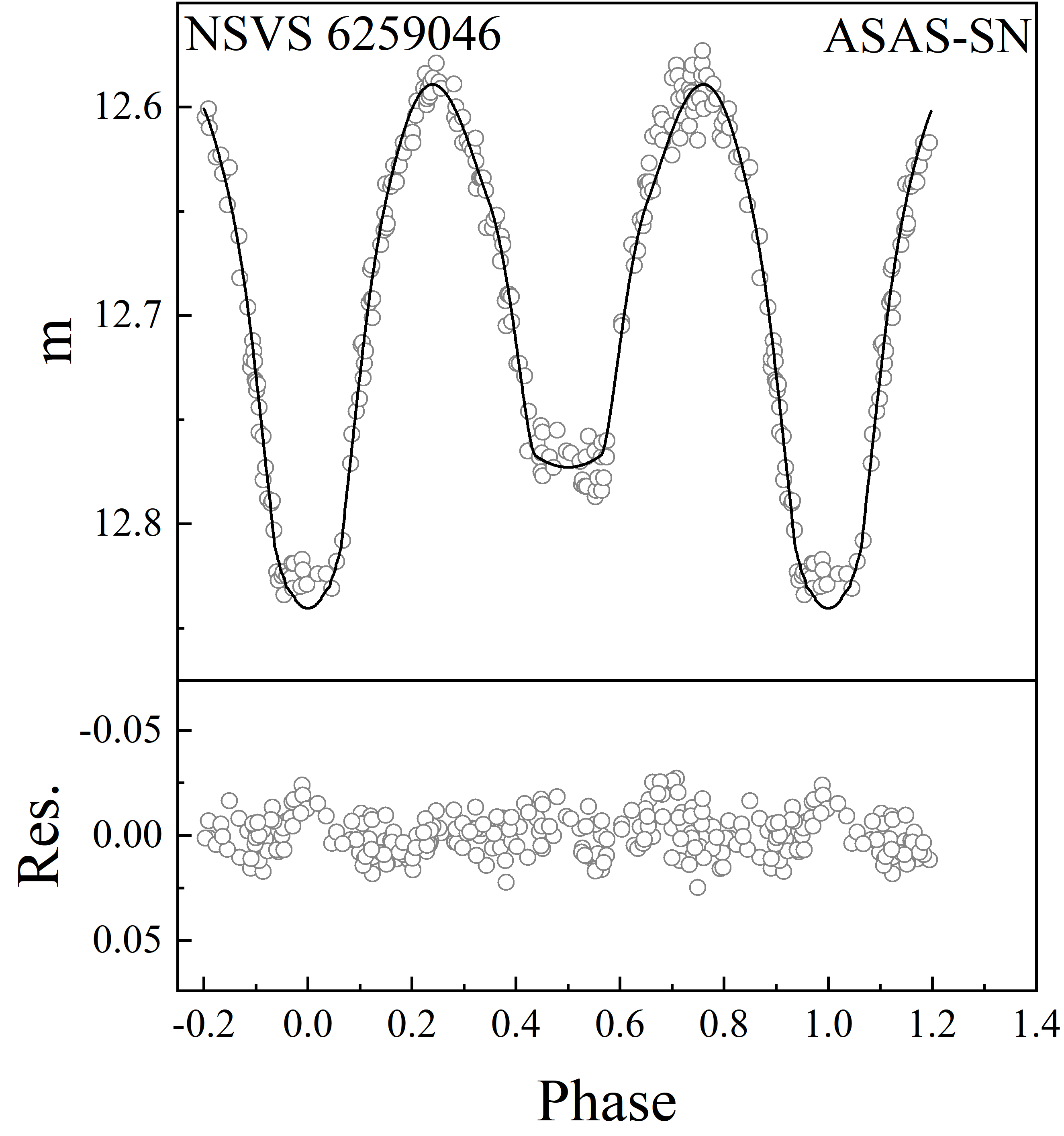}}\subfigure{
\includegraphics[width=5.75cm,height = 5.75cm]{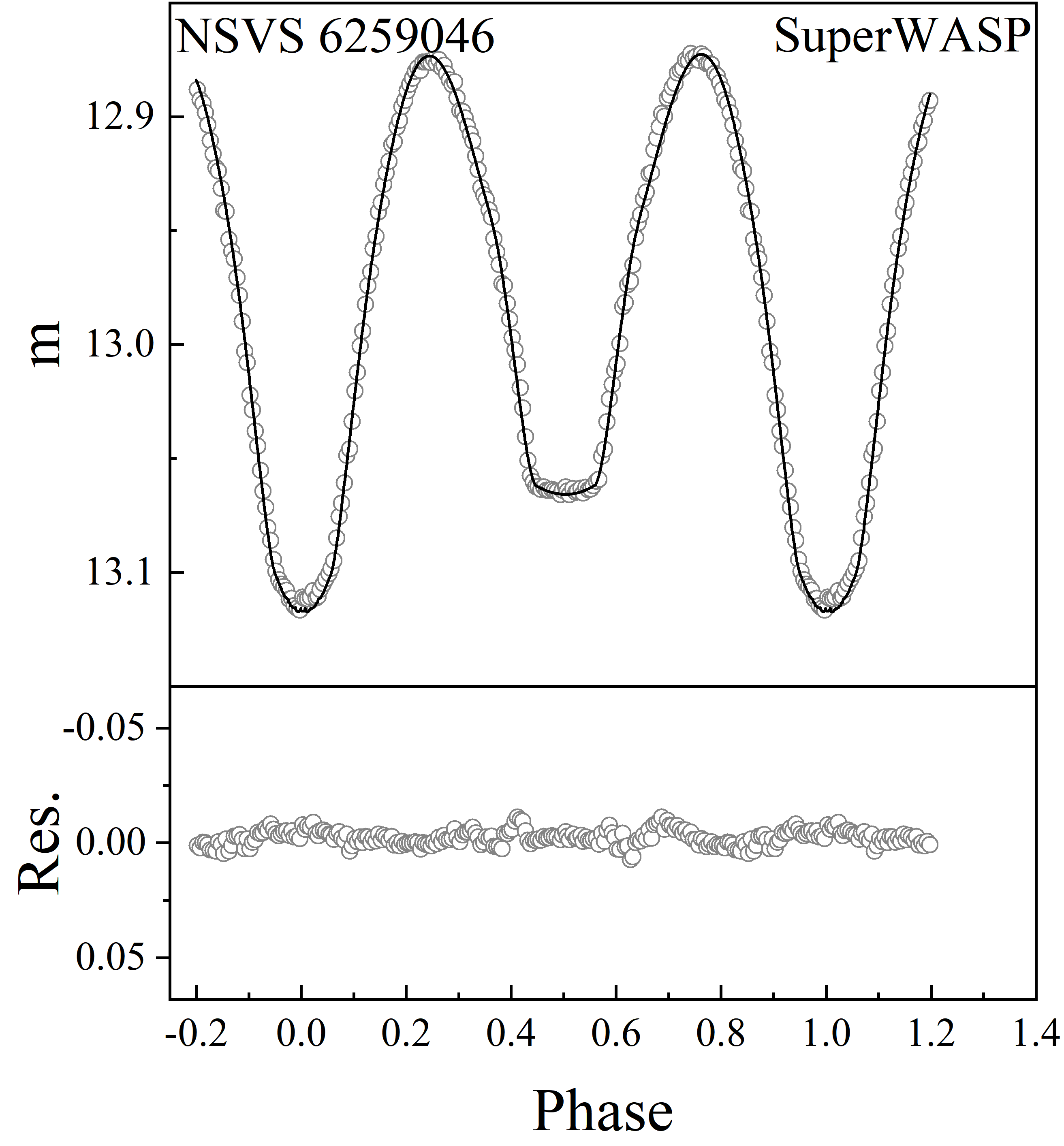}}\subfigure{
\includegraphics[width=5.75cm,height = 5.75cm]{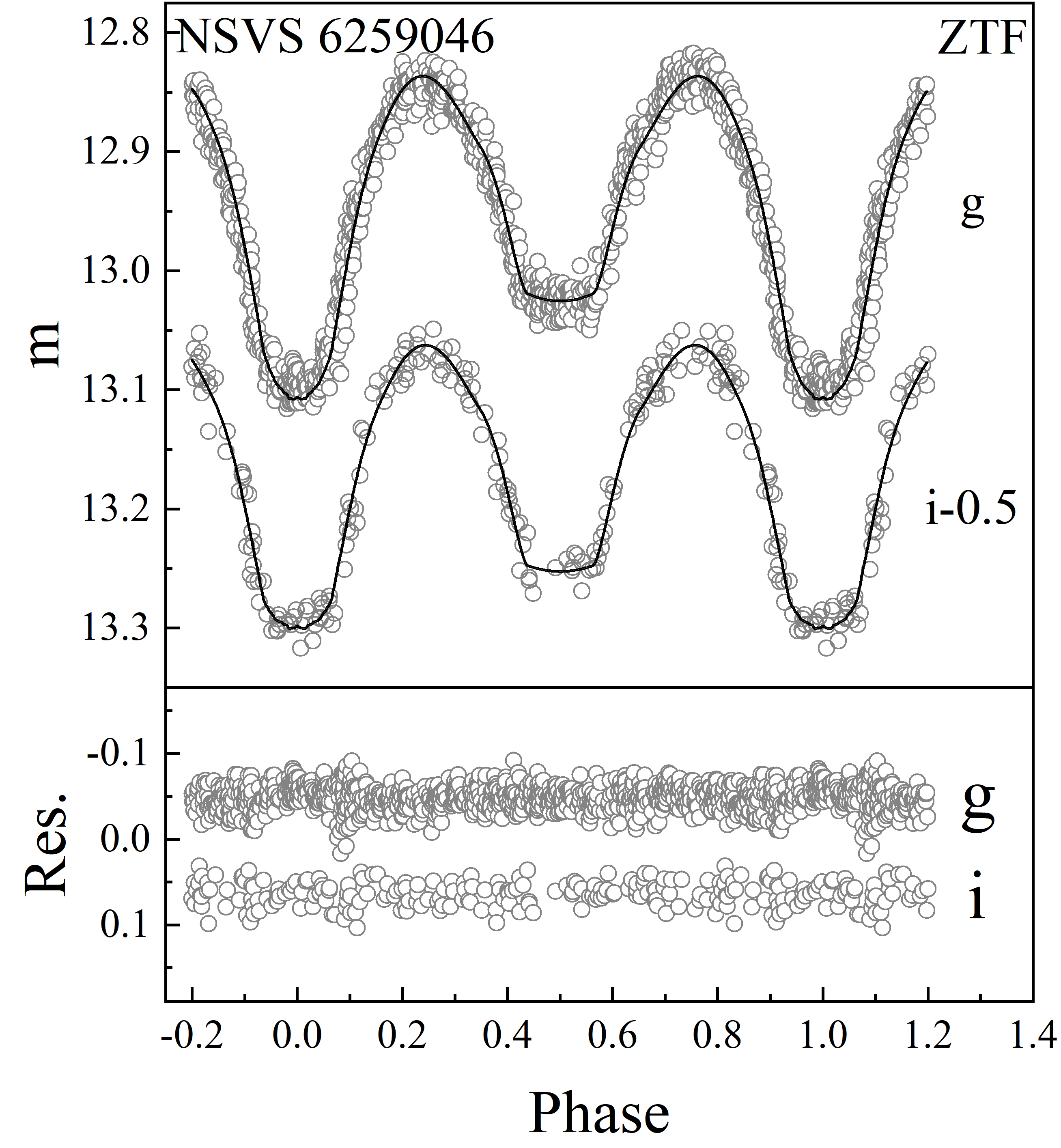}}
\caption{(Continued)}
\end{figure*}

\begin{table*}
    \centering
    \caption{Photometric solutions of V0699 Cep}
    \label{Table:Photometric solutions of V0699 Cep}
    \begin{tabular}{lccc}
        \toprule
        Parameter & NEXT & TESS & ASAS-SN \\
        \midrule
        \(T_1\) (K) & \(6660 \pm 31\) & \(6662 \pm 31\) & \(6666 \pm 44\) \\
        \(T_2\) (K) & \(6659 \pm 56\) & \(6651 \pm 57\) & \(6641 \pm 119\) \\
        \(q\) (\(M_1/M_2\)) & \(0.240 \pm 0.002\) & \(0.225 \pm 0.001\) & \(0.240 \pm 0.002\) \\
        \(i\) (\(^\circ\)) & \(80.3 \pm 0.2\) & \(78.8 \pm 0.1\) & \(77.1 \pm 3.0\) \\
        \(\Omega\) & \(2.274 \pm 0.003\) & \(2.229 \pm 0.003\) & \(2.286 \pm 0.017\) \\
        \(r_1\) & \(0.524 \pm 0.001\) & \(0.534 \pm 0.001\) & \(0.523 \pm 0.005\) \\
        \(r_2\) & \(0.282 \pm 0.004\) & \(0.284 \pm 0.003\) & \(0.281 \pm 0.007\) \\
        \(L_2/(L_1+L_2+L_3)\) (g) & \(0.187 \pm 0.030\) & - & - \\
        \(L_3/(L_1+L_2+L_3)\) (g) & \(0.163 \pm 0.005\) & - & - \\
        \(L_2/(L_1+L_2+L_3)\) (r) & \(0.177 \pm 0.036\) & - & - \\
        \(L_3/(L_1+L_2+L_3)\) (r) & \(0.205 \pm 0.004\) & - & - \\
        \(L_2/(L_1+L_2+L_3)\) (i) & \(0.173 \pm 0.039\) & - & - \\
        \(L_3/(L_1+L_2+L_3)\) (i) & \(0.224 \pm 0.004\) & - & - \\
        \(L_2/(L_1+L_2+L_3)\) (TESS-Sector 16) & - & \(0.184 \pm 0.026\) & - \\
        \(L_3/(L_1+L_2+L_3)\) (TESS-Sector 16) & - & \(0.139 \pm 0.003\) & - \\
        \(L_2/(L_1+L_2+L_3)\) (TESS-Sector 17) & - & \(0.180 \pm 0.029\) & - \\
        \(L_3/(L_1+L_2+L_3)\) (TESS-Sector 17) & - & \(0.162 \pm 0.003\) & - \\
        \(L_2/(L_1+L_2+L_3)\) (V) & - & - & \(0.169 \pm 0.043\) \\
        \(L_3/(L_1+L_2+L_3)\) (V) & - & - & \(0.254 \pm 0.034\) \\
        \(f\) (\%) & \(37.3 \pm 2.2\) & \(45.1 \pm 1.9\) & - \\
        Cool Spot & Star 2 & - & - \\
        \(\theta^\circ\) & \(135 \pm 1\) & - & - \\
        \(\lambda^\circ\) & \(288 \pm 3\) & - & - \\
        \(r_{\text{spot}}\) & \(33 \pm 1\) & - & - \\
        \(T_{\text{spot}}\) & \(0.787 \pm 0.014\) & - & - \\
        \bottomrule
    \end{tabular}
\end{table*}

\begin{table*}
    \centering
    \caption{Photometric solutions of V0844 Aur}
    \label{Table:Photometric solutions of V0844 Aur}
    \begin{tabular}{lcccccc}
        \toprule
        V0844 Aur & NEXT & TESS & ASAS-SN & SuperWASP \\
        \midrule
        \(T_1\) (K) & \(6642 \pm 239\) & \(6537 \pm 244\) & \(6540 \pm 244\) & \(6556 \pm 241\) \\
        \(T_2\) (K) & \(6130 \pm 408\) & \(6604 \pm 448\) & \(6591 \pm 443\) & \(6521 \pm 440\) \\
        \(q\) (\(M_1/M_2\)) & \(0.198 \pm 0.001\) & \(0.193 \pm 0.001\) & \(0.198 \pm 0.001\) & \(0.177 \pm 0.003\) \\
        \(i\) (\(^\circ\)) & \(78.8 \pm 0.2\) & \(78.4 \pm 0.2\) & \(77.1 \pm 0.6\) & \(77.5 \pm 0.6\) \\
        \(\Omega\) & \(2.180 \pm 0.004\) & \(2.159 \pm 0.002\) & \(2.144 \pm 0.005\) & \(2.122 \pm 0.008\) \\
        \(r_1\) & \(0.539 \pm 0.001\) & \(0.545 \pm 0.001\) & \(0.546 \pm 0.002\) & \(0.551 \pm 0.002\) \\
        \(r_2\) & \(0.267 \pm 0.004\) & \(0.269 \pm 0.003\) & \(0.262 \pm 0.002\) & \(0.264 \pm 0.011\) \\
        \(L_2 / (L_1 + L_2)\) (B) & \(0.129 \pm 0.002\) & - & - & - \\
        \(L_2 / (L_1 + L_2)\) (V) & \(0.144 \pm 0.001\) & - & \(0.183 \pm 0.002\) & - \\
        \(L_2 / (L_1 + L_2)\) (g) & \(0.135 \pm 0.001\) & - & - & - \\
        \(L_2 / (L_1 + L_2)\) (r) & \(0.150 \pm 0.001\) & - & - & - \\
        \(L_2 / (L_1 + L_2)\) (i) & \(0.158 \pm 0.001\) & - & - & - \\
        \(L_2 / (L_1 + L_2)\) (TESS-Sector 19) & - & \(0.198 \pm 0.001\) & - & - \\
        \(L_2 / (L_1 + L_2)\) (TESS-Sector 59) & - & \(0.198 \pm 0.001\) & - & - \\
        \(L_2 / (L_1 + L_2)\) (TESS-Sector 73) & - & \(0.198 \pm 0.001\) & - & - \\
        \(L_2 / (L_1 + L_2)\) (SuperWASP) & - & - & - & \(0.180 \pm 0.002\) \\
        \(f\) (\%) & \(37.7 \pm 2.9\%\) & \(45.4 \pm 1.8\%\) & \(40.0 \pm 4.3\%\) & \(46.9 \pm 7.0\%\) \\
        Cool Spot & Star 1 & - & - & - \\
        \(\theta^\circ\) & \(125 \pm 1\) & - & - & - \\
        \(\lambda^\circ\) & \(179 \pm 0\) & - & - & - \\
        \(r_{\text{spot}}\) & \(22 \pm 0\) & - & - & - \\
        \(T_{\text{spot}}\) & \(0.738 \pm 0.008\) & - & - & - \\
        \bottomrule
    \end{tabular}
\end{table*}

\begin{table*}
    \centering
    \caption{Photometric solutions of NSVS 6259046}
    \label{Table:Photometric solutions of NSVS 6259046}
    \begin{tabular}{lcccccc}
        \toprule
        NSVS 6259046 & \multicolumn{2}{c}{NEXT} & TESS & ASAS-SN & SuperWASP & ZTF \\
        \midrule
        \(T_1\) (K) & \multicolumn{2}{c}{\(6677 \pm 22\)} & \(6699 \pm 20\) & \(6698 \pm 21\) & \(6655 \pm 19\) & \(6699 \pm 24\) \\
        \(T_2\) (K) & \multicolumn{2}{c}{\(6360 \pm 21\)} & \(6140 \pm 18\) & \(6360 \pm 21\) & \(6528 \pm 19\) & \(6127 \pm 22\) \\
        \(q\) (\(M_1/M_2\)) & \multicolumn{2}{c}{\(0.090 \pm 0.001\)} & \(0.091 \pm 0.000\) & \(0.090 \pm 0.001\) & \(0.064 \pm 0.000\) & \(0.093 \pm 0.001\) \\
        \(i\) (\(^\circ\)) & \multicolumn{2}{c}{\(85.3 \pm 0.2\)} & \(83.8 \pm 0.1\) & \(84.0 \pm 1.0\) & \(75.1 \pm 0.2\) & \(83.8 \pm 0.3\) \\
        \(\Omega\) & \multicolumn{2}{c}{\(1.889 \pm 0.003\)} & \(1.903 \pm 0.001\) & \(1.910 \pm 0.002\) & \(1.798 \pm 0.000\) & \(1.921 \pm 0.005\) \\
        \(r_1\) & \multicolumn{2}{c}{\(0.606 \pm 0.001\)} & \(0.600 \pm 0.001\) & \(0.597 \pm 0.001\) & \(0.637 \pm 0.000\) & \(0.594 \pm 0.002\) \\
        \(r_2\) & \multicolumn{2}{c}{\(0.221 \pm 0.008\)} & \(0.215 \pm 0.003\) & \(0.207 \pm 0.001\) & \(0.216 \pm 0.001\) & \(0.210 \pm 0.008\) \\
        \(L_2 / (L_1 + L_2)\) (g) & \multicolumn{2}{c}{\(0.090 \pm 0.001\)} & - & - & - & \(0.072 \pm 0.001\) \\
        \(L_2 / (L_1 + L_2)\) (r) & \multicolumn{2}{c}{\(0.096 \pm 0.001\)} & - & - & - & \(0.086 \pm 0.001\) \\
        \(L_2 / (L_1 + L_2)\) (i) & \multicolumn{2}{c}{\(0.099 \pm 0.001\)} & - & - & - & - \\
        \(L_2 / (L_1 + L_2)\) (SuperWASP) & \multicolumn{2}{c}{-} & - & - & \(0.086 \pm 0.001\) & - \\
        \(L_2 / (L_1 + L_2)\) (TESS-Sector 56) & \multicolumn{2}{c}{-} & \(0.087 \pm 0.000\) & - & - & - \\
        \(L_2 / (L_1 + L_2)\) (V) & \multicolumn{2}{c}{-} & - & \(0.074 \pm 0.002\) & - & - \\
        \(f\) (\%) & \multicolumn{2}{c}{\(64.4 \pm 4.7\%\)} & \(46.7 \pm 2.3\%\) & - & - & - \\
        Cool and Hot Spot & Star 1 (Cool Spot) & Star 2 (Hot Spot) & - & - & Star 2 (Cool Spot) & - \\
        \(\theta^\circ\) & \(174 \pm 1\) & \(40 \pm 4\) & - & - & \(43 \pm 2\) & - \\
        \(\lambda^\circ\) & \(140 \pm 7\) & \(151 \pm 8\) & - & - & \(178 \pm 2\) & - \\
        \(r_{\text{spot}}\) & \(33 \pm 1\) & \(33 \pm 1\) & - & - & \(32 \pm 1\) & - \\
        \(T_{\text{spot}}\) & \(0.802 \pm 0.011\) & \(1.202 \pm 0.023\) & - & - & \(0.802 \pm 0.015\) & - \\
        \bottomrule
    \end{tabular}
\end{table*}

\begin{table*}
    \centering
    \caption{Photometric solutions of V0508 And}
    \label{Table:Photometric solutions of V0508 And}
    \begin{tabular}{lccccc}
        \toprule
        Parameter & NEXT & TESS & ASAS-SN & SuperWASP & ZTF \\
        \midrule
        \(T_1\) (K) & \(6624 \pm 17\) & \(6626 \pm 15\) & \(6624 \pm 16\) & \(6620 \pm 16\) & \(6623 \pm 23\) \\
        \(T_2\) (K) & \(6292 \pm 37\) & \(6233 \pm 28\) & \(6237 \pm 44\) & \(6279 \pm 30\) & \(6251 \pm 57\) \\
        \(q\) (\(M_1/M_2\)) & \(0.094 \pm 0.001\) & \(0.088 \pm 0.000\) & \(0.094 \pm 0.001\) & \(0.080 \pm 0.000\) & \(0.085 \pm 0.002\) \\
        \(i\) (\(^\circ\)) & \(84.3 \pm 0.2\) & \(83.2 \pm 0.1\) & \(85.0 \pm 0.2\) & \(79.3 \pm 0.2\) & \(85.1 \pm 0.4\) \\
        \(\Omega\) & \(1.915 \pm 0.003\) & \(1.909 \pm 0.000\) & \(1.927 \pm 0.002\) & \(1.877 \pm 0.002\) & \(1.901 \pm 0.010\) \\
        \(r_1\) & \(0.597 \pm 0.001\) & \(0.596 \pm 0.000\) & \(0.589 \pm 0.001\) & \(0.607 \pm 0.001\) & \(0.598 \pm 0.004\) \\
        \(r_2\) & \(0.216 \pm 0.006\) & \(0.204 \pm 0.001\) & \(0.198 \pm 0.001\) & \(0.203 \pm 0.005\) & \(0.202 \pm 0.019\) \\
        \(L_2 / (L_1 + L_2)\) (g) & \(0.090 \pm 0.001\) & - & - & - & \(0.074 \pm 0.003\) \\
        \(L_2 / (L_1 + L_2)\) (r) & \(0.096 \pm 0.001\) & - & - & - & - \\
        \(L_2 / (L_1 + L_2)\) (i) & \(0.099 \pm 0.002\) & - & - & - & - \\
        \(L_2 / (L_1 + L_2)\) (TESS) & - & \(0.088 \pm 0.000\) & - & - & - \\
        \(L_2 / (L_1 + L_2)\) (V) & - & - & \(0.079 \pm 0.001\) & - & - \\
        \(L_2 / (L_1 + L_2)\) (SuperWASP) & - & - & - & \(0.077 \pm 0.001\) & - \\
        \(f\) (\%) & \(43.2 \pm 4.1\%\) & \(22.4 \pm 0.9\%\) & - & \(37.0 \pm 4.1\%\) & - \\
        \bottomrule
    \end{tabular}
\end{table*}

\section{\textit{O - C} Analysis}
\begin{figure*}
\centering  
\subfigure{
\includegraphics[width=8cm,height = 8cm]{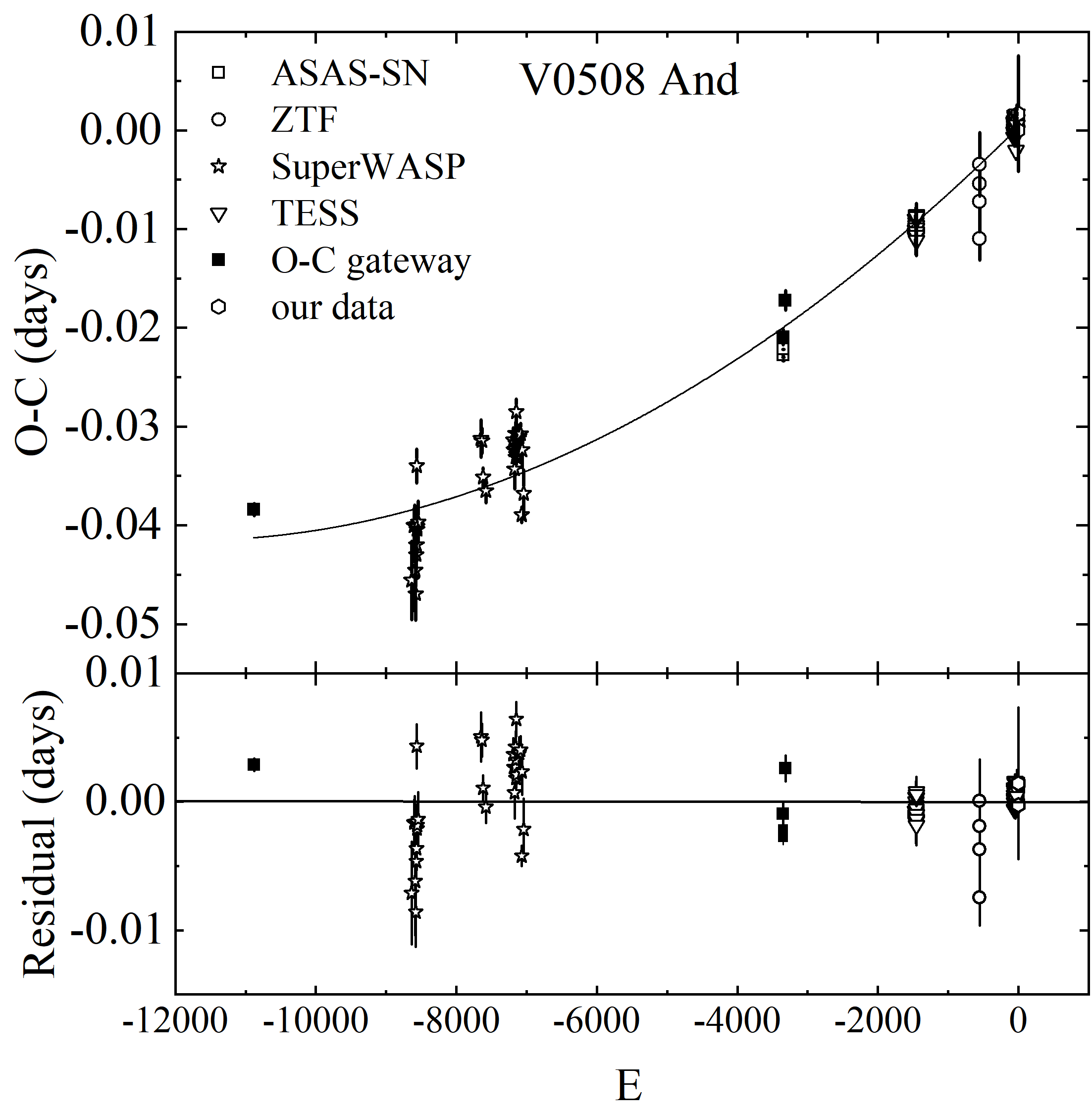}}\subfigure{\includegraphics[width=8cm,height = 8cm]{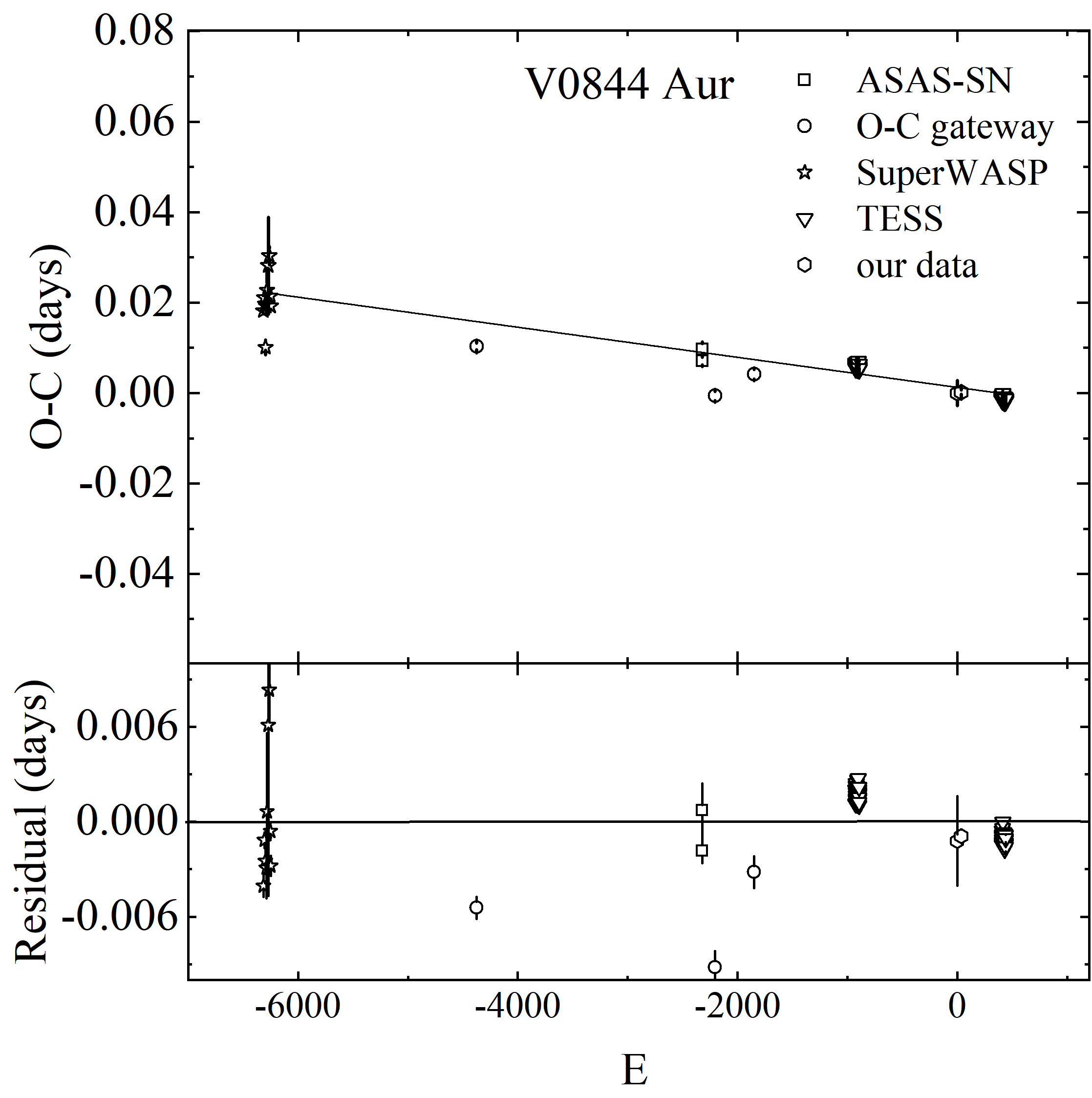}}
\subfigure{\includegraphics[width=8cm,height =8cm]{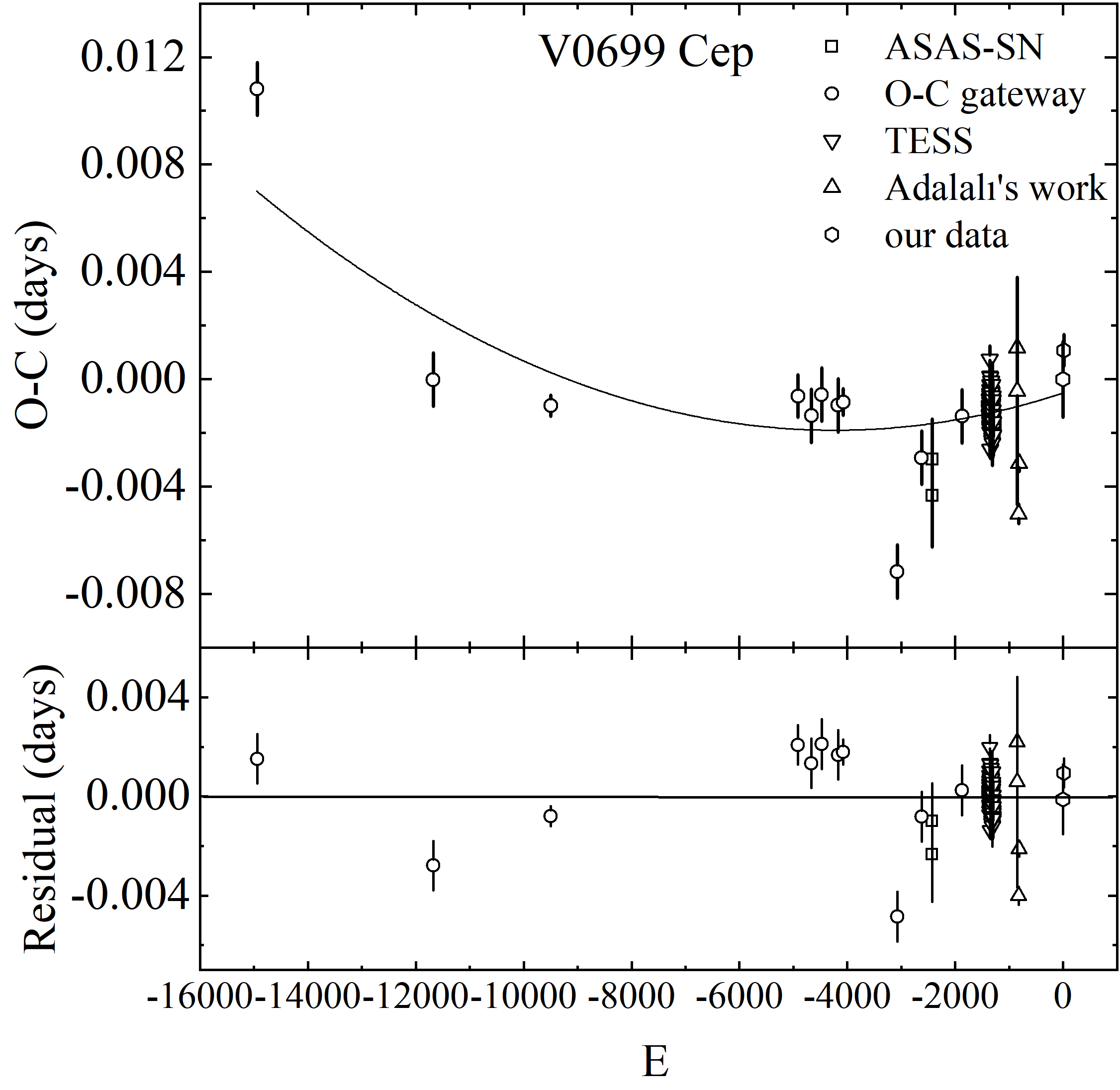}}\subfigure{\includegraphics[width=8cm,height = 8cm]{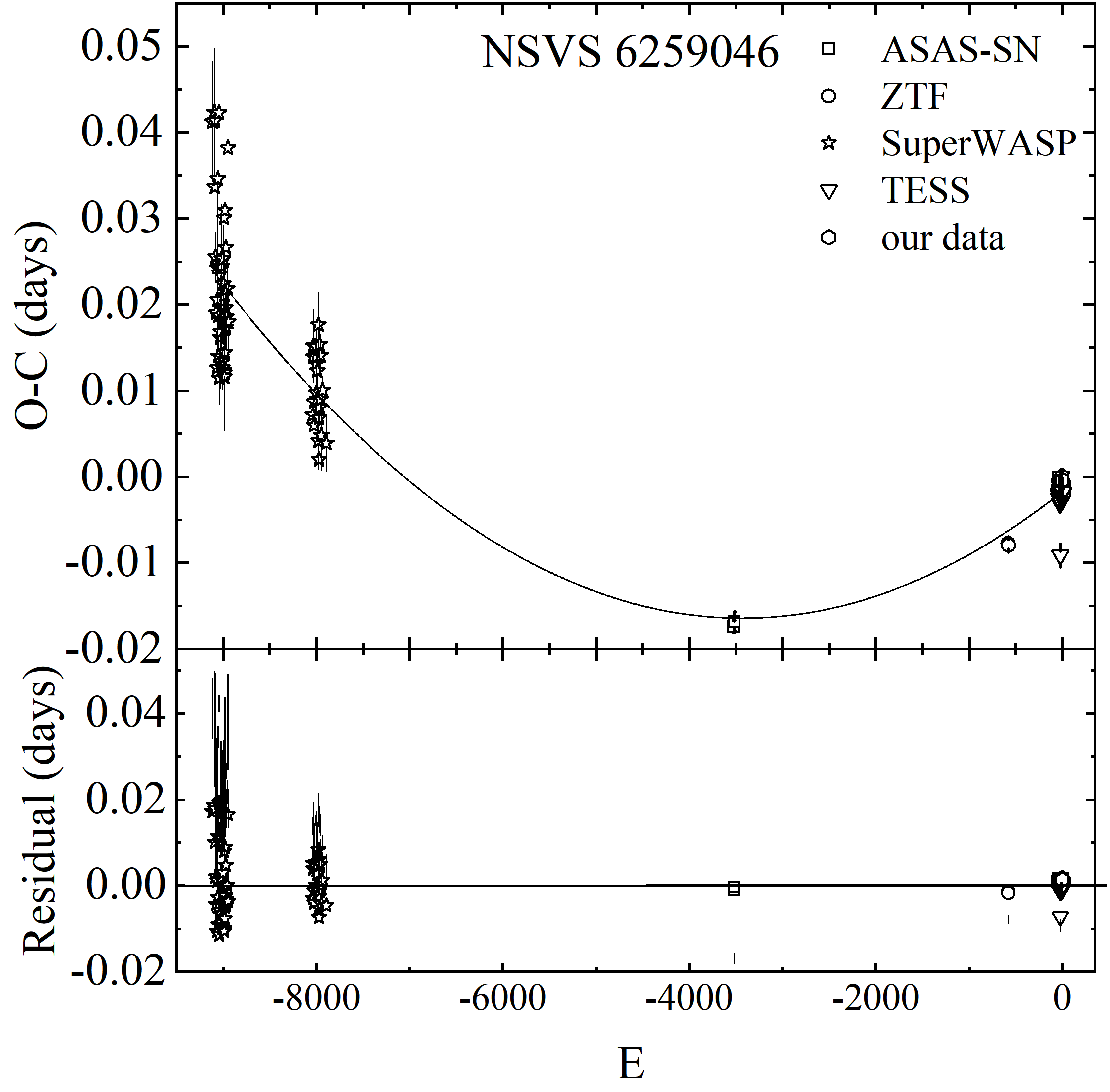}}
\caption{The upper illustration shows the $O - C$ diagram and the lower illustration refers to the residuals. Eclipsing times are from NEXT, ASAS-SN, CRTS, TESS, SuperWASP, ZTF, $O - C$ Gateway and \protect\cite{2024NewA..11202270A}.}
\label{Figure:The O–C diagrams of the four targets}
\end{figure*}

Studying changes in orbital periods is crucial for understanding the dynamic interactions of binary systems, including the potential existence of a third body and how material transfers between the two components \citep{2006JASS...23..105K,2014A&A...570A..25D,2016ApJ...817..133Z,2019ApJ...871L..10E,2018PASP..130g4201L,2019AJ....157..207L,2019ApJ...877...75P}. We gather as many times of minimum light as possible from our observations and photometric surveys, including data from NEXT, ASAS-SN, CRTS, TESS, SuperWASP, ZTF and $O - C$ Gateway.\footnote{\href{https://var.astro.cz/en}{https://var.astro.cz/en}} Previous studies \citep{2024NewA..11202270A} also analyzed V0699 Cep, and we have collected 4 eclipsing times from them.
For NEXT, TESS and SuperWASP data, eclipsing times are calculated directly using the K–W method \citep{1956BAN....12..327K}. A technique called period shift method \citep{2020AJ....159..189L} is applied to determine the eclipsing times for CRTS, ASAS-SN, and ZTF, whose data points are relatively scarce and do not form a complete eclipsing curve. We convert the calculated eclipsing minimum from each survey to Barycentric Julian Date (BJD) through Time Utilities \citep{2010PASP..122..935E}.\footnote{\href{https://astroutils.astronomy.osu.edu/time/}{https://astroutils.astronomy.osu.edu/time/}} Then, the following equation is used to find the calculated times,
\begin{equation}
      Min.I=Min.I_0+P\times E,
\end{equation}
where $Min.I$ and $Min.I_0$ denote the observed minimum and initial epoch, respectively, and E represents the cycle number. Table \ref{Table:The Eclipsing times and O – C values of the four targets} presents all eclipsing times and $O - C$ values. Fig. \ref{Figure:The O–C diagrams of the four targets} displays the $O - C$ diagrams of the four targets. V0699 Cep, NSVS 6259046 and V0508 And show distinct parabolic trends, for which the following equation is used to fit their $O - C$ curves,
\begin{equation}
      O-C=\varDelta Min.I_0+\varDelta P\times E+\frac{\beta}{2}\times E^2,
\end{equation}
where $\Delta Min.I_0$ represents the adjustment made to the initial epoch, $\Delta P$ denotes a correction applied to the orbital period, and $\beta$ signifies the long-term orbital period changes rate. The outcomes are listed in Table \ref{Table:Fitting parameters of $O-C$ diagrams}. The positive $\beta$ indicates that the orbital periods of V0699 Cep, NSVS 6259046 and V0508 And are all increasing in the long term. Conversely, there is no evidence toward a parabolic tendency for V0844 Aur, we linearly fit its $O - C$ curve.

\begin{table*}
\centering
\caption{The Eclipsing times and $O - C$ values of the four targets.}
\label{Table:The Eclipsing times and O – C values of the four targets}
\begin{tblr}{
  column{even} = {c},
  column{3} = {c},
  column{5} = {c},
  column{7} = {c},
  hline{1,3,11} = {-}{},
}
Targets & BJD         & E       & $O - C$       & Residual    & Source \\
        & (2400000+)  &         & (days)    &    (days)     &           \\
V0508 And & 51464.04070 & -10880  & -0.03839  & 0.00040 & O-C gateway    \\
          & … & …    & …  & … & …      \\
V0844 Aur & 54382.68684 & -6315.5 & 0.01816 & 0.00068     & SuperWASP    \\
          & … & …   & …  & …     & …    \\
V0699 Cep & 47804.39825 & -14941.5 & 0.01082 & 0.00100     & O-C gateway  \\
          & … & …   & …  & …     & …      \\
NSVS 6259046 & 53153.69265 & -9117 & 0.04125 & 0.00700     & SuperWASP    \\
          & … & …   & …  & …     & …      \\
\end{tblr}
\begin{tablenotes}   
        \footnotesize               	  
        \item[] Note. This table is available in its entirety in machine-readable form.	  
\end{tablenotes}         
\end{table*}

\begin{table*}
    \centering
    \caption{Fitting parameters including rates of period variation and mass transfer of the four CBs by $O - C$ analysis.}
    \label{Table:Fitting parameters of $O-C$ diagrams}
    \begin{tabular}{lccccccc}
        \hline
        Targets & $\Delta Min.I_0$ & Error & $\Delta P$ & Error & $\beta$ & Error & $\frac{dM_1}{dt}$ \\
        & ($\times 10^{-4}$ d) & ($\times 10^{-4}$ d) & ($\times 10^{-7}$ d) & ($\times 10^{-7}$ d) & ($\times 10^{-7}$ d yr$^{-1}$) & ($\times 10^{-7}$ d yr$^{-1}$) & ($\times 10^{-8} M_{\odot}$ yr$^{-1}$) \\
        \hline
        V0508 And & 2.7 & 2.6 & 70.5     & 2.8     & 2.8  & 0.3  & 2.9 \\
        V0844 Aur & 12.2  & 2.0 & $-33.3$   & 1.1    & --       & --       & -- \\
        V0699 Cep & $-5.3$  & 2.7 & 6.6 & 2.2  & 0.7  & 0.2  & 1.9 \\
        NSVS 6259046 & $-17.6$ & 5.9 & 85.7     & 10.8     & 12.4  & 1.2  & 12.6 \\
        \hline
    \end{tabular}
\end{table*}

\section{Spectral Analysis}
\begin{figure*}
\centering
\subfigure{
\includegraphics[width=6cm,height = 4cm]{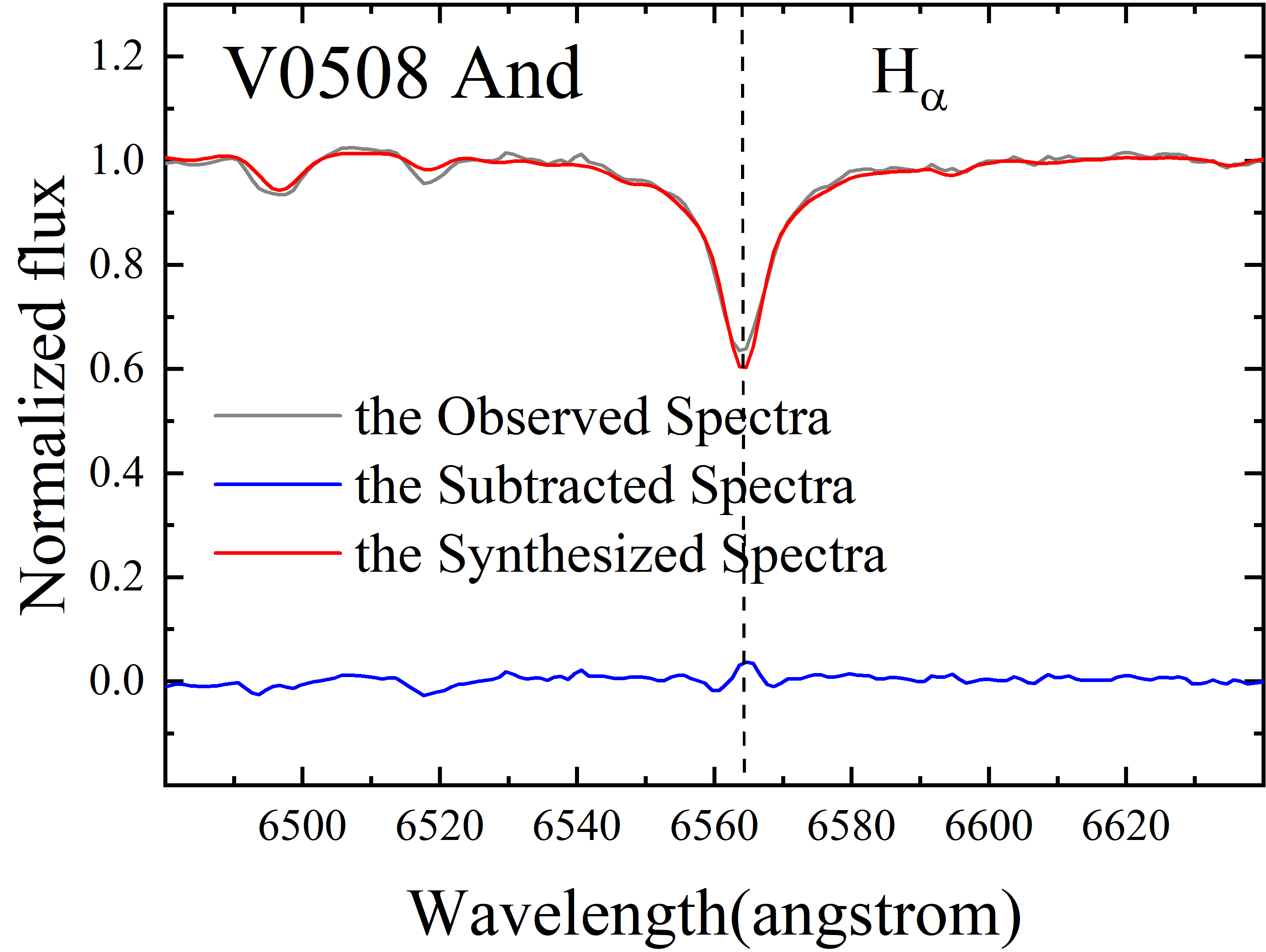}}\subfigure{
\includegraphics[width=6cm,height = 4cm]{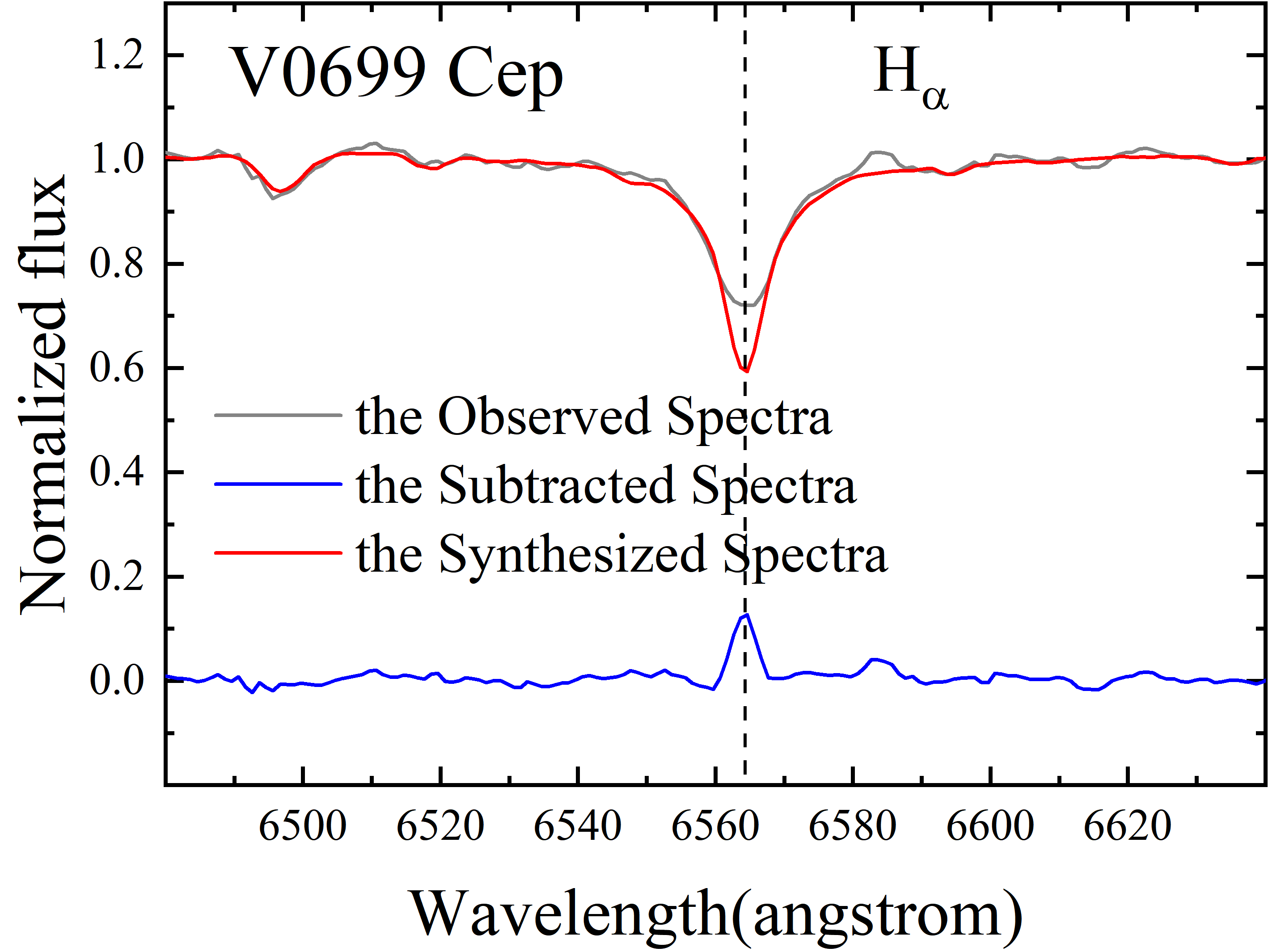}}\subfigure{
\includegraphics[width=6cm,height = 4cm]{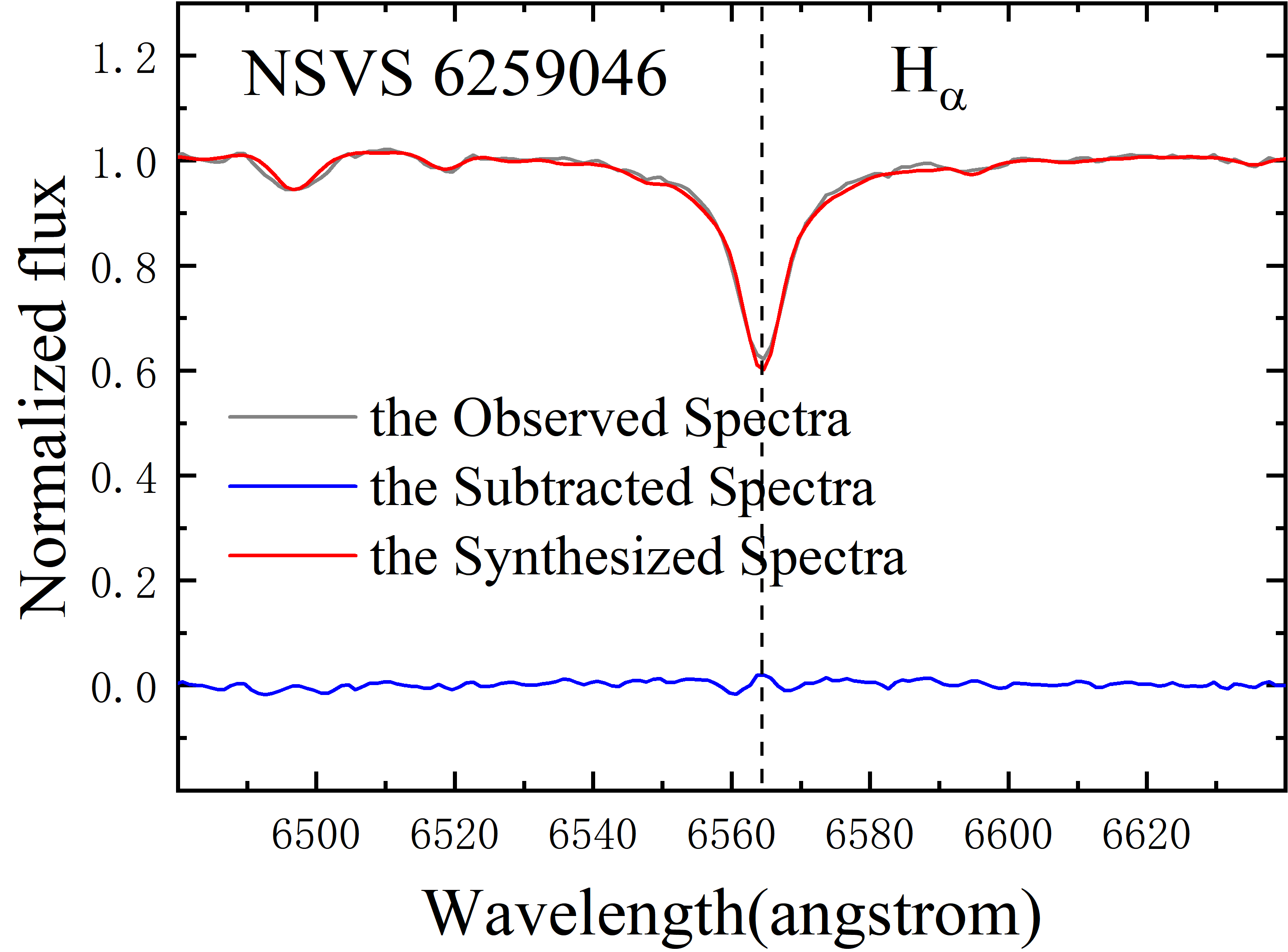}}
\caption{The normalized observed, synthesized, and subtracted spectra near the location of $H_{\alpha}$ observed by LAMOST.}
\label{Figure:The spectroscopic investigation of our targets}
\end{figure*}

The emission lines in the spectrum such as $H_{\alpha}$, $H_{\beta}$, Ca~\textnormal{II} H\&K and IRT, etc. can serve as reliable indicators of chromospheric activity \citep{1978ApJ...226..379W,1984OrGeo...5..217B,1995A&AS..114..287M}. According to Table \ref{Table:Spectral observations of our targets from LAMOST}, we choose V0699 Cep, NSVS 6259046 and V0508 And, whose signal-to-noise ratios are high enough for us to analyze. The observed spectrum of a star, however, generally exhibits absorption in the $H_{\alpha}$ line, which is attributed to the stellar photosphere's absorption. Thus, we use the spectral subtraction method according to \cite{1985ApJ...295..162B}, removing photospheric spectrum from the observed spectrum to obtain pure chromospheric spectrum. First, we select spectra of two inactive stars for each system from the catalog of \cite{2021ApJS..256...14Z}, ensuring that the temperature difference between the inactive stars and the target was less than 200 K. One spectrum is used as the primary template, while the other served as the secondary template. After downloading and normalizing the spectra, we apply the STARMOD code \citep{1985ApJ...295..162B} to generate synthetic spectrum for the two inactive stars, considering radial velocities, spin angular velocities, and the luminosity ratios of the components. Next we obtain the subtracted spectra from the synthetic spectra and the target spectra, and the results are shown in Fig. \ref{Figure:The spectroscopic investigation of our targets}. For NSVS 6259046 and V0508 And, there is no $H_\alpha$ emission line, which indicates no chromospheric activity for these two targets. Only V0699 Cep shows clear $H_{\alpha}$ emission lines, suggesting it possesses chromospheric activity. So we calculate its equivalent width (EW) of the $H_{\alpha}$ line using package of the \href{https://iraf.noirlab.edu/}{IRAF} to further quantify its chromospheric activity. The EW is found out to be 0.41\,\AA. The EW of $H_{\alpha}$ for other CBs with $T_{\scriptsize \text{eff}}$ higher than 5900 K are presented in Table \ref{Table:The Halpha EW of other early type low mass-ratio CBs}. The comparative analysis shows that V0699 Cep displays a relatively high degree of chromospheric activity among early-type stars.

\begin{table*}
\centering
\caption{The $H_{\alpha}$ EW of other CBs with $T_{\scriptsize \text{eff}}$ higher than 5900 K. We find that V0699 Cep exhibits a larger $H_{\alpha}$ EW compared to other CBs with $T_{\scriptsize \text{eff}}$ higher than 5900 K.}
\label{Table:The Halpha EW of other early type low mass-ratio CBs}
\begin{tabular}{lcccccc}
\toprule
Star & UT Date & SNR & Spectral type & $T_{\mathrm{eff}}$ & $H_{\alpha}$ & Ref. \\
     &         &     &               & (K)                & (\AA)        &  \\
\midrule
J001422.52+415331.1                  & 2013 Nov 14 & 279 & F0 & 6568($\pm$9)  &  0.294($\pm$0.006) & (1)\\
J022733.60+360447.1                  & 2017 Jan 06 & 174 & F5 & 6477($\pm$9)  &  0.252($\pm$0.003) & (1)\\
J040927.64+463400.6                  & 2012 Dec 21 & 157 & A7 & 7106($\pm$25) &  0.761($\pm$0.005) & (2)\\
J042640.22+370047.7                  & 2013 Feb 02 & 2 & -- & --  & -- & (1)\\
                                     & 2013 Oct 04 & 55 & F7 & 5936($\pm$41)  & 0.285($\pm$0.021) & \\
                                     & 2013 Oct 04 & 89 & F6 & 6054($\pm$18)  & 0.223($\pm$0.011) & \\
                                     & 2015 Jan 12 & 148 & F6 & 6115($\pm$14) & 0.368($\pm$0.037) & \\
J054950.33+584208.9                  & 2013 Feb 18 & 71 & F0 & 6988($\pm$30)  & 0.239($\pm$0.019) & (1)\\
                                     & 2013 Dec 27 & 131 & F0 & 6987($\pm$15) & 0.223($\pm$0.009) & \\
J062623.89+275644.2                  & 2012 Feb 29 & 172 & F0 & 7072($\pm$16) & 0.011($\pm$0.004) & (2)\\
J110658.46+511200.1                  & 2013 Dec 21 & 203 & F5 & 6252($\pm$12) & 0.372($\pm$0.022) & (1)\\
                                     & 2015 Dec 30 & 233 & F6 & 6255($\pm$11) & 0.323($\pm$0.030) & \\
                                     & 2016 Jan 28 & 121 & F7 & 6195($\pm$20) & 0.200($\pm$0.018) & \\
J115742.23+074820.3                  & 2014 May 22 & 56 & F5 & 5997($\pm$47)  & 0.202($\pm$0.013) & (1)\\
J153433.52+122516.7                  & 2015 Feb 13 & 14 & F6 & 6077($\pm$366) & 0.638($\pm$0.065) & (1)\\
J190254.60+395028.6                  & 2020 Sep 21 & 55 & F5 & 6454($\pm$49)  & 0.240($\pm$0.006) & (2)\\
J192211.74+492834.2                  & 2016 Dec 11 & 386 & F0 & 7307($\pm$7)  & 0.007($\pm$0.004) & (2)\\                  
J192728.90+421943.1                  & 2013 May 20 & 106 & F5  & 6619($\pm$28) & 0.104($\pm$0.005) & (2)\\
J233332.90+180429.9                  & 2014 Oct 13 & 73 & F0 & 6728($\pm$27)   & 0.301($\pm$0.026) & (1)\\
HAT 307-0007476                      & 2015 Dec 20 & 20 & F0 & 6843($\pm$68)   & 0.373($\pm$0.016) & (3)\\
                                     & 2015 Dec 27 & 24 & F0 & 6821($\pm$64)   & 0.258($\pm$0.031) & \\
                                     & 2016 Jan 12 & 30 & A9 & 6820($\pm$155)  & 0.296($\pm$0.025) & \\
                                     & 2016 Nov 04 & 56 & F0 & 6895($\pm$35)   & 0.303($\pm$0.005) & \\
V342 UMa                             & 2022 Feb 24 & 22 & F7 & 5923($\pm$109)  & 0.096($\pm$0.048) & (4)\\
\bottomrule
\end{tabular}

\begin{tablenotes}   
        \footnotesize               	  
        \item[] References. (1)\cite{2022AJ....164..202L},
(2)\cite{2025ApJS..280...26S},
(3)\cite{2025MNRAS.537.2258L},
(4)\cite{2025MNRAS.537.3366Z}.
      \end{tablenotes}         
\end{table*}

\section{Discussion and Conclusion}
We conduct a photometric and spectroscopic investigation of four CBs with period longer than 0.7 days. The reliability of our photometric results is supported by the characteristic flat secondary eclipsing minima, as earlier studies have demonstrated that in the absence of large unidentified third light, the photometric and spectroscopic methods yield nearly equal mass ratios for totally eclipsing contact binaries \citep{2003yCat.5119....0P,2005Ap&SS.296..221T,2021AJ....162...13L}.

The contact degree (\( f \)) serves as a key criterion for classifying contact binary systems into deep, medium, and shallow contact categories. A system is classified as deep contact if \( f \geq 50\% \), medium contact if \( 25\% \leq f < 50\% \), and shallow contact if \( f < 25\% \) \citep{2005AJ....130..224Q,2023MNRAS.519.5760L}. V0699 Cep (\( f \) = 37.3\%), V0844 Aur (\( f \) = 37.7\%) and V0508 And (\( f \) = 43.2\%) are found to be medium-contact binaries, while NSVS 6259046 (\( f \) = 64.4\%) is found to be a deep-contact binary. The four targets are classified as A-type CBs due to the positive correlation between the mass and temperature of the primary and secondary components \citep{1970VA.....12..217B}. Althrough chromospheric activity usually exists in the kind of systems with spectral type later than F \citep{2008MNRAS.389.1722E}, we find obvious chromospheric activity on V0699 Cep. 

\subsection{O'Connell Effect and Magnetic Activity}
During the analysis of the light curves, we find that V0844 Aur, V0699 Cep and NSVS 6259046 exhibit O'Connell effect, with the fraction of the magnitude difference between the two maxima relative to the total amplitude (depth of primary minimum) being 2.7\%, 11.6\%, and 7.9\%, respectively. In late-type CBs, the asymmetry in the light-curve maxima, known as the O’Connell effect, is typically explained by the presence of dark spots. This phenomenon is attributed to rapid differential rotation and the large convective zone \citep{2011AN....332..607P}. Theoretically, most early-type CBs display symmetric light curves. However, several studies have shown that, despite the absence of convective zones, a significant O'Connell effect can still be observed in some early-type CBs \citep{2011AN....332..607P,2022ApJS..262...10K}. All of our four targets are early-type systems, three of them exhibit O'Connell effect. For V0844 Aur and V0699 Cep, the 0.25-phase maximum is brighter than the 0.75-phase maximum in the NEXT data. As for NSVS 6259046, the 0.75-phase maximum is brighter than the 0.25-phase maximum in the NEXT and SuperWASP data. But in the TESS data, their maxima are all equal. We conclude that these three targets all show variable O’Connell effect. Due to the poor quality of ASAS-SN and ZTF data, the O’Connell effect could not be identified in these datasets. Overall, we infer that these targets exhibit strong magnetic activity. Our study provides evidence supporting that even in early-type systems lacking convective zones, the O'Connell effect can still be present. The relatively weaker O'Connell effect observed in V0844 Aur and NSVS 6259046 compared to V0699 Cep suggests that their magnetic activity may also be less intense. This is consistent with our finding that chromospheric activity is only detected in V0699 Cep. Moreover, the photometric and spectroscopic observations were not conducted simultaneously; therefore, it is possible that the O'Connell effect was detected photometrically while no chromospheric activity was observed spectroscopically.

\subsection{Statistics of 217 CBs and Absolute Parameters Calculation}
\begin{figure}
 \includegraphics[width=\columnwidth]{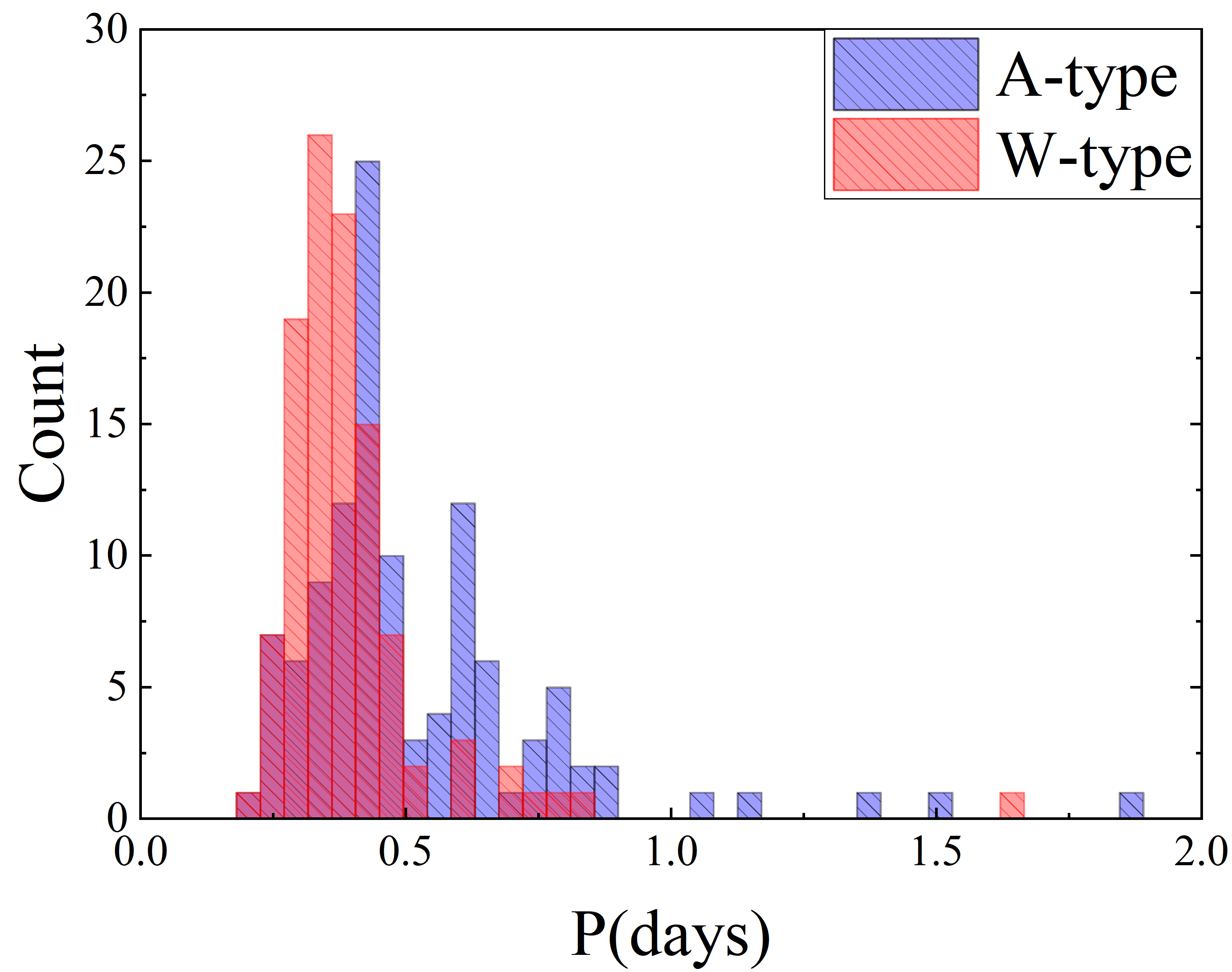}
 \caption{The orbital period distribution of A-type and W-type CBs from the statistical analysis of 217 systems.}
 \label{Figure:The Period distribution of A-type and W-type CBs}
\end{figure}

We compile a dataset of 217 CBs with both reliable radial velocity and photometric observations shown in Table \ref{Table:The 217 CBs with Spectroscopic and Photometric Observations}. We divide A-type and W-type CBs into two groups and investigate their period distributions separately, which is shown in Fig. \ref{Figure:The Period distribution of A-type and W-type CBs}. We find that A-type systems systematically have longer orbital periods than W-type systems.

To assess the evolutionary status of the our targets, we calculate their absolute physical parameters. In order to achieve that, we first calculate the distance between the two components by using the orbital period and semimajor axis (\( P \)–\( a \)) empirical relationship. However, most of the previous work established this relationship based on CBs with periods shorter than 0.7 days \citep[e.g.][]{2024PASP..136b4201P},
\begin{equation}
 a\ =\ \left( 0.372_{-0.114}^{+0.113} \right) +\left( 5.914_{-0.298}^{+0.272} \right) \ \times \ P.
\end{equation}
Among our four systems, none have an orbital period shorter than 0.7 days. To determine their absolute parameters more accurately, we use 44 CBs in Table \ref{Table:The 217 CBs with Spectroscopic and Photometric Observations} to determine a new \( P \)–\( a \) relationship for long-period contact binaries. Given the limited number of CBs with periods longer than 0.7 days, we perform the fitting using systems with period of 0.5–1.0 range to enhance the reliability of the fit.

Regarding the fitting approach, since Kepler's third law is a power law relationship between $a$ and $P$, we apply a logarithmic transformation to them. According to the diagram, we perform a linear regression in log-log space.

\begin{figure*}
 \includegraphics[width=12cm,height = 10cm]{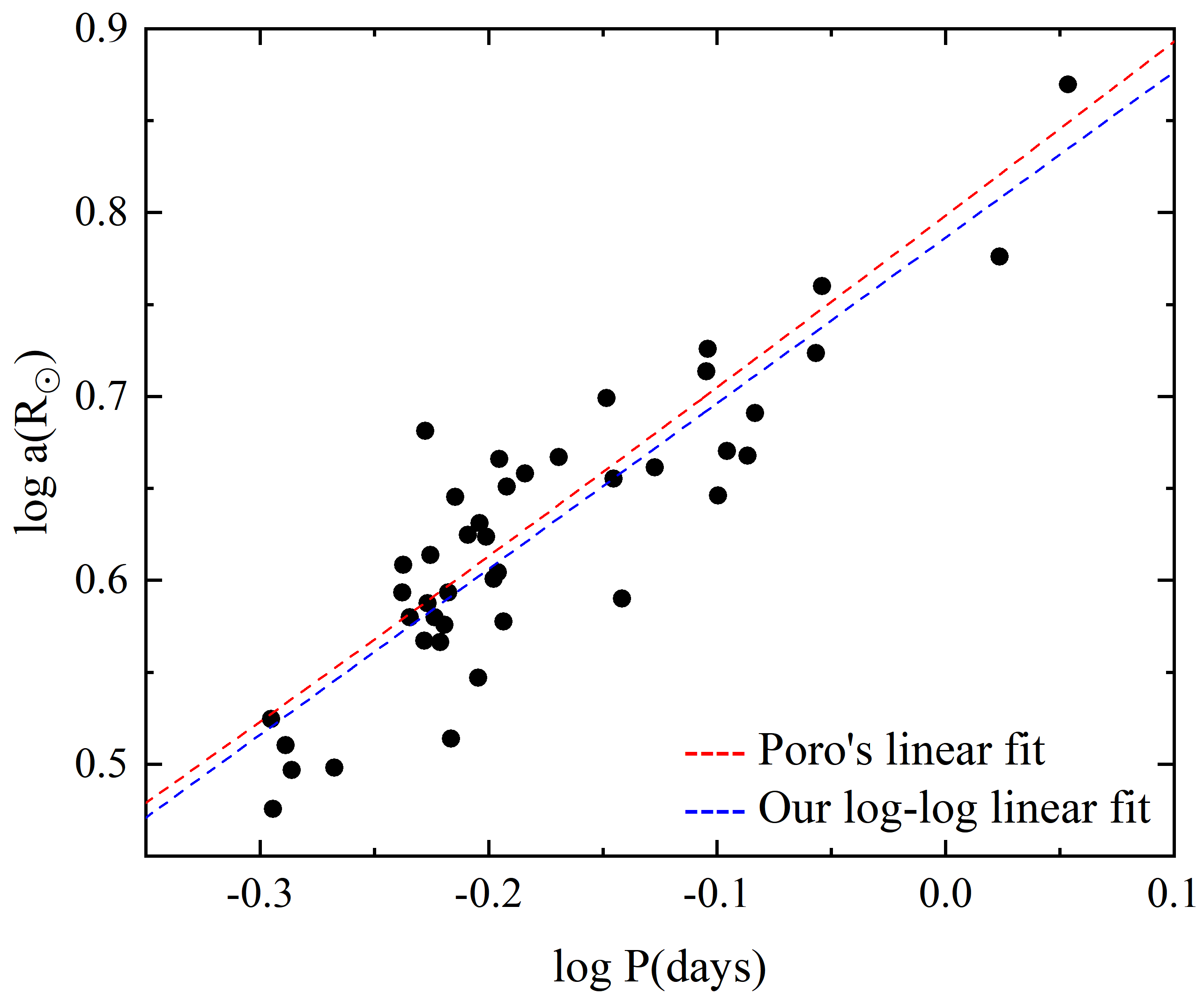}
 \caption{Black solid dots represent a sample of 44 CBs with P > 0.5 days selected from 217 systems. Dashed line represents the relationship between the period \( P \) and the semimajor axis \( a \). Blue represents our linear fit in log-log space using this sample. Red represents the linear relationship established by \protect\cite{2024PASP..136b4201P} using samples with P < 0.7 days, which is expressed in logarithmic form and incorporated into the figure. We have calculated the residual sum of squares for the two methods across the 44 long-period CBs. The residual sum of squares is 0.0582 for our log-log fit, and 0.0611 for Poro's linear fit. With the smaller residual sum of squares, the log-log fit proves to be more suitable for the long-period regime.}
 \label{Figure:Relationship between the period P and the semimajor axis a}
\end{figure*}

A relationship is determined in Fig. \ref{Figure:Relationship between the period P and the semimajor axis a}, which is shown as follows:
\begin{equation}
\log a\,\,=\,\,\left( 0.786\pm 0.014 \right) +\left( 0.901\pm 0.072 \right) \,\,\times \,\,\log P.
 \label{Equation:P–a log-log}
\end{equation}

A direct comparison between our newly derived relation and that of \cite{2024PASP..136b4201P} is carried out. We quantitatively evaluate the performance by calculating the residual sum of squares (RSS) for the two models against the 44 long-period systems. The result shows that our log-log fit achieves a smaller RSS of 0.0582, while the relation from \cite{2024PASP..136b4201P} produces a larger RSS of 0.0611. Their relation, while well-suited for its original short-period sample (\(P < 0.5\) days), shows systematic deviations when applied to the long-period regime (\(P > 0.5\) days). The smaller RSS value indicates that the log-log fit is a more suitable relation for long-period regime.

Therefore, we calculate the semimajor axes of our four long period binary systems using this linear fit in log-log space. The relative radii and semimajor axis are then used to determine \( R_1 \) and \( R_2 \). According to Kepler’s third law, the total mass of the system is given by:
\begin{equation}
M_1 + M_2 = \frac{0.0134 a^3}{P^2}.
\end{equation}
Then we use the mass ratio \( q \) to determine the primary mass \( M_1 \) and the secondary mass \( M_2 \). \( M \), \( a \), and \( P \) are expressed in solar masses (\( M_\odot \)), solar radii (\( R_\odot \)), and days, respectively.
Next, the luminosities \( L_1 \) and \( L_2 \) are derived using the Stefan-Boltzmann law:
\begin{equation}
L = 4\pi\sigma T^4 R^2.
\end{equation}
The absolute parameters of our targets are summarized in Table \ref{Table:Absolute parameters of the four targets}.

\begin{table*}
    \centering
    \caption{Absolute parameters of the four targets by the \( P \)–\( a \) empirical relation method and the Gaia distance method.}\textsuperscript{a}
    \label{Table:Absolute parameters of the four targets}
    \resizebox{\textwidth}{!}{ 
    \begin{tabular}{lcccccccccc}
        \toprule
        Targets & RUWE & $A_v$ & Gaia Distance & $a$ & $M_1$ & $M_2$ & $R_1$ & $R_2$ & $L_1$ & $L_2$ \\
        & & & (kpc) & ($R_\odot$) & ($M_\odot$) & ($M_\odot$) & ($R_\odot$) & ($R_\odot$) & ($L_\odot$) & ($L_\odot$)  \\
        \midrule
        V0508 And(\( P \)–\( a \) empirical relation)    &  --  &  --  &  --  & 4.86(±0.18) & 2.34(±0.26) & 0.22(±0.03) & 2.90(±0.11) & 1.05(±0.07) & 14.52(±1.12) & 1.55(±0.20)  \\
        V0508 And(Gaia distance)                         & 1.26 & 0.13 & 0.98 & 4.78(±0.08) & 2.22(±0.12) & 0.21(±0.01) & 2.85(±0.05) & 0.99(±0.04) & 14.05(±0.48) & 1.47(±0.07)  \\
        V0844 Aur(\( P \)–\( a \) empirical relation)    &  --  &  --  &  --  & 5.13(±0.18) & 2.23(±0.24) & 0.44(±0.05) & 2.76(±0.10) & 1.37(±0.07) & 13.32(±0.99) & 2.37(±0.30)  \\
        V0844 Aur(Gaia distance)                         & 1.02 & 0.76 & 0.91 & 5.96(±0.21) & 3.50(±0.36) & 0.69(±0.08) & 3.21(±0.12) & 1.59(±0.08) & 18.15(±0.47) & 3.04(±0.11)  \\
        V0699 Cep(\( P \)–\( a \) empirical relation)    &  --  &  --  &  --  & 5.03(±0.18) & 2.12(±0.23) & 0.51(±0.06)  & 2.63(±0.10) & 1.42(±0.07) & 12.23(±0.91) & 3.54(±0.36) \\
        V0699 Cep(Gaia distance)                         & 1.78 & 0.82 & 0.75 & 5.91(±0.10) & 3.45(±0.17) & 0.83(±0.05)  & 3.10(±0.06) & 1.67(±0.05) & 16.94(±0.57) & 4.86(±0.27) \\
        NSVS 6259046(\( P \)–\( a \) empirical relation) &  --  &  --  &  --  & 4.63(±0.18) & 2.26(±0.27) & 0.20(±0.03)  & 2.81(±0.12) & 1.02(±0.08) & 14.03(±1.15) & 1.54(±0.23) \\
        NSVS 6259046(Gaia distance)                      & 0.96 & 0.43 & 1.12 & 4.67(±0.08) & 2.32(±0.12) & 0.21(±0.01)  & 2.66(±0.05) & 1.03(±0.05) & 14.33(±0.46) & 1.51(±0.06) \\
        \bottomrule
    \end{tabular}
    }
    \parbox{\textwidth}{\textit{Note.} \textsuperscript{a} Errors were estimated in terms of error transfer formulae.}
\end{table*}

The orbital periods of V0508 And, V0699 Cep, and NSVS 6259046 are increasing continuously, while that of V0844 Aur may remain unchanged at present. As the Applegate mechanism is generally associated with periodic or quasi-periodic $O - C$ variations \citep{1992ApJ...385..621A,2016A&A...587A..34V,2018MNRAS.476.5274L}, and our targets show no evidence of such periodic variations, the observed long-term period change is more consistent with a mass transfer scenario. The long-term increase in orbital period could be caused by mass transfer from the less massive star to the more massive one \citep{2011AJ....142..124Z,2021RAA....21..174K,2022MNRAS.517.1928G,2022AJ....164..202L,2022NewA...9501800G,2023PASJ...75..701M,2024ApJ...961...97Z,2024MNRAS.527.6406L,2025MNRAS.537.2258L}. Based on this assumption, we use the following equation to estimate the rate of mass transfer:
\begin{equation}
\frac{dM_1}{dt} = \frac{M_1 M_2}{3P (M_1 - M_2)} \times \frac{dP}{dt}.
\end{equation}
The final values are displayed in Table \ref{Table:Fitting parameters of $O-C$ diagrams}.

In addition to the \( P \)–\( a \) empirical relation method, there is another approach that utilizes Gaia distances to calculate absolute parameters.

Following the research by \citet{2021AJ....162...13L}, we employ the equation:
\begin{equation}
M_v = m_{v\max} - 5 \log D + 5 - A_v
\end{equation}
to calculate the absolute magnitudes ($ M_v $), where $ A_v $ represents the extinction coefficients, $ m_{v\max} $ denotes the brightest V-band magnitude from the ASAS-SN theoretical curve, and $ D $ is the distance provided by Gaia DR3 \citep{2016A&A...595A...1G,2023A&A...674A...1G}. The bolometric corrections ($ BC_v $) are then derived through interpolation from the table given by \citet{2013ApJS..208....9P}. The absolute bolometric magnitude $ M_{\text{bol}} $ is given by:
\begin{equation}
M_{\text{bol}} = M_v + BC_v.
\end{equation}
From the relation
\begin{equation}
M_{\text{bol}} = -2.5 \log \left( \frac{L_{\text{total}}}{L_{\odot}} \right) + 4.75,
\end{equation}
we derive the luminosities for the primary and secondary components ($ L_{1,2} $). The equation:
\begin{equation}
M_{\text{bol}} = -2.5 \log \left[ (A \times r_1)^2 \left( \frac{T_1}{T_\odot} \right)^4 + (A \times r_2)^2 \left( \frac{T_2}{T_\odot} \right)^4 \right] + 4.75
\end{equation}
is applied to compute the distance ($ A $) between the two components, where $ r_1 $ and $ r_2 $ are the relative radii of the primary and secondary components from the photometric solutions. Using
\begin{equation}
\frac{R_1}{R_\odot} = r_1 \times A \quad \text{and} \quad \frac{R_2}{R_\odot} = r_2 \times A,
\end{equation}
the radii of the primary and secondary components ($ R_{1,2} $) in solar units are determined. Finally, by using Kepler’s third law
\begin{equation}
M_1 + M_2 = 0.0134 \frac{A^3}{P^2},
\end{equation}
where $ P $ is in days and $ A $ is in solar units ($ R_\odot $), the total mass of the system is obtained. The mass ratio from the W-D program is then used to estimate the masses of the primary and secondary ($ M_{1,2} $). The absolute parameters along with their errors are presented in Table \ref{Table:Absolute parameters of the four targets}.

However, to ensure the accuracy of the calculated absolute physical parameters, the RUWE must remain below 1.4 \citep{2024PASP..136b4201P} and $A_v$ should not exceed approximately 0.4 \citep{2024NewA..11002227P}. We have listed the RUWE and $A_v$ of each target in Table \ref{Table:The initial and stability parameters of the four targets}. Due to that V0699 Cep exhibits a RUWE > 1.4 and Av > 0.4, while V0844 Aur also shows Av > 0.4, the absolute physical parameters of V0699 Cep and V0844 Aur calculated using the Gaia distance may be less accurate than those obtained with the \( P \)–\( a \) empirical relation. Meanwhile, the results from the Gaia distance method for V0508 And and NSVS 6259046 are consistent with those obtained from the \( P \)–\( a \) empirical relation method. Therefore, the absolute parameters given by the \( P \)–\( a \) empirical relation are chosen as the final results.

\begin{table*}
    \centering
    \caption{The initial mass, age, orbital angular momentum and stability parameter of the four targets.}
    \label{Table:The initial and stability parameters of the four targets}
    \begin{tabular}{lcccccc}
        \toprule
        Targets & $M_{1i}$ & $M_{2i}$ & $q_{\text{ini}}$ & $t$ & $J_o$ & $\frac{J_s}{J_o}$ \\
        & (M$_\odot$) & (M$_\odot$) & & (Gyr) & (cgs) &  \\
        \midrule
        V0508 And & 1.66 & 2.24 & 1.35 & 2.59 & $4.307 \times 10^{51}$ & 0.152 \\
        V0844 Aur & 1.62 & 2.25 & 1.39 & 2.38 & $8.240 \times 10^{51}$ & 0.077 \\
        V0699 Cep & 1.49 & 2.40 & 1.62 & 1.91 & $9.024 \times 10^{51}$ & 0.067 \\
        NSVS 6259046 & 1.57 & 2.26 & 1.43 & 2.56 & $3.809 \times 10^{51}$ & 0.162 \\
        \bottomrule
    \end{tabular}
\end{table*}

\subsection{Initial Mass and Age}
We determine the initial mass and mass ratio of each component using the equations proposed by \cite{2013MNRAS.430.2029Y}:
\begin{equation}
    M_{2i} = M_2 + \Delta M = M_2 + 2.50(M_L - M_2 - 0.07)^{0.64},
\end{equation}
where $M_2$ denotes the present mass of the secondary star, while $M_L$ is obtained from:
\begin{equation}
    M_L = \left(\frac{L_2}{1.49}\right)^{1/4.216}.
\end{equation}
The initial mass of the primary star is derived as:
\begin{equation}
    M_{1i} = M_1 - (\Delta M - M_{\text{lost}}) = M_1 - \Delta M (1 - \gamma),
\end{equation}
where $M_1$ represents the current mass of the primary, and the total mass lost by the system is denoted as $M_{\text{lost}}$. The parameter $\gamma$ is defined by:
\begin{equation}
    \gamma = \frac{M_{\text{lost}}}{\Delta M}.
\end{equation}
$\gamma = 0$ corresponds to the conservative case and $\gamma = 1$ corresponds to the case that the mass loss by the primary star is 0. In our calculations, we adopt $\gamma = 0.664$, following \cite{2013MNRAS.430.2029Y}.
\begin{equation}
    q_{\text{ini}} = \frac{M_{2i}}{M_{1i}},
\end{equation}
Here, $M_{1i}$ and $M_{2i}$ are the present masses of the primary and secondary stars, respectively, while $q_{\text{ini}}$ represents their initial mass ratio.
Table \ref{Table:The initial and stability parameters of the four targets} reveals that unlike the current situation, the secondary star had a much higher mass than the primary initially. It evolved faster and expanded beyond its Roche lobe, it began to lose mass to its companion \citep{2013MNRAS.430.2029Y}. This mass transfer caused the two stars to switch roles. Later, both stars grew large enough to go beyond their Roche lobes and formed a shared envelope \citep{2013MNRAS.430.2029Y}.
\cite{2013MNRAS.430.2029Y} also found that A-subtype CBs typically have initial secondary masses in the range of $1.7 M_\odot < M_{2i} < 2.6 M_\odot$, and initial primary masses in the range of $0.3 M_\odot < M_{1i} < 1.5 M_\odot$, assuming $\gamma = 0.664$. The initial masses derived for our targets align with these values.

To estimate the system's age, we apply the age determination model from \cite{2014MNRAS.437..185Y}:
\begin{equation}
    \overline{M_2} = \frac{M_{2i} + M_L}{2},
\end{equation}
\begin{equation}
    t_{\text{MS}} = 10^{\left( \frac{M}{M_\odot} \right)^{-4.05}} \times \left( 5.60 \times 10^{-3} \frac{M}{M_\odot} + 3.993 \right)^{3.16} + 0.042,
\end{equation}
\begin{equation}
    t \approx t_{\text{MS}}(M_{2i}) + t_{\text{MS}}(M_2),
\end{equation}
where $t_{\text{MS}}$ represents the main sequence lifetime of a star.
The computed results are listed in Table \ref{Table:The initial and stability parameters of the four targets}. Additionally, \cite{2014MNRAS.437..185Y} established the mean age of A-subtype contact binaries as:
\begin{equation}
    \text{age} = 4.37 \pm 1.23 \quad \text{Gyr}.
\end{equation}
Approximately, the ages of our targets fall between 2 and 3 Gyrs, shorter than the mean age, as our targets are long-period CBs, which are more massive and evolve at a faster rate compared to other short period CBs. The notable decrease in mass ratio over a relatively short time indicates extensive mass transfer throughout their evolution.

\subsection{Evolutionary Status and Stability}
\begin{figure*}
\centering  
\subfigure{
\includegraphics[width=8.5cm,height = 6.75cm]{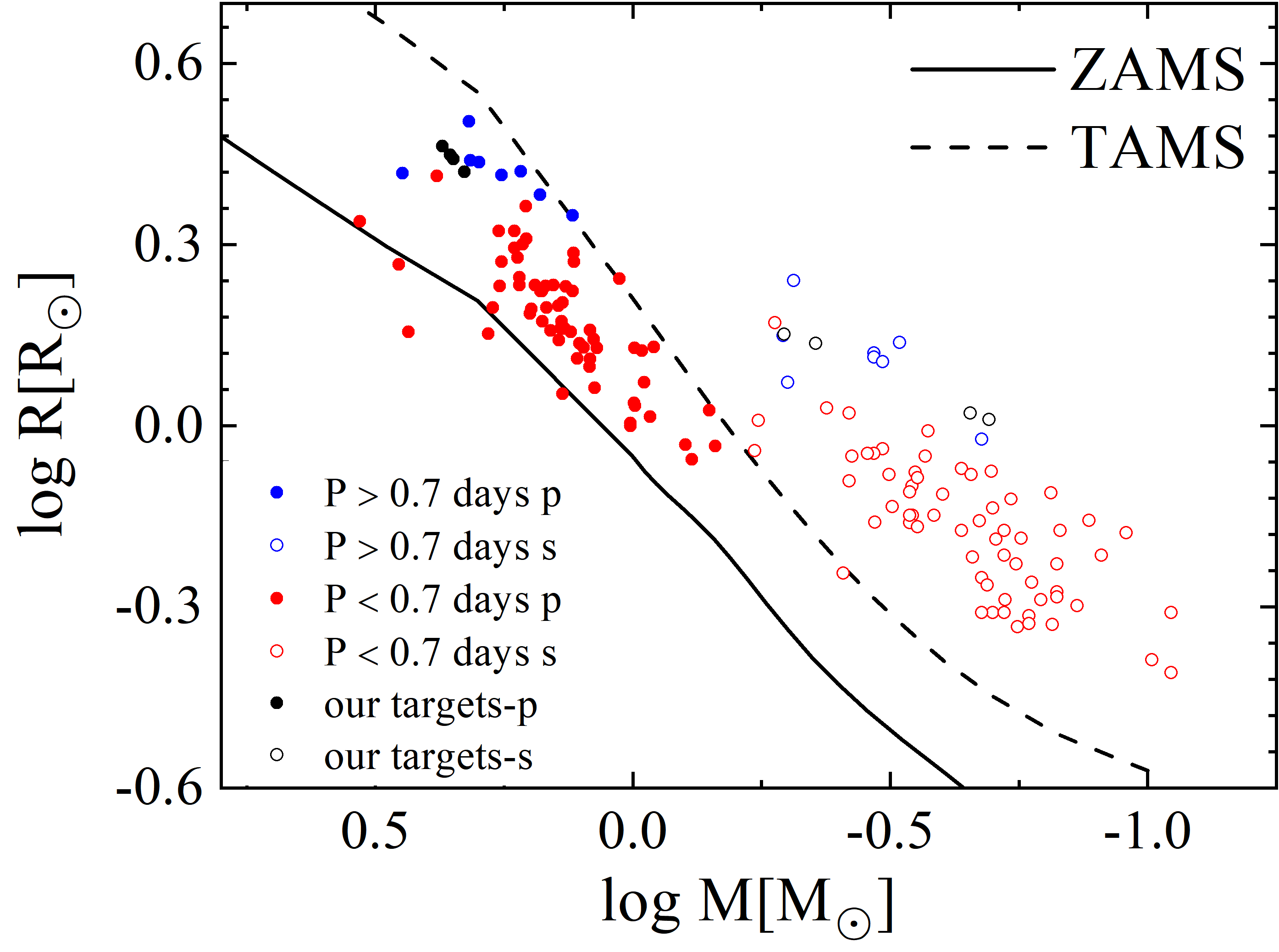}}\subfigure{
\includegraphics[width=8.5cm,height = 6.75cm]{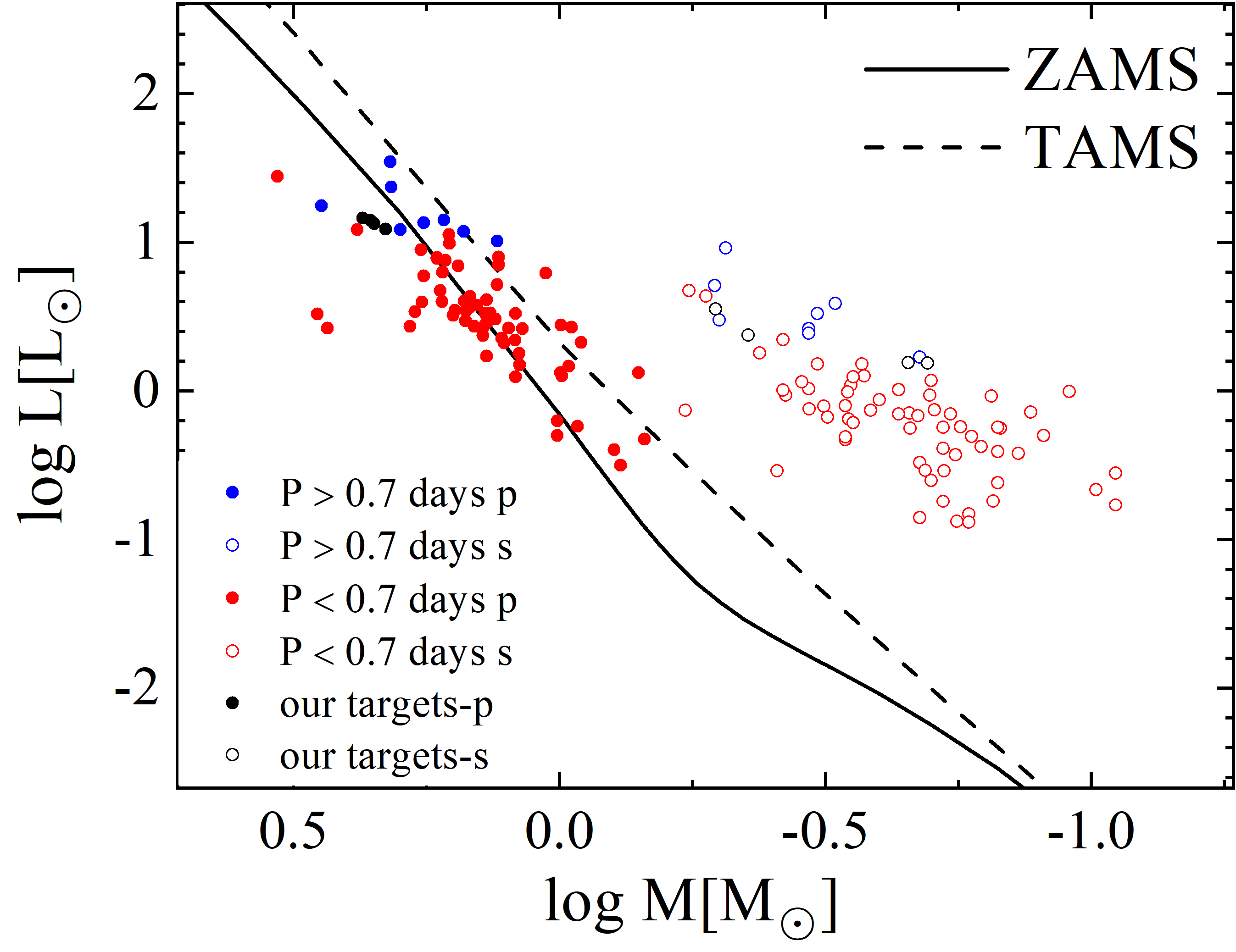}}
\caption{The \( M \)–\( R \) diagram and the \( M \)–\( L \) diagram. Solid dots represent the primary stars, while hollow circles correspond to the secondary ones. Blue represents systems with period longer than 0.7 days, red represents systems with period lower than 0.7 days, black represents our targets. The solid and dashed lines indicate the zero-age main sequence (ZAMS) and terminal-age main sequence (TAMS), respectively, as modeled by \protect\cite{2002MNRAS.329..897H}.}
\label{Figure:The M-R diagram and the M-L diagram}
\end{figure*}

With the parameters displayed in Table \ref{Table:The 217 CBs with Spectroscopic and Photometric Observations}, we plot systems with mass ratio lower than 0.25 and our four low-mass ratio systems in the mass–luminosity (\( M \)–\( L \)) and mass–radius (\( M \)–\( R \)) diagrams, which are shown in Fig. \ref{Figure:The M-R diagram and the M-L diagram}. We also plot the zero-age main sequence (ZAMS) and terminal-age main sequence (TAMS) modeled by \cite{2002MNRAS.329..897H} in the diagrams, respectively. All primary stars are clustered along the main sequence, positioned between ZAMS and TAMS in the diagram. In contrast, the lower-mass secondary stars appear oversized and overluminosity, placing them above the TAMS.

There exists a clear and consistent difference between the observed properties of their secondary stars and what we would expect from single stars. This key characteristic was first identified by \cite{1941ApJ....93..133K}.
Two main explanations have been proposed to account for this difference. First is the possibility that the secondary star is more evolved than the primary, causing it to expand beyond its main-sequence radius, which was the assumption underlying the work of \cite{2013MNRAS.430.2029Y}. Second is the possibility that the secondary is not evolved, but that the transfer of energy from the primary has altered its radius. This is an idea that goes back to \cite{1941ApJ....93..133K} and which was the assumption underlying the models of minimum mass ratio in \cite{2022ApJS..262...12K}.

Given the same mass, long-period systems exhibit greater radii and luminosities in both components than short-period systems, implicating that the longer period primaries must be a little more evolved than their short period counterparts.

\begin{figure}
 \includegraphics[width=\columnwidth]{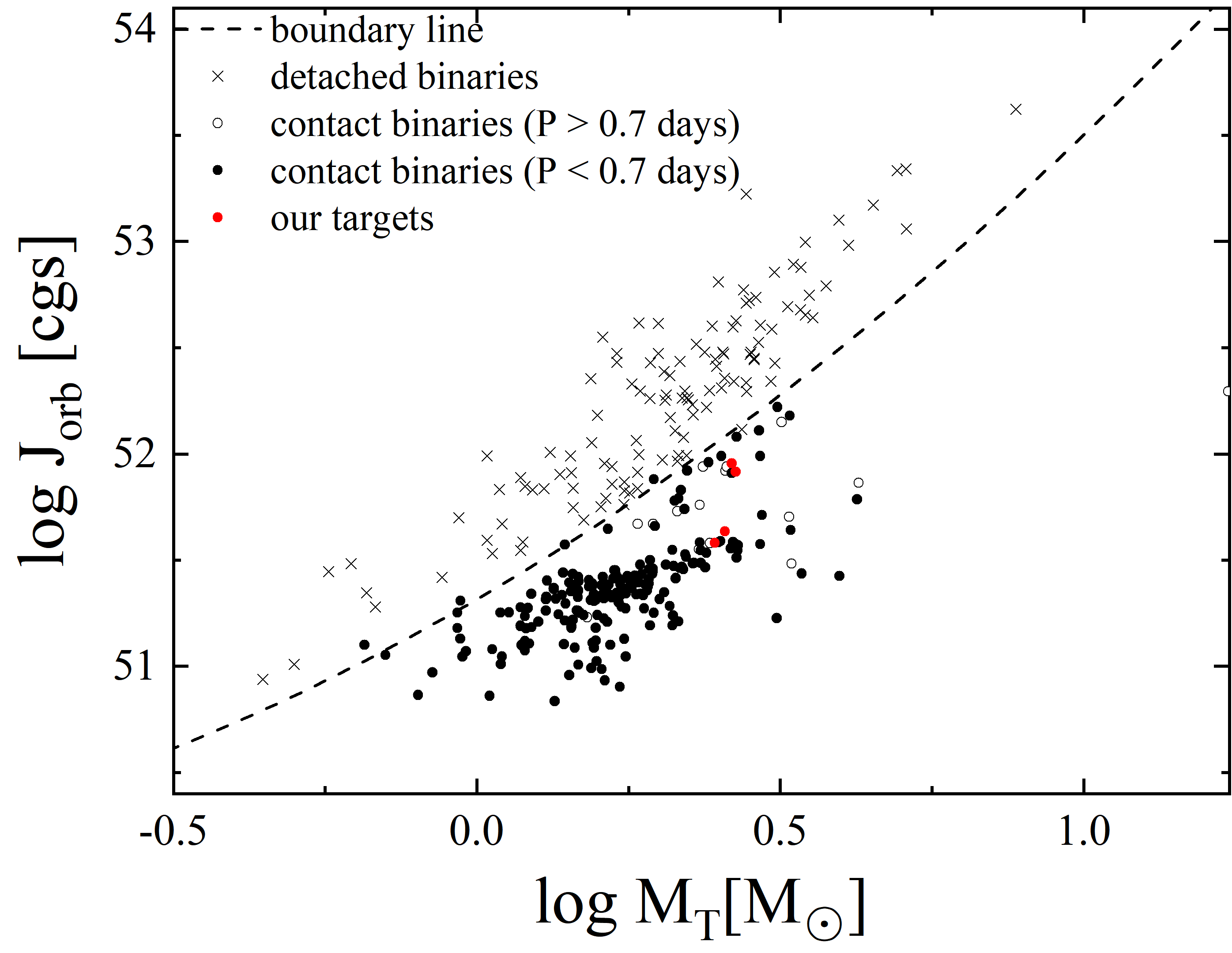}
 \caption{Relationship between orbital angular momentum and total mass for binary systems. The boundary ($J_{\text{lim}}$) and the detached binaries, shown as gray crosses, are sourced from \protect\cite{2006MNRAS.373.1483E}. Hollow circles represent CBs with periods longer than 0.7 days, while solid circles signify CBs with periods lower than 0.7 days, which are displayed on Table \ref{Table:The 217 CBs with Spectroscopic and Photometric Observations}. Solid red circles represent the four targets analyzed in this study.}
 \label{Figure:Relationship between orbital angular momentum and total mass for binary systems}
\end{figure}

To better understand how CBs form and evolve, we calculate their the orbital angular momentum. 
With the binary system’s total mass $M_T$ (expressed in solar masses), period $P$ (measured in days), and mass ratio $q$, the orbital angular momentum $J_o$ of the selected contact binaries is determined. This calculation follows the equation derived by \cite{2013AJ....146..157C}:
\begin{equation}
    J_o = 1.24 \times 10^{52} \times M_T^{3/5} \times P^{1/3} \times q \times (1+q)^{-2}.
\end{equation}
We visualize the relationship between orbital angular momentum ($J_o$) and system’s total mass in Fig. \ref{Figure:Relationship between orbital angular momentum and total mass for binary systems}. This figure encompasses both detached and contact binary systems and a boundary separating them \citep{2006MNRAS.373.1483E}. The vast majority of contact binary systems are positioned below this boundary, while detached binaries are found above it.
This distribution implies that the values of $J_o$ for contact binaries are generally lower than those for detached binaries of equal mass. The transition from detached to contact binaries likely occurs through angular momentum loss mechanisms, particularly via stellar winds, contributing to shorter orbital periods \citep{1966AnAp...29..331H,2007AJ....134.1769Q,2015ApJ...798L..42Q,2019MNRAS.485.4588L}.

According to Darwin’s instability theory \citep{1980A&A....92..167H}, the secondary component in a binary system is unable to maintain synchronous rotation with the primary due to tidal effects. As a result, the system may progress toward a merger when $\frac{J_{\text{spin}}}{J_{\text{orb}}} \geq \frac{1}{3}$, where $J_{\text{spin}}$ represents spin angular momentum, and $J_{\text{orb}}$ refers to orbital angular momentum. To study the stability of CBs, we need to calculate the ratio of spin angular momentum to orbital angular momentum ($J_s/J_o$), using the following equation \citep{2015AJ....150...69Y}: 
\begin{equation}
    \frac{J_s}{J_o} = \frac{q}{(1+q)} \left( k_1^2 r_1^2 + k_2^2 r_2^2 q \right),
\end{equation}
where $r_1$ and $r_2$ denote the relative radii of the primary and secondary components, while $k_1^2$ and $k_2^2$ represent the dimensionless gyration radii of the two stars.

\cite{1995ApJ...444L..41R} showed that assuming \(k_1^2 = 0.06\) leads to a minimum mass ratio of \(q \approx 0.09\) for stability. However, the actual values of \(k_1^2\) and \(k_2^2\) depend on the internal structure of the primary star. For our sample, we estimate \(k_1\) using the relations from \cite{2009A&A...494..209L} for rotating and tidally distorted ZAMS stars: \(k_1 = 0.5391 - 0.2504 \times M_1\) for \(0.4M_{\odot} < M_1 < 1.4M_{\odot}\), and \(k_1 = 0.1523 + 0.0141 \times M_1\) for \(M_1 > 1.4M_{\odot}\). For the secondary component, we adopt \(k_2^2 = 0.205\) from \cite{2007MNRAS.377.1635A}, as the secondary component has a very low mass
and is therefore likely to be fully convective.

Using these theoretical estimates, we compute the angular momentum ratio \(J_s/J_o\) for our four targets and list in Table \ref{Table:The initial and stability parameters of the four targets}. All values are found to be below the stability limit of \(1/3\) \citep{1980A&A....92..167H}, confirming that these systems are currently stable.

In conclusion, photometric and spectroscopic studies of four CBs with period longer than 0.7 days, are performed. Three targets exhibit O'Connell effect. Obvious chromospheric activity is found on V0699 Cep. $O - C$ analysis shows no long-term variation in V0844 Aur and the increasing orbital periods in other three targets. Their absolute parameters, initial masses and ages are calculated. Their evolutionary status and stability are discussed. 217 CBs with both reliable radial velocity and photometric observations are collected. Differences in \( M \)–\( R \) and \( M \)–\( L \) relations between long-period and short-period systems are investigated, with 0.7 days serving as the main boundary between the two groups.

\section*{Acknowledgements}
Thanks the anonymous referee very much for the constructive and insightful criticisms and suggestions to improve our manuscript.
This work is supported by the National Natural Science Foundation of China (NSFC) (Nos. 12273018, 12573033), the Joint Research Fund in Astronomy (No. U1931103) under cooperative agreement between NSFC and Chinese Academy of Sciences (CAS), the Shandong Provincial Natural Science Foundation of China (Grants ZR2021QA079, ZR2025MS81), the Taishan Scholars Young Expert Program of Shandong Province, the Qilu Young Researcher Project of Shandong University, the Young Data Scientist Project of the National Astronomical Data Center, the Cultivation Project for LAMOST Scientific Payoff and Research Achievement of CAMS-CAS. The calculations in this work were carried out at Supercomputing Center of Shandong University, Weihai.

The spectral data were provided by Guoshoujing Telescope (the Large Sky Area Multi-Object Fiber Spectroscopic Telescope; LAMOST), which is a National Major Scientific Project built by the Chinese Academy of Sciences. Funding for the project has been provided by the National Development and Reform Commission. LAMOST is operated and managed by the National Astronomical Observatories, Chinese Academy of Sciences.

Based on observations obtained with the Samuel Oschin 48-inch Telescope at the Palomar Observatory as part of the Zwicky Transient Facility project. ZTF is supported by the National Science Foundation under grant No. AST-1440341 and a collaboration including Caltech, IPAC, the Weizmann Institute for Science, the Oskar Klein Center at Stockholm University, the University of Maryland, the University of Washington, Deutsches Elektronen-Synchrotron and Humboldt University, Los Alamos National Laboratories, the TANGO Consortium of Taiwan, the University of Wisconsin at Milwaukee, and Lawrence Berkeley National Laboratories. Operations are conducted by COO, IPAC, and UW.

This paper makes use of data from the DR1 of the WASP data (\cite{2010A&A...520L..10B}) as provided by the WASP consortium, and the computing and storage facilities at the CERIT Scientific Cloud, reg. no. CZ.1.05/3.2.00/08.0144, which is operated by Masaryk University, Czech Republic.

This work includes data collected by the TESS mission. Funding for the TESS mission is provided by NASA Science Mission Directorate. We acknowledge the TESS team for its support of this work.

We thank Las Cumbres Observatory and its staff for their continued support of ASAS-SN. ASAS-SN is funded in part by the Gordon and Betty Moore Foundation through grants GBMF5490 and GBMF10501 to the Ohio State University, and also funded in part by the Alfred P. Sloan Foundation grant G-2021-14192.

This work has made use of data from the European Space Agency (ESA) mission Gaia (\href{https://www.cosmos.esa.int/gaia}{https://www.cosmos.esa.int/gaia}), processed by the Gaia Data Processing and Analysis Consortium (DPAC; \href{https://www.cosmos.esa.int/web/gaia/dpac/consortium}{https://www.cosmos.esa.int/web/gaia/dpac/consortium}). Funding for the DPAC has been provided by national institutions, in particular the institutions participating in the Gaia Multilateral Agreement.

\section*{Data Availability}
Here is the data link:

ASAS-SN data is available at

\href{https://asas-sn.osu.edu/variables/lookup}{https://asas-sn.osu.edu/variables/lookup}. 

CRTS data is available at 

\href{http://crts.caltech.edu}{http://crts.caltech.edu}.

SuperWASP data are available at 

\href{https://wasp.cerit-sc.cz/search}{https://wasp.cerit-sc.cz/search}.

TESS data is available at 

\href{http://archive.stsci.edu/tess/bulk_downloads.html}{http://archive.stsci.edu/tess/bulk\_downloads.html}. 

ZTF data is available at 

\href{https://irsa.ipac.caltech.edu/cgi-bin/ZTF}{https://irsa.ipac.caltech.edu/cgi-bin/ZTF}.

\begin{appendix}
\section{The Physical Parameters of 217 CBs}
Table \ref{Table:The 217 CBs with Spectroscopic and Photometric Observations}. The 217 CBs with Spectroscopic and Photometric Observations is listed in the appendix.

Supplementary data are available at \href{https://academic.oup.com/mnras}{\textit{MNRAS}} online.

Table \ref{Table:The photometric observation obtained by NEXT}. The photometric observation obtained by NEXT.

Table \ref{Table:The Eclipsing times and O – C values of the four targets}. The Eclipsing times and $O - C$ values of the four targets.

\begin{table*}
\centering
\caption{The 217 CBs with Spectroscopic and Photometric Observations.}
\label{Table:The 217 CBs with Spectroscopic and Photometric Observations}
\resizebox{\textwidth}{!}{ 
\begin{tabular}{lccccccccccccccccccc}
    \toprule
    System & Type & $P$ (days) & $V_{\max}$ (mag) & $q_{\mathrm{sp}}$ & $q_{\mathrm{ph}}$ & $i$ (°) & $a$ ($R_{\odot}$) & $f$ (\%) & $T_1$ (K) & $T_2$ (K) & $M_1$ ($M_{\odot}$) & $M_2$ ($M_{\odot}$) & $R_1$ ($R_{\odot}$) & $R_2$ ($R_{\odot}$) & $L_1$ ($L_{\odot}$) & $L_2$ ($L_{\odot}$) & $J_{\mathrm{spin}}/J_{\mathrm{orb}}$ & $\log J_o$ (cgs) & Ref. \\
    \midrule
LY Aur                       & W    & 4.0020   & 6.68       & 0.528 & 0.528 & 88    & 36.10   & --      & 31000  & 31150  & 25.50   & 14.00   & 16.10   & 12.60   & 213796  & 134896  & 0.1890   & 54.31       & (1),(2)             \\
V382 Cyg                     & A    & 1.8855   & 8.33       & 0.742 & 0.677 & 85    & 23.45   & 9       & 36000  & 34415  & 27.90   & 20.80   & 9.70    & 8.50    & 141905  & 89949   & 0.1665   & 54.38       & (3)                 \\
SV Cen                       & W    & 1.6581   & 8.71       & 0.707 & --    & 82    & 15.30   & 92      & 14050  & 23000  & 9.60    & 7.70    & 7.40    & 6.80    & 11700   & 1900    & 0.1159   & 52.29       & (4),(5)             \\
V606 Cen                     & A    & 1.4951   & 9.35       & 0.527 & 0.541 & 87    & 15.58   & 4       & 29200  & 21770  & 14.70   & 7.96    & 6.83    & 5.19    & 30400   & 5440    & 0.1067   & 53.77       & (6)                 \\
TU Mus                       & A    & 1.3870   & 8.25       & 0.678 & 0.652 & 78    & 17.70   & 5       & 38700  & 33200  & 16.70   & 10.40   & 7.20    & 5.70    & 104712  & 35481   & 0.0973   & 53.91       & (7)                 \\
ATO J343.6450+34.6956        & A    & 1.1316   & --         & 0.632 & --    & 37    & 7.41    & 11      & 6862   & 6604   & 2.60    & 1.66    & 3.18    & 2.59    & 20.08   & 11.48   & 0.0580   & 51.86       & (8)                 \\
\bottomrule
    \end{tabular}
}
\end{table*}

\addtocounter{table}{-1}
\begin{table*}
\centering
\caption{The 217 CBs with Spectroscopic and Photometric Observations. (Continued).}
\resizebox{\textwidth}{!}{ 
    \begin{tabular}{lccccccccccccccccccc}
    \toprule
    System & Type & $P$ (days) & $V_{\max}$ (mag) & $q_{\mathrm{sp}}$ & $q_{\mathrm{ph}}$ & $i$ (°) & $a$ ($R_{\odot}$) & $f$ (\%) & $T_1$ (K) & $T_2$ (K) & $M_1$ ($M_{\odot}$) & $M_2$ ($M_{\odot}$) & $R_1$ ($R_{\odot}$) & $R_2$ ($R_{\odot}$) & $L_1$ ($L_{\odot}$) & $L_2$ ($L_{\odot}$) & $J_{\mathrm{spin}}/J_{\mathrm{orb}}$ & $\log J_o$ (cgs) & Ref. \\
    \midrule
FI Boo                       & W    & 0.3900   & --         & 0.373 & --    & 38    & 2.55    & 49      & 5420   & 5746   & 1.07    & 0.40    & 1.28    & 0.85    & 1.27    & 0.71    & 0.0995   & 51.35       & (109)               \\
TYC 1174-344-1               & A    & 0.3887   & 11.03      & 0.187 & --    & 88    & 2.64    & 52      & 6500   & 6357   & 1.38    & 0.26    & 1.45    & 0.71    & 3.31    & 0.74    & 0.0893   & 51.21       & (110)               \\
AU Ser                       & A    & 0.3865   & --         & 0.692 & --    & 82    & 2.52    & 4       & 5140   & 4986   & 0.85    & 0.59    & 1.04    & 0.88    & 0.68    & 0.43    & 0.0866   & 51.43       & (111)               \\
BI CVn                       & W    & 0.3842   & --         & 0.413 & --    & 71    & 2.92    & 31      & 6093   & 6170   & 1.61    & 0.66    & 1.40    & 0.96    & 2.44    & 1.19    & 0.0554   & 51.48       & (112)               \\
V546 And                     & W    & 0.3830   & --         & 0.254 & --    & 86    & 2.46    & 30      & 6240   & 6517   & 1.08    & 0.28    & 1.23    & 0.66    & 2.05    & 0.71    & 0.1071   & 51.24       & (113)               \\
TX Cnc                       & W    & 0.3829   & --         & 0.455 & --    & 62    & 2.76    & 17      & 6121   & 6250   & 1.32    & 0.60    & 1.27    & 0.89    & 2.03    & 1.09    & 0.0604   & 51.46       & (40)                \\
\bottomrule
    \end{tabular}
}
\end{table*}

\FloatBarrier
\afterpage{
\begin{tablenotes}   
        \footnotesize               	  
        \item[] References. (1)\cite{2014NewA...26..112Z},
(2)\cite{1994Obs...114..107S},
(3)\cite{2013AJ....145....9Y},
(4)\cite{1992AJ....103..960R},
(5)\cite{1982A&A...110..246D},
(6)\cite{2022ApJ...924...30L},
(7)\cite{2003AJ....126.2988T},
(8)\cite{2025ApJ...979...69W},
(9)\cite{2013AJ....145...80D},
(10)\cite{2011AN....332..607P},
(11)\cite{2014PASP..126..121C},
(12)\cite{2003AJ....125.3258R},
(13)\cite{2006AcA....56..127G},
(14)\cite{2016AJ....151...67Z},
(15)\cite{2007AJ....133..169D},
(16)\cite{2018AJ....156..199S},
(17)\cite{2002AJ....124.1738R},
(18)\cite{1998A&A...336..920R},
(19)\cite{2004A&A...417..725Y},
(20)\cite{2005ApJ...629.1055Y},
(21)\cite{2015PASP..127..742A},
(22)\cite{2011NewA...16...12C},
(23)\cite{2006AJ....132..769P},
(24)\cite{2018RAA....18...20T},
(25)\cite{2019AJ....157..111Y},
(26)\cite{1987MNRAS.226..899B},
(27)\cite{2010AcA....60..305S},
(28)\cite{2013NewA...22...57L},
(29)\cite{2005AcA....55..389Z},
(30)\cite{2000A&A...356..603C},
(31)\cite{2002IBVS.5258....1P},
(32)\cite{2001AJ....122.1974R},
(33)\cite{2011RAA....11.1158Y},
(34)\cite{2005AJ....130..767R},
(35)\cite{2005AJ....129.2806Z},
(36)\cite{2015PASJ...67...98Z},
(37)\cite{2008AJ....136..586R},
(38)\cite{2018PASP..130c4201L},
(39)\cite{2014AJ....147...91L},
(40)\cite{2011MNRAS.412.1787D},
(41)\cite{2012NewA...17..143O},
(42)\cite{2021SerAJ.203...29P},
(43)\cite{2004AcA....54..299Z},
(44)\cite{2005PASJ...57..983Y},
(45)\cite{2011AJ....141..207P},
(46)\cite{2009AJ....137.3655P},
(47)\cite{2014AJ....148..126C},
(48)\cite{2015AJ....149..168P},
(49)\cite{1986MNRAS.223..581H},
(50)\cite{2001CoSka..31....5C},
(51)\cite{2005NewA...10..163A},
(52)\cite{2003A&A...412..465K},
(53)\cite{1993A&A...278..463G},
(54)\cite{2000A&AS..147..243W},
(55)\cite{2013PASJ...65....1P},
(56)\cite{2012NewA...17..603E},
(57)\cite{2022MNRAS.512.1244C},
(58)\cite{2012NewA...17..673Z},
(59)\cite{2016RAA....16....2L},
(60)\cite{2004AcA....54..195B},
(61)\cite{2005AcA....55..123G},
(62)\cite{2015Ap&SS.357...59P},
(63)\cite{2004Ap&SS.291...21W},
(64)\cite{2020MNRAS.493.1565D},
(65)\cite{2016NewA...46...73E},
(66)\cite{2015AJ....149..194L},
(67)\cite{2011PASP..123...34L},
(68)\cite{2019RAA....19...10Y},
(69)\cite{2017ApJ...840....1M},
(70)\cite{2009NewA...14..461O},
(71)\cite{2016PASJ...68..102X},
(72)\cite{2004PASP..116..826Y},
(73)\cite{2010JASS...27...69O},
(74)\cite{2008NewA...13..468S},
(75)\cite{2005AJ....130..224Q},
(76)\cite{2018Ap&SS.363...34S},
(77)\cite{2016ApJ...817..133Z},
(78)\cite{2016PASA...33...43S},
(79)\cite{2014NewA...31...56U},
(80)\cite{2004AJ....127.1712P},
(81)\cite{1999IBVS.4702....1G},
(82)\cite{2005ARBl...20..193G},
(83)\cite{2009AJ....137.3646P},
(84)\cite{2006IBVS.5687....1T},
(85)\cite{2017ASPC..510..372H},
(86)\cite{2012RAA....12..419Y},
(87)\cite{1994A&A...289..827G},
(88)\cite{2003AJ....126.1555K},
(89)\cite{2010A&A...514A..36C},
(90)\cite{2002A&A...387..240O},
(91)\cite{2011AN....332..690G},
(92)\cite{2009NewA...14..321E},
(93)\cite{1996A&AS..118..453L},
(94)\cite{1988AJ.....95..894Y},
(95)\cite{2003AJ....126.1960Y},
(96)\cite{2006LPIBu.107...13D},
(97)\cite{2016NewA...47...57G},
(98)\cite{2011A&A...525A..66D},
(99)\cite{2019AJ....158..186K},
(100)\cite{2018RAA....18...46K},
(101)\cite{2010MNRAS.408..464Z},
(102)\cite{2010A&A...519A..78Z},
(103)\cite{1993AJ....106..361L},
(104)\cite{2009Ap&SS.321...19L},
(105)\cite{2019PASP..131h4202L},
(106)\cite{2025AJ....169..139X},
(107)\cite{2007A&A...465..943S},
(108)\cite{2018MNRAS.479.3197A},
(109)\cite{2013AJ....146..157C},
(110)\cite{2011NewA...16..242G},
(111)\cite{2018IBVS.6256....1A},
(112)\cite{2014NewA...29...57N},
(113)\cite{2015NewA...36..100G},
(114)\cite{2007ASPC..362...82O},
(115)\cite{2015IBVS.6134....1N},
(116)\cite{2017AJ....154..260P},
(117)\cite{2005Ap&SS.296..305S},
(118)\cite{2018AJ....156...77K},
(119)\cite{2021MNRAS.506.4251Z},
(120)\cite{2019AJ....157...73K},
(121)\cite{2015NewA...34..271Y},
(122)\cite{2001A&A...367..840D},
(123)\cite{2013NewA...21...46L},
(124)\cite{1993ApJ...407..237B},
(125)\cite{1976PASP...88..936L},
(126)\cite{2007AJ....133..255L},
(127)\cite{2017AJ....154...99L},
(128)\cite{2013AJ....146...35Y},
(129)\cite{2016AstBu..71...64G},
(130)\cite{2004MNRAS.348.1321B},
(131)\cite{2009AJ....138.1465H},
(132)\cite{2020NewA...7801354K},
(133)\cite{2019AJ....157..207L},
(134)\cite{2006NewA...12..192E},
(135)\cite{2013SerAJ.186...47C},
(136)\cite{2015AJ....149..120L},
(137)\cite{2010IBVS.5951....1N},
(138)\cite{1986AJ.....92..666K},
(139)\cite{2018NewA...62...46K},
(140)\cite{2016Ap&SS.361...63L},
(141)\cite{1990A&A...231..365L},
(142)\cite{2015AJ....149...62X},
(143)\cite{1984A&AS...58..405M},
(144)\cite{2011AJ....141..147L},
(145)\cite{2015NewA...34..262H},
(146)\cite{2019IBVS.6266....1N},
(147)\cite{1991ApJ...374..307L},
(148)\cite{2014NewA...30...64L},
(149)\cite{2018AN....339..472K},
(150)\cite{1999A&AS..136..139Y},
(151)\cite{2015NewA...41...26G},
(152)\cite{2015NewA...39....9G},
(153)\cite{2020MNRAS.497.3381G},
(154)\cite{2014NewA...31...14O},
(155)\cite{2013AJ....145...39Z},
(156)\cite{2012PASJ...64...48L},
(157)\cite{2001CoSka..31..129V},
(158)\cite{2018AcA....68..159A},
(159)\cite{2011AJ....142...99C},
(160)\cite{1995AcA....45..753K},
(161)\cite{2001MNRAS.328..635Q},
(162)\cite{2019PASJ...71...34S},
(163)\cite{2016NewA...46...31G},
(164)\cite{2015NewA...41...17L},
(165)\cite{2018A&A...612A..91M},
(166)\cite{2021MNRAS.501.2897G},
(167)\cite{2009AJ....138..540Y},
(168)\cite{2019NewA...68...20K},
(169)\cite{1989A&A...211...81H},
(170)\cite{2001AJ....122..402L},
(171)\cite{2013RAA....13.1330K},
(172)\cite{2020MNRAS.497.3493Z},
(173)\cite{1995ApJ...455..300H},
(174)\cite{2019RAA....19...97Y},
(175)\cite{2014A&A...563A..34L},
(176)\cite{2019NewA...68...51M},
(177)\cite{2004AJ....128.2997S},
(178)\cite{2004MNRAS.347..307Z},
(179)\cite{2015A&A...578A.103L},
(180)\cite{2011AN....332..626K},
(181)\cite{2010NewA...15..155Y}.
      \end{tablenotes}         
}

\FloatBarrier

\end{appendix}

\bibliographystyle{mnras}
\bibliography{example} 





\bsp	
\label{lastpage}
\end{document}